\documentstyle[prb,aps,psfig,special]{revtex}

\newcommand{\be}{\begin{equation}}
\newcommand{\ee}{\end{equation}}
\newcommand{\bea}{\begin{eqnarray}}
\newcommand{\eea}{\end{eqnarray}}

\newcommand{\lsim}{\stackrel{\scriptscriptstyle 
<}{\scriptscriptstyle \sim}}
\newcommand{\gsim}{\stackrel{\scriptscriptstyle
>}{\scriptscriptstyle \sim}}
\newcommand{\toh}{{\textstyle {1\over 2}}}
\newcommand{\half}{{\textstyle {1\over 2}}}

\newcommand{\Eq}[1]{Eq.~(\ref{#1})}
\newcommand{\Eqs}[1]{Eqs.~(\ref{#1})}
\newcommand{\sect}[1]{section~\ref{#1}}
\newcommand{\fig}[1]{Fig.~\ref{#1}}
\newcommand{\sfrac}[2]{\mbox{$\textstyle {#1 \over #2}$}}
\newcommand{\sss}{\scriptscriptstyle }
\newcommand{\kb}{k_{\scriptscriptstyle B}}
\newcommand{\Tk}{T_{\scriptscriptstyle K}}
\newcommand{\Vk}{V_{\scriptscriptstyle K}}
\newcommand{\ve}{\varepsilon}
\newcommand{\vf}{v_{\scriptscriptstyle F}}

\newcommand{\sL}{{\scriptscriptstyle L}}
\newcommand{\sR}{{\scriptscriptstyle R}}

\newcommand{\sRL}{{\scriptscriptstyle RL}}
\newcommand{\lk}{\lambda_{\scriptscriptstyle K}}

\newcommand{\Ue}{\overline U^{\scriptscriptstyle (e)}}
\newcommand{\Uo}{\overline U^{\scriptscriptstyle (o)}}
\newcommand{\J}{{\cal J}}
\newcommand{\lpsi}{\overline{\psi}}
\newcommand{\uv}{\underline v}
\newcommand{\balpha}{\mbox{\boldmath $\alpha$}}
\newcommand{\bbeta}{\mbox{\boldmath $\beta$}}
\newcommand{\nfl}{non-Fermi-liquid}
\newcommand{\nflr}{non-Fermi-liquid regime}

\newcommand{\SS}{scattering states}
\newcommand{\SA}{scattering amplitudes}

\begin{document} 

%\include{labelsI}
%\include{labelsIII}

%\columnsep0.4truecm

%\flushbottom
\draft
%\preprint{cond-mat/9702049, http://xxx.lanl.gov/ps/cond-mat/9702049}

\title{The 2-Channel Kondo Model II: CFT Calculation of Non-Equilibrium
Conductance through a Nanoconstriction containing 2-Channel
Kondo Impurities}

\author{Jan von Delft${}^{1,}$\cite{newadd}, 
A. W. W. Ludwig${}^2$, Vinay Ambegaokar${}^1$}
\address{
%${}^1$Institut f\"ur Theoretische Festk\"orperphysik, 
%Universit\"at Karlsruhe, 76128 Karlsruhe, Germany \\
${}^1$Laboratory of Atomic and Solid State Physics, Cornell University, 
Ithaca, NY 14853, USA \\
${}^2$ University of California, Santa Barbara, CA 93106 , USA
}

\date{September 9, 1998. Published in Ann.\ Phys.\
{\bf 273}, 175-241 (1999).}
\maketitle

%\begin{center}
%\today
%\end{center}

\begin{abstract}
Recent experiments by Ralph and Buhrman  on zero-bias
anomalies in quenched Cu nanoconstrictions
(reviewed in a preceding paper, I)
are in accord with the assumption that   the interaction
between electrons  and nearly degenerate  two-level systems
in the constriction can be described,
for sufficiently small voltages and temperatures ($V,T < \Tk$),
by the 2-channel Kondo (2CK) model. Motivated by these experiments, 
we introduce a  generalization of the 2CK model, which we call 
the nanoconstriction 2-channel Kondo model (NTKM),
that takes into account the complications arising from the non-equilibrium
electron distribution in the nanoconstriction.
We calculate the conductance $G(V,T)$ 
of the constriction in the weakly non-equilibrium regime
of $V,T \ll \Tk$ by combining concepts from
Hershfield's $Y$-operator formulation of non-equilibrium
problems and Affleck and Ludwig's exact conformal
field theory (CFT) solution   of the 2CK problem
(CFT technicalities will be discussed in a subsequent paper, III).
Finally, we extract from the conductance 
a universal scaling curve $\Gamma(v)$ and compare it with
experiment. Combining our results  with those of Hettler, Kroha
and Hershfield, we conclude that the NTKM achieves quantitative
agreement with the experimental scaling data.
\end{abstract}
\pacs{PACS numbers: 72.15.Qm,
72.10.Fk,
63.50.+x,
71.25.Mg}

\narrowtext
%\newpage
%\noindent
%%%\twocolumn[\hsize\textwidth\columnwidth\hsize\csname%
%%%@twocolumnfalse\endcsname%   [-bracket closes after e-mail ...

%{\bf Reference page}

%\noindent\
%Running head:\quad 
%2CK II: Constriction Conductance 

%\noindent
%Corresponding author: \\
%Jan von Delft\\
%Institut f\"ur Theoretische Festk\"orperphysik\\
%Universit\"at Karlsruhe\\ 76128 Karlsruhe\\
%Germany\\
%\\
%Tel: (0721) 608 6363 \quad or \quad 608 6363\\
% \\
%Fax: (0721) 698150
%\\
%\\
%e-mail: vondelft@tfp.physik.uni-karlsruhe.de 
%%%]

%\newpage

%%%\twocolumn[\hsize\textwidth\columnwidth\hsize\csname%
%%%@twocolumnfalse\endcsname%   [-bracket closes at  \vspace*{1cm}]
%\setcounter{page}{2}

\tableofcontents

%%%\vspace*{1cm}]

\section{Introduction}

This is the second in a series of three papers (I,II,III) \cite{I,II,III}
devoted to the 2-channel Kondo model (2CK). 
In the preceding paper (I),
we gave a detailed review of 
a possible experimental realization of this model, namely
the experiments by Ralph and Buhrman (RB)
\cite{RB92,Ralph93,RLvDB94,RB95} on non-magnetic zero-bias anomalies
(ZBAs) in Cu nanoconstrictions, and related experiments
by  Upadhyay, Louie and Buhrman \cite{ULB96} on Ti constrictions. 
The experimental facts were summarized in the
form of fifteen important properties of the data, nine
for Cu and six for Ti constrictions, 
[see (Cu.1) to (Cu.9) and (Ti.1) to (Ti.6), 
in section IV of I). The main conclusion
of paper~I was that all experimental facts are in accord
with the assumption that the ZBA is caused by the scattering
of electrons off  nearly degenerate two-level systems (TLS),
with whom they interact according
to the non-magnetic Kondo model of Zawadowski \cite{Zaw80,VZ83},
which renormalizes to the 2CK model at sufficiently low temperatures.
(See Appendices~\ref{ch:bk} and \ref{app:bk} for background on
Zawadowski's  model.)

In the present paper (II), we focus on property (Cu.6) for
the Cu constrictions:
in the so-called {\em weakly non-equilibrium}\/ 
or {\em non-Fermi-liquid regime}\/
of sufficiently small voltages
and temperatures ($ V \ll \Vk$ and $T \ll \Tk$,
but arbitrary ratio $v = e V/\kb T$)
where $\Vk$ and $\Tk$ are experimentally determined cross-over scales)
 the conductance $G(V,T)$  was found to satisfy 
the following scaling relation \cite{RLvDB94}:
\be
\label{IIscaling}
       {G(V,T) - G(0,T) \over  T^{\alpha}} =  F (v) \; ,
\ee
with scaling exponent $\alpha = \half$. This was
interpreted as strong evidence that the samples fall
in the low-temperature regime of the 2CK 
model, because its conformal field theory (CFT) solution 
by Affleck and Ludwig (AL) \cite{AL93,Lud94a} suggests
precisely such a scaling form near its $T=0$ fixed point,
and correctly predicts that $\alpha = \half$, as observed. 

If this interpretation is correct, it would imply 
that RB had directly observed non-Fermi-liquid behavior, because
in the 2CK model, the exponent $\alpha = \half$
is one of the signatures of non-Fermi-liquid physics
(for a Fermi liquid, $\alpha = 2$). 
Thus RB's experiments attracted a lot of interest,
because non-Fermi-liquid behavior,  so treasured by theorists,
 has been rather difficult to demonstrate unambiguously experimentally. 

However, it is of course quite conceivable that the scaling 
behavior can also be accounted for by some other theory.
Indeed, Wingreen, Altshuler and Meir  \cite[(a)]{WAM95}
have pointed out that an exponent of $\alpha = \half$
also arises within an alternative interpretation of
the experiment, based not on 2CK physics but the physics of 
disorder (We believe, though, that their scenario contradicts other 
important experimental facts, see Appendix~A.1 of Paper~I.
Moreover, the agreement between the experimental and theoretical  scaling
curves found in Ref.~\cite{WAM95}(a) was the result of a recently
discovered error in the analysis \cite{Wingreen-priv};
once the error is corrected, the agreement ceases.)

It is therefore desirable to develop additional quantitative criteria
for comparing the experiment to various theories. Now,
in paper~I it was shown that a sample-independent
scaling function $\Gamma (v)$ could be extracted from
the sample-dependent scaling function $F(v)$ of \Eq{IIscaling}.
According to the  2CK interpretation, this $\Gamma (v)$ should
be a universal scaling function,  a fingerprint of the 2CK fixed
point, independent of sample-specific details.
A very stringent quantitative test of any theory for 
the RB experiment would therefore be to calculate 
$\Gamma (v)$, and compare it to experiment. 

The present paper is devoted to this task.
 $\Gamma(v)$ is calculated analytically within the framework of
the 2CK model and its exact CFT solution by AL,
and the results are compared to the RB experiment. 
When combined with recent numerical results
of Hettler, Kroha and Hershfield {\em et al.} \cite{HKH94,HKH95}, 
agreement with the experimental scaling curve is obtained,
thus lending further quantitative support to the 2CK interpretation
for the Cu constrictions.

In order to describe the scattering of electrons off
two-level systems {\em in a  nanoconstriction geometry,}\/ 
we introduce a generalization of the 2CK model,
which we call the {\em nanoconstriction two-channel Kondo model (NTKM),}\/
that takes into account the complications arising from the non-equilibrium
electron distribution in the nanoconstriction. The generalization
consists of labelling the electrons by an additional {\em species}\/
index $\sigma = (L,R)$, which denotes 
their direction of incidence (toward the left or right
for electrons injected from the right or left lead).

In equilibrium ($V=0$), our NTKM reduces to the 2CK model.
Therefore, for $T \ll \Tk$, it 
displays the same non-Fermi-liquid behavior as the latter.
When the voltage is turned on, by continuity {\em there
must exist a regime in which the voltage is still 
sufficiently small (namely $V\ll \Tk$) that non-Fermi-liquid behavior
persists despite}\/ $V\neq 0$. We shall call this $T,V \ll \Tk$ regime
the {\em non-Fermi-liquid regime,}\/ and associate
it with the scaling regime of (Cu.6) identified in the experiment.
At higher voltates  ($V\gsim \Tk)$,  the non-Fermi-liquid behavior 
is destroyed. Therefore, we shall
focus exclusively on the case   $V\ll \Tk$ in this paper,
and accordingly the acronym NTKM will henceforth be understood
to stand for ``the nanconstriction 2-channel Kondo model
in the \nflr''.

The \nflr\ has to be treated by non-perturbative methods. 
The method we  use combines ideas from CFT with concepts
from Hershfield's $Y$-operator formulation of
non-equilibrium problems \cite{Hers93}. 
We show that all one needs to  calculate
the current using Hershfield's formalism are 
certain scattering amplitudes, 
to be denoted by $ \tilde U_{\eta \eta'}(\ve')$. 
We assume that in the \nflr, the scattering amplitudes
are essentially independent of $V$ (since
$V$-dependent corrections are of order $V/\Tk \ll 1$
and hence negligible (they are discussed in Appendix~\ref{app:Vcorr}).
We  then show that 
the $V=0$ values of the scattering amplitudes
can be extracted from an equilibrium Green's function
$G_{\eta \eta'} (\tau, ix; \tau' , ix') = 
- \langle T \psi_\eta (\tau, i x)\psi_{\eta'}^\dagger (\tau, ix')
\rangle$ that  is known exactly from CFT.
Once the $ \tilde U_{\eta \eta'}(\ve')$ are known,
it is straightforward to calculate the non-linear
current  $I(V,T)$ through  the constriction, and extract from
it the scaling function $\Gamma (v)$.

%This procedure makes the implicit assumption
%Thus, the only effect of $V \neq 0$ considered here is 
%that the occupation of the  electrons injected toward the
%scatterer non-equilibrium distribution,
%which (going beyond AL's equilibrium theory)
%we specify by using Hershfield's formalism.

No knowledge of CFT is required to read the present 
paper (excepting Appendix~\ref{app:Vcorr}), 
because the only step for which CFT
is really needed, namely the calculation of 
$G_{\eta \eta'} $, is carried out in paper~III, and here we only
cite the needed results. (Actually, the calculation
of $G_{\eta \eta'} $, too, can be done without CFT,
since it has been shown very recently using
abelian bosonization that AL's Green's
functions can be obtained  without using CFT \cite{vDZ,Ye};
this will be discussed in paper III.)

The outline of this paper is as follows:
In section~\ref{sec:poorVV} we introduce the NTKM,
and in section~\ref{sec:Y-Kondo} outline our 
strategy for solving it by a combination of
CFT methods with Hershfield's $Y$-operator approach.
This strategy is implemented in section~\ref{sec:implementation},
where the scattering states are calculated.
The current and scaling function are calculated in 
section~\ref{sec:Gscaling}. Our results for $\Gamma(v)$ are compared to
experiment and the NCA results of Hettler, Kroha and Hershfield
in section~\ref{sec:expcompare}, and our conclusions summarized
in section~\ref{sec:concc}. 

More than half of the paper is  taken up by appendices.
The lengthier ones 
(\ref{sec:sharvin},\ref{sec:VZ},\ref{app:bk},\ref{sec:criticism},%
\ref{sec:non-eqapp},\ref{NCA}) summarize, for the sake of convenience,
background material that is assumed known in the main text;
the others (\ref{naivepert},\ref{sec:simpex},\ref{app:Vcorr})
contain original work related to the main text.
In Appendix~\ref{sec:sharvin}, we recall some standard results from the
semi-classical theory of non-equilibrium transport
through a ballistic nanoconstriction. 
Appendices~\ref{sec:VZ} and~\ref{app:bk} provide  a brief review of 
the recent series of papers by  Zar\'and and Zawadowski on the
(bulk) non-magnetic Kondo model and its renormalization
toward  the 2CK model at low temperatures.
Recent criticism of their conclusions are discussed
in Appendix~\ref{sec:criticism}.
In Appendix~\ref{naivepert} we compare our CFT results
with those from the poor man's scaling approach
in the limit of large channel number  ($k \to \infty$), in 
which the latter approach becomes exact. 
Hershfield's $Y$-operator formalism is briefly
reviewed in Appendix~\ref{sec:non-eqapp}.
Appendix~\ref{sec:simpex} illustrates the general scattering
states formalism developed in sections~\ref{sec:implementation} and
\ref{currentcalc} with an simple example.
In Appendix~\ref{NCA} we give some background on 
the NCA calculations of Hettler, Kroha and Hershfield.
Finally,  in Appendix~\ref{app:Vcorr} we discuss 
$V/\Tk$-correction to our results.

\section{The Nanoconstriction Non-Magnetic Kondo Model}
\label{sec:poorVV}
\label{sec:pcmodel} 

In this section we introduce a new model,
to be called the {\em nanoconstriction
two-channel Kondo model}\/  (NTKM), to describe
the interaction of conduction electrons with a TLS in the nanoconstriction.
We shall take as guideline the
results of Zawadowski and coworkers, who introduced the
non-magnetic Kondo Hamiltonian to describe the TLS-electron
interaction (summarized in Appendix~\ref{ch:bk})
and showed that under renormalization it flows 
towards the \nfl\ fixed point of the 2CK model
(in a way summarized in Appendix~\ref{app:bk}).\footnote{
It should be pointed out that the question as to
whether a realistic TLS-electron system will reach 
the 2CK \nfl\ regime under renormalization is currently controversial
\cite{MF95,WAM95,ZZ96a,ZZ96b} (see Appendix~\ref{sec:criticism}). 
In the present paper, though,
we do not attempt  to clarify any of the
controversial issues. We simply take the view that
it would be useful to know what the scaling curve would look
like if the system indeed does reach the 2CK \nfl\ regime,
and hence do the calculation, assuming it does.}
However, we shall not be interested in the details of
the renormalization process from some bare to some effective
model. Instead our attitude here 
 is that of phenomenologists: since the detailed
microscopic nature of the presumed TLSs is unknown,
so too is the ``correct'' microscopic, bare Hamiltonian.
The best one can hope for is to find a phenomenological
Hamiltonian that satisfactorily accounts for the observed phenomena.
As argued at length  in paper~I,
the 2CK model with energy splitting $\Delta \simeq 0$
passes this test on a qualitative
level. We regard this as sufficient justification to 
use 2CK ideas  as a basis for quantitative calculations, 
in order to test whether quantitiative  agreement with experiment
can be achieved. 

The NTKM that we shall write down is the simplest model we can 
think of that contains the non-Fermi-liquid physics
of the  2CK model, but also accounts for
  the complications brought about by a nanoconstriction
geometry relative to the bulk situation. 
We introduce it as a phenomenological Ansatz, 
without attempting to provide a detailed microscopic derivation. 
%(though we  believe that this would be a worthwhile endeavor, since 
%our results show quantitative agreement between the NTKM
%and experiment).
Since our aim is to calculate a {\em universal}\/ curve, characteristic of the
2CK model but experimentally found to be sample-independent, 
we believe that such lack of attention to microscopic details 
has experimental justification.

The main complications arising in a nanoconstriction 
geometry relative to the bulk case
are,  firstly, that one has to distinguish between electrons leaving
and entering the $L$ and $R$ leads, and secondly, that 
the application of a voltage induces a 
non-equilibrium electron distribution in the nanoconstriction.

We thus have to deal with
a non-equilibrium problem with  non-trivial interactions.
The standard procedure   (due to Kadanoff and Baym \cite{KB62})
 for defining such a problem requires
conceptual care and may for clarity be organized into six steps:
\\
\phantom{(S3)}
First the problem is defined in the
absence of interactions, by defining
\\ (S1) 
a free Hamiltonian $H_o$ with eigenstates $\{ | \ve \eta \rangle_o \}$, 
\\(S2) 
a free density matrix $\rho_o$ governing their non-equilibrium  occupation, 
\\ (S3)
and the physical quantities of interest, in our case the
current $I$ (with expectation value\\
%\phantom{(S3)}
$\langle I \rangle  = \mbox{Tr} \rho_o I  /  \mbox{Tr} \rho_o \;  $
in the absence of interactions).
\\
%\phantom{(S4)} 
Then the interactions are switched on, by defining
\\ (S4)  an interaction Hamiltonian $H_{int}$, 
\\ (S5) and the full density matrix $\rho$, which governs the
non-equilibrium occupation of \label{S5}states 
\\ 
%\phantom{(S5)} 
for the fully interacting
system. (Typically, this is done by adiabatically switching on
\\ %\phantom{(S5)} 
$H_{int}$,  and keeping track of how the initial
$\rho_o$ develops  into a final $\rho$.) \\
(S6) Expectation values are calculated
according  to \vspace{2mm}$\langle I \rangle  
= \mbox{Tr} \rho I  /  \mbox{Tr} \rho \; . $

In this section, we address steps (S1) to (S4).
[(S5) and (S6)  are discussed in sections~\ref{sec:implementation}
\ref{sec:Gscaling}, respectively]. 
We also explain,  within the poor man's scaling approach, 
why the flow towards the \nfl\ regime is not disrupted
by $V \neq 0$ as long as $V  \ll \Tk$. 
 
\subsection{Free Hamiltonian $H_o$}
\label{defnonequibmodel}

We consider a single TLS  at the center of the
nanoconstriction (see Fig.~1 of paper~I
for a scetch of the nanoconstrictions used in the RB experiment).
We consider only those modes of electrons that contribute to the ZBA,
i.e.\ that interact with this TLS when passing through the nanoconstricion.

To describe these electrons, 
we imagine that the  ``free nanoconstriction Schr\"odinger
equation'' for free electrons and some random
static impurities but no TLS-electron interaction, 
with boundary conditions that
all electron wave-functions vanish on the metal-insulator
boundary, has already been solved (impossible in practice,
but not in principle).
This provides us [step (S1)] with a complete set
of single-particle eigenstates 
$\{ |\ve, \eta \rangle_o 
= c_{o \varepsilon \eta}^{\dagger} |0 \rangle \}$ (where $| 0 \rangle$ =
vacuum), in terms of which $H_o$ is diagonal:
\be
\label{eq4.Hoe}
        H_o =
        \sum_{\eta} \! \int_{-D}^D
         \!\!  d \varepsilon \: \ve
        \; c_{o \varepsilon \eta}^{\dagger} c_{o \varepsilon \eta} \, .
\ee
Here the continuous energy label $\ve$ is taken to lie in
a band of width $2D$, symmetric about the equilibrium Fermi energy
(at $\ve = 0$), with constant\footnote{
Very recent work by Zar\'and and Udvardi \protect\cite{ZU96}\protect\
has shown that using a constant density of states is
probably less realistic in a 
nanoconstriction than in the bulk (where it is standard),
because the local density of states fluctuates strongly as a function of
$r$ and $\ve$. This is the kind of complication that our
phenomenological approach has to ignore.}
 density of states\footnote{
Since the density of states
diverges for infinite systems, the expectation values of
some operators, e.g.\ the current \mbox{[}e.g.\ see 
footnote~\protect\ref{divexp}\protect\
and Eq.~(\protect\ref{eq4.Ib}\protect)\mbox{]},
have to be evaluated
in a finite system with a discrete energy spectrum. In such 
cases, we use the replacement rules\label{replacementsfinite}:
$
        \int \! d \ve \longrightarrow N_o^{-1} \sum_\ve \; ,
$
  and
$
        \delta (\ve - \ve') \longrightarrow N_o \delta_{\ve \ve'}
        \; .
$}
$N_o$. The latter
has been absorbed into the normalization of the 
$c_{o \ve \eta}^\dagger$'s, which we take as 
\be
\label{eq4.cnorm}
        \{ c_{o \varepsilon \eta}, c_{o \varepsilon \eta}^{\dagger} \}
        = \delta_{\eta \eta'} \delta (\ve - \ve') \; .
%\qquad (\mbox{or}\quad = \delta_{\eta \eta'} N_o \delta_{\ve \ve'}
%\quad \mbox{in a finite system})\; .
\ee

The label $\eta$ collectively denotes a set of discrete
quantum numbers,  $\eta \equiv (\sigma, \alpha, i)$
= (species,pseudo-spin,channel)-index, which have the following
meaning:
 $i = \uparrow, \downarrow$
is the  electron's Pauli spin, which will be seen below to
play the role of {\em channel-}\/index in the NTKM.
$\alpha = 1,2$ is a discrete \label{p:discreten}
{\em pseudo-spin index,}\/ the nanoconstriction analogue
of Vlad\'ar and Zawadowksi's  ``angular'' index $\alpha$
[see e.g.\ Eq.~(2.36) of the first paper of \cite{VZ83};
in \cite{HKH94}, $\alpha$ was called a ``parity'' index].
It labels those two sets of free states 
$\{ | \ve, \sigma, 1, i \rangle \}$ and 
$\{ | \ve, \sigma, 2, i \rangle \}$ that in the \nflr\ will couple
most strongly to the TLS.
For example, if the free wave-functions were
expanded in terms of angular harmonics,
$\alpha=1,2$ would label two complicated linear combinations
of  $Y_{l,m} (\theta, \phi)$ functions.
Strictly speaking $\alpha$ 
can take on a large number of discrete values, but we ignore all
but two, in the spirit
of Zawadowski's bulk result \cite{VZ83} that the others decouple
from the impurity when the temperature is lowered and the system
flows toward a \nfl\ fixed point with an effective 
electron pseudo-spin of $\half$. (The modes we ignore contribute to
the background conductance, but not to the ZBA.)

Finally,  $\sigma = (+,-) = (L,R)$,
the {\em species index,}\/ denotes
the direction of propagation of the incident electron:
 $\sigma=L=+$ for left-moving electrons, 
incident toward the left from $z = + \infty$ in 
the right lead; $\sigma = R = -$ for
right-moving electrons, incident toward the right from $z = -\infty$ in the
left lead.
(For example,  in spherical coordinates  the asymptotic behavior of 
the incident (or transmitted) parts of the wave-function  of
both $L$- and $R$-movers will be proportional to 
 $e^{- i k r}/r$ (or $e^{i k r}/r$) as $r \to \infty$.)
The nanoconstriction geometry necessitates this distinction
between $L$- and $R$-movers
(not needed  in the bulk case), firstly because $L$- and $R$-movers
originate from different leads, which are at different chemical
potentials if $V \neq 0$, 
and secondly because they contribute with different sign to
the  current.

\subsection{The free density matrix $\rho_o$}
\label{sec:nonequilnano}
\label{sec:freedensity}

We now turn to step (S2), the definition of $\rho_o$, the free density
matrix for $H_{int} = 0$ but arbitrary voltage.  The right and left
leads have chemical potentials (measured relative to the equilibrium
chemical potential $\mu$) of $+ eV/2 $ and $ - eV/2$,
respectively.\footnote{Our figures and arguments are given
  for the case $eV > 0$. We take $e = - | e |$ and hence $V= -|V|$.
  With $\mu_\pm = \pm e V/2$ for $R/L$ leads, there then is a net flow
  of electrons from right to left, \label{f:curpos}and the current to
  the right is positive.} As input, we use a standard
result from the semi-classical theory of non-equilibrium transport of
electrons through a ballistic nanoconstriction \cite{JvGW80}
(summarized in Appendix~\ref{app:Vtransport}): At the center of the
constriction, the distribution of occupied electron states in momentum
space is highly anisotropic (see Fig.~\ref{fig:fkr} in
Appendix~\ref{app:Vtransport}). It consists of {\em two}\/ sectors, to
be denoted by $L$ or $R$, that contain the momenta of all electrons
that are {\em incident as $L$ or $R$-movers,}\/ i.e.\ are injected
from the $R$ or $L$ leads. Consequently, the Fermi energies of the
$L/R$ sectors are equal to those of the $R/L$ leads, namely $\mu_\pm =
\pm \half eV$.

We formalize these standard results by associating
the $L/R$ sectors with the {\em species}\/
 quantum number $\sigma = L/R = \pm$ introduced
above (correspondingly $\mu_\eta$ will stand for
$\mu_\pm$), and adopting the following form 
for the free density matrix $\rho_o$:
\be 
\label{eq4.rhoo}
        \qquad \rho_o 
        \equiv  e^{-\beta [{H}_o - {Y}_o]}  \; ,
      \qquad 
\langle O \rangle_o \equiv {\mbox{Tr} \rho_o  O
         \over \mbox{Tr} \rho_o  } \; ,
\ee
where the $Y_o$-operator is defined by
\be
\label{eq4.Yo}
   {Y}_o   \equiv \half e V \left(  N_L - N_R \right) \: = \: 
         \sum_{\eta }   \mu_{\eta}
        \int \! d \varepsilon \,
        c^{\dagger}_{o \varepsilon \eta}  {c}_{o \varepsilon \eta} \; .
\ee
Here $  N_L$ and $  N_R$ denote the
total number of $L$- and $R$-moving electrons.\footnote{
\label{divexp}To evaluate 
$\langle c^\dagger_{o \ve \eta}   c_{o \ve' \eta} \rangle$
for $\varepsilon = \varepsilon'$, we have to give meaning to 
$\delta (\ve - \ve)$ of \protect\Eq{bk.Fermi}, 
which seems to diverge because we took the thermodynamic limit
of an infinitely large system.
We do this by replacing $\delta (\ve - \ve)$ by the corresponding
finite-system expression of $N_o \delta_{\ve \ve'}$
[see footnote~\protect\ref{replacementsfinite}\protect], i.e.\  we use
$
        \langle c^\dagger_{o \ve \eta} c_{o \ve \eta} \rangle
        = f (\ve, \eta)  N_o \; .
$
\label{f:diagonalm}}
It follows that (in the absence of interactions)
\bea
\label{bk.Fermi}
&&\langle c^\dagger_{o \ve \eta} (\tau)  c_{o \ve' \eta'} (\tau')  \rangle
= e^{\ve (\tau - \tau')}
f (\ve, \eta) \delta_{\eta \eta'} \delta (\ve - \ve') \; ,
\label{fermifunction}
\\
\nonumber
&& \mbox{where} \quad       f  (\ve, \eta) \equiv 
        {1  \over e^{\beta (\ve - \mu_\eta)} + 1 } \; .
\eea

\subsection{The free current through the nanoconstriction}
\label{freecurrent}

The ZBA arises from backscattering by the TLS of electrons
that would otherwise have passed through the constriction.
Thus, we assume that they would contribute
one unit $e^2/h$ of conductance
if the interaction were turned off. (More generally, 
one could use ${\cal T}_\eta e^2/h$, where ${\cal T}_\eta$
is a transmission coefficient, but this only affects
the (non-universal) amplitude of the ZBA.)
Thus, we may define [step (S3)] our current operator
simply as the difference between the number of electrons transmitted
as $L$- or $R$-movers: 
\be
\label{geometrical basis statescurrent}
        \hat  I =  {|e| \over N_o h }
       \sum_{\eta } \! \int \! \! d \ve
        \sigma 
        \; c_{o \varepsilon \eta}^{\dagger} c_{o \varepsilon \eta} \; .
\ee
Our signs are chosen such that 
$\langle \hat I \rangle_o >0$ if the net flow of electrons is from
right to left, while the  prefactor $|e|/h N_o$ is needed,
because of our choice of normalization, to 
obtain${}^{\ref{f:diagonalm}}$ a conductance of $e^2/ h$ per channel.

\subsection{The nanoconstriction 2CK interaction}
\label{nanointeraction}

We now come to step (S4), the specification
of the electron-TLS interaction, for which we make the
following phenomenological Ansatz:
\bea
\label{bk.HeffV}
        H_{int} &=&
        \int \! \! d \ve \! \!  \int \! \! d \ve' \!
        \sum_{\eta \eta' } c^{\dagger}_{o \ve \eta} 
        \, V_{\eta \eta'} \,
        c_{o \ve' \eta'}       , 
\label{effectiveVV}
\\ \nonumber
        V_{\eta \eta'} &\equiv & 
        v_{\sss K}
        v_{\sigma \sigma'} \delta_{i i'}
        \left( \half  \vec \sigma_{\alpha \alpha'}
        \, \cdot \vec S \right) \; ,
\eea
Here  $\vec S$ is the TLS pseudo-spin operator
acting in the two-dimensional Hilbert space of the 
TLS. Following the assumption (A2) of section~V.D of I,
%~\ref{sec:asymmetry} and VI.C of paper I,
we henceforth assume that $\Delta$, the TLS excitation
energy,  is the smallest energy scale in the problem, 
and set $\Delta = 0$. 

As far as the pseudospin and channel indices $\alpha$ and $i$ are concerned, 
$H_{int}$ is simply the isotropic 2CK Hamiltonian
to which, according to Zawadowski's analysis for a bulk system,
a realistic TLS coupled to electrons will flow at sufficiently low
temperatures. However, we introduced
an extra  Hermitian  $2 \!\times \! 2$
matrix $v_{\sigma \sigma'}$, which enables
an incident electron, say a $L$-mover, to be 
scattered into either a $L$- or
a $R$-mover, independent of whether 
its pseudo-spin index $\alpha$ and that of 
the TLS do or do not flip.\footnote{\label{f:tunnel}
Note that the interaction of \Eqs{effectiveVV} and (\ref{cl.VKroha})
is reminiscent of the tunneling Hamiltonian $H_{tun}$ in the
standard problem  of electrons tunneling through an insulating barrier
that separates two electronic baths: 
the off-diagonal components of $v_{\sigma \sigma'} $
transfers an electron from one bath to the other, with the 
implicit assumption that this does not disturb the
thermal distribution of electrons in the baths 
significantly.}
  In general, $v_{\sigma \sigma'}$  can be any Hermitian  
matrix, but, for reasons given below, it is actually sufficient 
to consider  only the very simple case 
\be
\label{cl.VKroha}
        v_{\sigma \sigma'} = \toh
        \left( \lower.8ex \hbox{ $
        {\stackrel{\scriptstyle 1}{\scriptstyle \vphantom{|^2} 1}}
        {\stackrel{\scriptstyle \phantom{\;-}1}{\scriptstyle
        \vphantom{|^2} \phantom{\;-}1}}
        $ } \right)_{\! \! \sigma \sigma'} \; .
\ee
Note that with this choice, our model is equivalent (after
a Schrieffer-Wolf transformation) to a model
recently studied by Hettler {\em et. al.} \cite{HKH94,HKH95}
using numerical NCA techniques, with whose 
results we shall compare our own (see section~\ref{subsec:expcompare}). 

The Hamiltonian introduced above is strictly speaking
not a 2CK Hamiltonian, since
$\sigma = \pm$ and $i = \uparrow, \downarrow$
give {\em four}\/ different combinations of indices
that do not Kondo-couple to the impurity. However,
 it can be mapped onto
a 2-channel model by making a unitary transformation,
\be 
\label{cl.N}   %% {eq4.lpsi}
        \bar c_{o \ve \bar \eta} =  N_{\bar \eta \eta} c_{o \ve \eta}
        \; , \qquad N_{\bar \eta \eta} \equiv 
        N_{\bar \sigma \sigma} \delta_{\bar \alpha \alpha}
        \delta_{\bar i i}  \; , 
\ee
chosen such that it diagonalizes $v_{\sigma \sigma'}$.
 For our present choice (\ref{cl.VKroha})
for $v_{\sigma \sigma'}$, $N_{\bar \sigma \sigma}$
is given by 
\be
\label{cl.Nexp} 
        N_{\bar \sigma \sigma} = 
        {\textstyle \frac{1}{\sqrt{2}}}
        \left( \lower.8ex \hbox{ $
        {\stackrel{\scriptstyle 1}{\scriptstyle \vphantom{|^2} 1}}
        {\stackrel{\scriptstyle \phantom{\; \;-}1}{\scriptstyle
        \vphantom{|^2} \; \; -1}}
        $ } \right)_{\! \! \bar \sigma \sigma}
        \; , \qquad
        \left( N v N^{-1} \right)_{\bar \sigma \bar \sigma'} =
        \left( \lower.8ex \hbox{ $
        {\stackrel{\scriptstyle 1}{\scriptstyle \vphantom{|^2} 0}}
        {\stackrel{\scriptstyle \phantom{\;-}0}{\scriptstyle
        \vphantom{|^2} \phantom{\;-}0}}
        $ } \right)_{\! \! \bar \sigma \bar \sigma'} \:\: .
\ee
We shall refer to the operators $c_{o \ve \eta}$ as 
$L/R$ operators and the $\bar c_{o \ve \bar \eta}$ as {\em even/odd}\/
operators, and always put a bar over all indices and matrices
refering to the even/odd basis. 
In the even-odd basis, the interaction becomes
\bea
\label{cl.NHk}
       H_{int} &=&
        \int \! \! d \ve \! \int \! \! d \ve' \!
        \sum_{\bar \eta, \bar \eta'} 
        \bar c^{\dagger}_{o \ve \bar \eta} \bar c_{o \ve' \bar \eta'} 
        \,  \overline V_{\bar \eta \bar \eta'} , 
\\ \nonumber 
        \overline V_{\bar \eta \bar \eta'} &=& v_{\sss K} 
        \delta_{\bar i \bar i'}
         \left( \begin{array}{cc}
                 {1 \over 2} \vec{\sigma}_{\bar \alpha \bar \alpha '} 
                \cdot  \vec S & \quad 0 \\
                0 & \quad 0
                \end{array}  \right)_{\bar \sigma \bar \sigma'}  .
\eea     
Thus, in the even/odd basis, one set
of channels, the {\em odd channels}\/
($\bar \sigma = o)$, completely decouples from
the impurity. The other set of channels,
the {\em even channels}\/ ($\bar \sigma = e$),
constitute a true 2CK problem,
which will eventually be responsible for  the non-Fermi liquid
behavior of the NTKM.

If one chooses
a more general $v_{\sigma \sigma'}$  than \Eq{cl.VKroha},
the odd channel will not completely decouple, but
(barring some accidental degeneracies) the
even and odd channels will always couple to the TLS  with different
strenghts. At low enough temperatures, 
 the one coupled more weakly can be assumed
to decouple completely (\`a la Zawadowski \cite{VZ83},
see section ~\ref{sec:2Dsubspace} of Appendix~\ref{ch:bk}),
leaving again a 2CK problem for the even channel.
This is the reason why it is sufficient
\label{p:Vsimple}
to take  $v_{\sigma \sigma'}$  as in (\ref{cl.VKroha}).

\subsection{Poor Man's Scaling Equations unaffected by $V$}
\label{sec:poorV}

The model we wrote down assumes that the NFL regime of the
TLS-electron system has already been
reached. However, one may wonder whether having  $V \neq 0$ 
would not prevent the TLS-electron system from reaching
the \nflr\ at all. That this is not the
case for  $V$ sufficiently small $(\ll \Tk)$ can be seen 
by the following poor man's scaling argument: 
Since the poor man's scaling equations
are derived by adjusting the cut-off from $D$ to $D'$,
which are both $\gg V,T$, 
they are independent of $V$ for the same
reason as that they are independent of $T$
(namely the change in coupling constants needed
to compensate $D \to D'$ does not  depend on
energies  $V$ and $T$ that are much smaller than $D$).
In other words, the scaling equations for $V\neq 0$
are the same as those for $V= 0$, meaning that the
initial RG flow is unaffected by $V \neq 0$.
Eventually, the RG flow is  cut off by either $V$ or
$T$, whichever is larger, resulting in an effective Hamiltonian
that depends on $V$ or $T$. 
 However, if both are $\ll \Tk$,
the RG flow will terminate in the close vicinity of
the  \nfl\ fixed point,  even if $V \neq 0$,
and the $V$ or $T$-dependence of the effective Hamiltonian
will be of order $V/\Tk$ or $T/ \Tk$, both $\ll 1$. 
This is the basis of our key assumption, stated in the introduction
and implicit in the Ansatz~(\ref{bk.HeffV}),
that for $V\neq 0$ but $V / \Tk \ll 1$, the \nflr\ is
governed by essentially the same effective Hamiltonian
as for $V= 0$.

\section{Outline of General Strategy}
\label{sec:Y-Kondo}

We now have to address step (S5)  of
the process of defining a fully interacting, non-equilibrium
problem, namely the definition of the full density matrix $\rho$
for $V \neq 0$ and $H_{int} \neq 0$.
In this section, the heart of this paper, we 
propose a strategy for doing this which
combines ideas from CFT 
with Hershfield's $Y$-operator formulation of non-equilibrium
problems.  The section is conceptual
in nature;  technical details follow in 
sections~\ref{sec:implementation} and~\ref{currentcalc}, and in paper~III.

\subsection{Hershfield's $Y$-operator approach to Non-Equilibrium Problems}
\label{sec:non-eq}

 Typically, the full $\rho$ is defined 
by adiabatically turning on $H_{int}$ and following
the evolution of the initial density matrix $\rho_o$ 
to a final $\rho$ (see Appendix~\ref{sec:non-eqapp}).
Expanding the time-evolution operator in powers of $H_{int}$,
one then generates a perturbation expansion that can be
handled using the Keldysh technique.

However, for the Kondo problem, perturbation theory breaks
down for $T< \Tk$, where many-body effects become important.
Therefore we shall adopt Hershfield's so-called $Y$-operator
 formulation of non-equilibrium problems \cite{Hers93},
which is in principle non-perturbative.

The main idea of Hershfield's approach (briefly summarized in
Appendix~\ref{sec:non-eqapp}), is as follows.  As the interaction $H_{int}$ is
adiabatically turned on, the density operator adiabatically evolves from its
initial form $\rho_o = e^{-\beta (H_o - Y_o)}$ into a final form that
Hershfield writes as $\rho \equiv e^{-\beta (H - Y)}$.  This defines the
operator $Y$, which is the adiabatically evolved version of $Y_o$ and is
conserved ($[Y,H] = 0$).  The formal similarity between $\rho$ and $\rho_o$
implies that when expressed in terms of $H$ and $Y$, the non-equilibrium
problem has been cast in a form that is formally equivalent to an equilibrium
problem.

This becomes particularly evident if one considers the
set of simultaneous eigenstates of $H$ and $Y$, which we shall
call the {\em scattering states}\/ and denote by
$\{|\ve  \eta \rangle  =  c_{\ve \eta}^\dagger |0\rangle \} $.
Loosely speaking, they  can be viewed as the states into 
which the free basis states $\{ | \ve \eta \rangle_o\}$ develop
as $H_{int}$ is turned on  (in the sense that $ c_{\ve \eta}^\dagger$
is some function of the $\{  c_{o \ve' \eta'}^\dagger \}$, 
which reduces to  $c_{o \ve \eta}^\dagger$ for $H_{int}= 0$). 
For scattering problems like the NTKM, 
in which a free electron is incident upon a scatterer
and scatters into something complicated, there evidently must be
a one-to-one correspondence between the states $|\ve \eta \rangle_o $ 
and  $|\ve  \eta \rangle $: the incident
parts of their wave-functions
$\langle \vec x |\ve \eta \rangle_o$ and $\langle \vec x |\ve \eta \rangle$
must be identical. (The outgoing parts, which 
contain scattering information, will of course be
different -- this will be made explicit
in \Eq{eq4.scatwave} below.)
This is why the free and scattering states can be labelled by the
same indices, and also have the same density of states.\footnote{
One might ask whether the very notion
of scattering states make sense for a dynamical impurity problem,
since the scatterer  is constantly
flipping its pseudo-spin. However,
in the CFT solution of Kondo problems, the impurity 
 completely disappears from the scene \label{p:impabsorbed}
(being absorbed in the definition of 
a new spin current,  see %%III \Eql{al.shiftx} of 
paper III).
Thus the theory contains only electron degrees  of freedom,
for which one {\em can}\/ meaningfully
introduce scattering states.}

Furthermore, for such scattering problems, 
 $H$ and $Y$ will have the following form:\footnote{For 
problems other than scattering problems,
 Eq.~(\protect\ref{eq4.Y}\protect) does not necessarily hold.}
\bea
\label{eq4.YH}
\label{eq4.YHa}
        H &=& \sum_{\eta } \int \! d \varepsilon \,
        \ve  c^{\dagger}_{ \varepsilon \eta} {c}_{ \varepsilon \eta}
        \; ,
\\
\label{eq4.Y}
        {Y}  & \equiv &
        \sum_{\eta } 
        \int \! d \varepsilon \,
        \mu_\eta c^{\dagger}_{ \varepsilon \eta} {c}_{ \varepsilon \eta} \;
        \qquad ( \neq Y_o) \; .
\eea
The form used here for $Y$ 
follows because $Y$ evolves from $Y_o$ as $H_{int}$
is  turned on, implying that
${Y}$ can be obtained from ${Y}_o$  by {\em replacing}\/  
the ${c_o}_{\varepsilon \eta}$ in \Eq{eq4.Yo} by the scattering-state 
operators $c_{\varepsilon \eta}$
into which the latter evolve \cite{Hers93}.
\Eq{eq4.YH} and (\ref{eq4.Y}) imply 
that non-equilibrium thermal expectation values of
the $c_{\varepsilon \eta}$'s have the standard form:
\bea
\label{eq4.ceVav}
\label{eq4.ceav}
& &        \langle c_{\varepsilon \eta}^{\dagger} (\tau)
        c_{\varepsilon' \eta'} (\tau') \rangle
        =  e^{\ve ( \tau - \tau') } f (\ve, \eta ) 
        \delta_{\eta \eta'} \delta (\ve - \ve' )  
\\ \nonumber
&& \mbox{where} \quad       f  (\ve, \eta) \equiv 
        {1  \over e^{\beta (\ve - \mu_\eta)} + 1 } \; .
\eea
This is precisely the same form as that
satisfied by the non-interacting $c_{o \ve \eta}$'s 
in the absence of interactions [see \Eq{bk.Fermi}].
The intuitive reason for this remarkably simple result is
clear: {\em the Boltzman weight of a scattering state
must be the same as that of the corresponding free state,
since the thermal equilibration that leads to the Boltzmann
factors happens deep inside the leads,
before the electrons are injected and scattered
by $H_{int}$}\/  (this of course remains true when $L$- and $R$ leads
have different chemical potentials -- all that happens for $V \neq 0$
is that the occupation probabilities pick up a  $V$-dependence
reflecting from which lead the electron was injected).

This result provides us with a very clear picture 
of how the current through a nanoconstriction 
should be calculated: when injecting
electrons from the leads into the constriction, 
the thermal weighting is done {\em precisely}\/ as for
free particles, i.e.\ an electron incident in the state
$|\ve' \eta' \rangle_o$ is injected with weight $f (\ve', \eta')$.
 For each such electron, one has to determine 
the {\em scattering  amplitude}\/
$\tilde U_{\eta \eta'} (\ve')$, i.e.\ the amplitude 
with which it emerges from the scattering 
process in the state $|\ve', \eta\rangle_o$
(where we assumed elastic scattering). 
These amplitudes (defined more explicitly below,
see section~\ref{defscatamp}) are the non-trivial ingredients of
the scattering states, which contain all relevant information
about the scattering process.\footnote{%%%%
\label{f:Malda} For a  many-body problem such as the Kondo problem,
complicated combinations of 
particle-hole excitations are created upon scattering,
which can {\em not}\/  simply be written as a linear combination 
$ \sum_\eta c_{o \ve' \eta}^\dagger  \tilde U_{\eta \eta'} (\ve')$
of single-particle excitations. 
However, it was shown by Maldacena and Ludwig \cite {ML95}
(see also \cite{vDZ}) that the scattering matrix for free
electrons incident on a Kondo impurity is unitary
if the single-particle Hilbert space 
of free-electron states $\{| \ve\eta\rangle_o\}$ 
is appropriately enlarged
to include ``Kondo excitations'' (called ``spinors''
by them, see %%III section~\protect\ref{sec:unitpar} 
%%III and Appendix~\protect\ref{app:spinor} of 
paper~III). 
This means that the  outgoing states
can be written as  linear combinations of free-electron
states $\{ | \ve\eta\rangle_{ o} \}$ and a new
set of Kondo excitation states $\{ | \ve\eta\rangle_{\tilde o} \}$.
The corresponding
set of  creation operators $\{ \tilde c_{o \ve \eta}^\dagger \}$
are complicated, highly  non-linear functions (not mere linear combinations)
of the $\{ c_{o \ve' \eta}^\dagger \}$  and 
will be discussed in paper~III. 
Thus, in the  formalism developed below, the unitary
transformation in \protect\Eq{eq4.unittrans} is implicitly
understood to act in the enlarged Hilbert space
of $\{  | \ve \eta\rangle_{o},
 | \ve\eta\rangle_{\tilde o} \}$ states,
and the collective index $\eta$ implicitly includes
another index $a = (f,k)$ to distinguish free from
Kondo states. However, this will only be made explicit in paper~III.}
 Once they  are known, it is straightforward to 
calculate the current as a thermally weighted sum
over transmission probabilities. 

Since expectation
values expressed in terms of scattering states are so
simple, it is useful to reexpress all physical operators
in terms of them. To this end, we define
%\footnote{
%Note
%that $U_{\eta' \eta} (\varepsilon',   \varepsilon )$ is not quite
%the same as the usual scattering matrix $S_{\eta \eta'}
%(\varepsilon', \ve)$, defined by $ | \ve \eta \rangle^{(+)}
%\equiv \sum_{ \eta'}  \! \int \!  d \ve'
% | \ve \eta' \rangle^{(-)}
%S_{\eta' \eta} (\ve' , \varepsilon) $ \cite[eq.~(19.48)]{Merz70}
%(in that language $| \ve \eta \rangle^{(+)}$
%corresponds to our $| \ve \eta \rangle$).
%$S_{\eta' \eta} (\ve' \varepsilon)$ maps two sets of eigenstates
%of the {\em full}\/ Hamiltonian onto each other,
%namely ``incoming'' states $ \{ | \ve \eta' \rangle^{(-)} \} $
%onto ``outgoing'' ones $\{ | \ve \eta \rangle^{(+)} \} $. In contrast,
%$U_{\eta'  \eta} (\varepsilon', \ve)$ maps 
%eigenstates $ \{ |\varepsilon' \eta' \rangle_o \} $ of $H_o$ onto
%``outgoing'' eigenstates $ \{ | \ve \eta \rangle \} $ of $H$.
%} 
$ U_{\eta' \eta} (\varepsilon',   \varepsilon )
        \equiv {}_o \langle {\varepsilon' \eta'} 
        |\varepsilon \eta \rangle$ 
to be the unitary transformation${}^{\ref{f:Malda}}$
that relates the scattering states  to the free basis states:
\bea
\label{eq4.unittrans}
        |\varepsilon \eta \rangle &=&
        \sum_{\eta'}  \! \int \!  d \ve' \;
        |\varepsilon' \eta' \rangle_o
        U_{\eta' \eta} (\varepsilon',   \varepsilon ) \; ,
\\
\label{eq4.utc}
        c_{\varepsilon \eta} & = &
        \sum_{\eta'}   \! \int \!  d \varepsilon' \;
        U^{\dagger}_{\eta \eta'} (\varepsilon,  \varepsilon' )
        c_{o \varepsilon' \eta'} \; ;
\\
\label{eq4.Sunit}
         \delta_{\eta \eta'} \delta(\ve - \ve') & = & 
        \sum_{\tilde \eta}  \! \int \!  d \tilde \varepsilon \;
        U^{\dagger}_{\eta \tilde \eta} (\varepsilon , \tilde \varepsilon )
        U_{\tilde \eta \eta'} (\tilde \varepsilon,  \varepsilon' ) \; .
\eea
For example, the current of \Eq{geometrical basis statescurrent}  
takes the form:
\bea
\label{transformedcurrent}
         I &= & {|e| \over N_o h }
       \sum_{\eta \eta' \eta'' } \! \int \! \! d \ve
         \! \int \! \! d \ve'  \! \int \! \! d \ve''
\\ \nonumber & &\times
        \mbox{Re} \left[ \sigma 
        U^\dagger _{\eta' \eta} (\varepsilon',   \varepsilon )
        U_{\eta \eta''} (\varepsilon,   \varepsilon'' )
        \; \langle
        c_{ \varepsilon' \eta'}^{\dagger} c_{ \varepsilon'' \eta''} 
        \rangle \right] \; .
\eea
The reality of $I$ is of course automatically ensured by the
hermiticity of the current operator, and the reminder Re[\, \, ] 
has been inserted merely for future convenience.

We shall show below that the $U_{\eta \eta'} (\varepsilon,   \varepsilon'
)$, and hence also the current,
are completely determined by the
$\tilde U_{\eta \eta'} (\ve')$. Unfortunately, Hershfield's
formalism gives no recipe for finding these
 explicitly for a given problem. 
Thus, the crucial question now becomes: how does one calculate the
scattering amplitudes?

\subsection{Equating CFT- and scattering-state
Green's Functions}
\label{sec:HeffII}

In general, finding the \SA\ is just as difficult as solving the problem by
other (e.g..\ Keldysh) methods. (For example, Keldysh methods are used to find
the scattering states of a closely related Kondo problem in Appendix~C of
Ref.~\cite{HKH95}.)  However, for $V=0$ the even sector of the NTKM is
equivalent to the 2CK model, which AL solved exactly using CFT
\cite{AL93,Lud94a,Aff90,AL91a,AL91b,AL92b,AL94}.  (This equivalence is shown
explicitly below, when we rewrite the model in field theoretical language, see
\Eq{finalHo} and (\ref{finalHscat}) below.) Therefore, we propose that the
\SS\ of the NTKM can be extracted from AL's results.  We now explain how this
can be done.

One of AL's central results is an explicit and exact
expression for the equilibrium Green's $ G_{\eta \eta'} 
= - \langle  \psi_{\eta}  \psi^{\dagger}_{\eta'} \rangle $
[defined explicitly in \Eq{eq4.GFtau}], which
 gives the amplitude 
that an incident $\eta'$-electron will emerge from the scattering
process as outgoing $\eta$-electron. Evidently, it
must contain information about the scattering amplitudes.
Indeed, we shall show that when the same equilibrium
Green's function  is  calculated explicitly using 
the scattering state formalism,
it is completely determined by $\tilde U_{\eta \eta'} (\ve')$.
Therefore,  {\em by equating the scattering-states
form for $ G_{\eta \eta'}$ to  the corresponding CFT result,
$\tilde U_{\eta \eta'} (\ve')$ can be extracted from the latter.}\/

Of course, this procedure only yields the 
$V=0$ value of $\tilde U_{\eta \eta'}$, whereas 
to calculate the nonequilibrium current, 
we actually need its $V\neq 0 $ values too. 
Moreover, it is clear that
in general $\tilde U_{\eta \eta'}$ {\em must}\/ 
depend on $V$, since if $V$ is  sufficiently large,
it is known to non-trivially affect the many-body physics of
the Kondo problem. For example,  for
$V \neq 0$, the difference in Fermi energies
of the $L$- and $R$ leads causes the Kondo peak in the density of states to 
split \cite{MWL93,WM94} into two
separate peaks (at energies $\mu \pm \half e V$,
see \fig{fig:NCAVT} of Appendix~\ref{NCA}, taken from \cite{HKH95}).
Moreover, the effective Hamiltonian in poor-man's scaling
approaches depends on $V$ if it is the largest low-energy
cut-off in the problem (see section~\ref{sec:poorV}),
and if $V$ is too large, it will cut off the
renormalization group flow towards the \nfl\ fixed point
before the \nflr\ is reached.  

However, such  $V$-induced effects should be negligible for
sufficiently small $V$. For example, 
when $V \ll \Tk $, the  splitting of the Kondo peak by $eV$ 
is negligible compared to its width, which is $\propto \Tk$.
Said in poor-man's scaling language, if $(T<) \, V \ll \Tk $, then 
$V \neq 0$ cuts off the renormalization group 
 flow  at a point sufficiently close to the \nfl\ fixed
point that  the physics should still governed by the 
latter. Hence, we propose that in the \nfl\ regime of 
$V \ll \Tk $, the $V$-dependence of the scattering amplitudes
 $\tilde U_{\eta \eta'}$ 
is negligible, and hence shall always use their $V=0$ values below.
(In a sense, the condition that this
procedure be valid can be regarded
as our definition of the ``\nflr''.) More formally, we assume
that $\tilde U_{\eta \eta'}$ can be expanded in powers of
$V/\Tk$, and use only the zeroth term. (In Appendix~\ref{app:Vcorr},
we show that  the leading $V/\Tk$ correction only produces a
subleading correction to the desired scaling function.)

The intuitive motivation for neglecting the 
$V$-dependence of the scattering amplitudes is based
on the assumption that the effect of $V \neq 0$ can 
be characterized as follows if  $V\ll \Tk$:
although the leads inject electrons into the \nfl\ state that,
since $V \neq 0$,  are able to probe its nature at energies 
different from   $\ve_{\sss F}$,  they only probe gently, 
i.e.\ they inject sufficiently
few that the \nfl\ state itself is not disrupted. Since the ``output'' of
this probing, namely the scattering amplitudes, 
depend non-linearly on $\ve$, the current will depend
non-linearly on $V$, too, even if $\tilde U_{\eta \eta'} (\ve')$ itself is
$V$-independent.

Another underlying assumption
of our proposed strategy is that the strong-coupling
or fixed-point fields $\psi_\eta (\tau, ix)$ 
occuring in the CFT treatment can  be expanded in terms of
a set of fermionic excitations
(though these are very complicated non-linear combinations
of the free ones, cf. footnote~\ref{f:Malda}), else it would not
make sense to equate a CFT Green's function
to one constructed from scattering states. That this 
is indeed the case will be shown in paper
III.%XXXX
%, section~\ref{ch:sp}. %XXXX

\section{Extracting Scattering States From CFT Results}
\label{sec:implementation}

To implement our strategy for finding $\tilde U_{\eta \eta'}$,
the first step is to rewrite the NTKM of
section~\ref{sec:pcmodel} in field theory language by introducing
a set of fields $\psi_\eta (ix)$. Then we define the Green's function 
$G_{\eta \eta'} =  - \langle T \psi_\eta \psi^\dagger_{\eta '} \rangle$,
and show that it is completely
determined by $\tilde U_{\eta \eta'}$
(which turns out to be its spectral function). 
Finally, we equate this $G_{\eta \eta'} $
to the corresponding exact CFT result of AL, 
 which allows us to obtain
the corresponding exact expression for  
$\tilde U_{\eta \eta'}$ explicitly.

\subsection{Transcription to Field Theory}
\label{sec:2dft}
\label{sec:tr2D}

To rewrite the  ``bare'' NTKM introduced in
section~\ref{sec:pcmodel} in field theory language, 
we introduce for each channel $\eta$ a
1-dimensional, second-quantized 
field $\psi_{\eta} (\tau, ix)$
(with $x \in [-l, l]$, $l \to \infty$) 
as a Fourier-integral over  all $\ve$:\footnote{
Strictly speaking,
the $\int \! d \ve$ integrals have
to be cut off, $\int_{-D}^{ D} \! d \varepsilon$,
at a bandwidth  $D$ satisfying $T,V \ll D$.
However, we take $D \to \infty$ (since the errors thus introduced
are of order $T/D, V/D \ll 1$ and hence negligible even for finite $D$).
This allows us to invert relations such as (\protect\ref{eq4.psiex}\protect)
straightforwardly.}
\bea
\label{eq4.psiex}
   &&     \psi_{\eta} (ix) \equiv 
        \case{1}{\sqrt{\hbar \vf}}
        \int_{- \infty}^{\infty} \!\! d \varepsilon \;
        e^{-i \varepsilon x/ \hbar \vf} \,
        c_{o \varepsilon \eta} \; , 
\\
\label{caspsi}
 &&       c_{o \varepsilon \eta}  = 
         \case{1}{\sqrt{\hbar \vf}}
        \lim_{l \to \infty} \int_{- l/2}^{l/2} \!
        {\textstyle {dx \over 2 \pi}}
        e^{i \varepsilon x / \hbar \vf}
        \psi_{\eta} (ix) \; ,
\\
\label{psinormalization}
      & &  \{  \psi_{\eta} (ix), \psi_{\eta'}^{\dagger} (ix) \}
       =  2 \pi \delta_{\eta \eta'} \delta (x - x') \; .
\eea
The factors of $\hbar$ and $ \vf$, inserted for dimensional
reasons, are henceforth set $=1$. 
%We shall almost always take the limit $l \to \infty$,
%but for the few cases for which a finite $l$ has to
%be used [e.g.\ \Eq{almostfinalcurrent}],
%we use the replacement rules of  \Eq{replacementsfinite},
%with $N_o = l/(h \vf)$.

Note that $\psi_{\eta} (ix)$ is {\em not}\/ the usual
electron field $\Psi (\vec x)$, which is constructed
from the actual (unknown) wave-functions 
$\langle \vec x | \ve \eta \rangle_o $
through $\Psi (\vec x) \equiv \sum_{\eta} \! \int \!\!  d \varepsilon
\langle \vec x | \ve \eta \rangle_o c_{o \ve \eta}$.
Instead, $\psi_{\eta} (ix)$ is best thought of
simply as the Fourier transform of $c_{o \varepsilon \eta}$,
this being a convenient way of rewriting
the problem in field-theoretical language. 
Nevertheless, the role of $x$ is strongly analogous to that
of the ``radial'' coordinate of the actual wave-function
$\Psi_{\ve \eta} (\vec x)$, and $\psi_\eta^\dagger (x)$
can be interpreted as the operator that creates an electron
with quantum numbers $\eta$ at ``position'' $x$. 

Using \Eq{caspsi},  $H_o$ and $H_{int}$   of
\Eqs{eq4.Hoe} and (\ref{bk.HeffV}) can be written as
\bea
\label{eq4.Hox}
         H_o  &=&
        \sum_{\eta \eta'} \! \int_{- \infty}^{\infty} \! \! 
        {\textstyle {dx \over 2 \pi}}
        \psi^{\dagger}_{\eta} (i x)
        i \partial_x 
        \psi_{\eta} (i x) \; , 
\\
\label{eq4.Hee}
        H_{int} & \equiv &
        \sum_{\eta \eta'} \! 
        \psi^{\dagger}_{\eta} (0)
         V_{\eta \eta'}
        \psi_{\eta'} (0) \; .
\eea

By simple Fourier transformation, we have hence
arrived at a 1+1-dimensional field theory, 
defined by \Eqs{eq4.Hox} and (\ref{eq4.Hee}).
The reason why this (and not a 3+1 dimensional theory)
resulted, is essentially that there is only {\em one}\/
continuous quantum number, namely $\ve$, in the problem,
with respect to which we can Fourier transform. 
This in turn is a result of the constriction geometry,
which defines a definite and unique origin,
and consequently a notion of a single ``radial'' coordinate
(in spherical coordinates it is the radius $r$),
to which our $x$ roughly corresponds.
Moreover, the fact that we assumed a constant density of
states and hence a linear dispersion implies that
the free fields are conformally invariant, which
is the key property required for the subsequent application
of AL's CFT methods. 

The Heisenberg equation of motion, 
\bea
\nonumber
         -  \partial_\tau \psi_{\eta} (\tau, ix) 
        &=& [ \psi_{\eta} (\tau, ix) , H_o + H_{int}] 
\\ \label{eq4.Schroedinger}
       & =&  \left( \delta_{\eta \eta'} i \partial_x 
        + {2 \pi} \delta(x)  V_{\eta \eta'} \right)
         \psi_{\eta'} (ix) \; .
\eea
shows that for all $x \neq 0$, the fields depend only
on $\tau + ix$.
[This is the reason for writing the argument of $\psi_{\eta}$ as $(ix)$
in \Eq{eq4.psiex}, since the $\tau$ dependence 
of $\psi_{\eta}$ can then simply
be obtained by analytic continuation ($ix  \to \tau+ix $).]
Consequently, {\em by construction}\/, 
{\em all}\/ fields are  ``mathematical left-movers'', 
incident from $x = \infty$ and  traveling toward
$x = - \infty$. The effect of the scattering term $H_{int}$
is to mix the different incident channels with each other
at $x= 0$, so that $\psi_\eta (\tau, ix)$ 
will differ from a free field only for $x < 0$. 
Thus, {\em we have turned
our problem into a one-dimensional scattering problem,}\/
with {\em all}\/ free fields incident from the right, and {\em all}\/
scattered ones outgoing to the left.% (see \fig{fig:scatt}).
This is in exact analogy to AL's treatment of the Kondo problem,
which in fact  was the motivation for introducing both physical
$L$- and $R$-movers as ``mathematical left-movers'' in \Eq{eq4.psiex}.   
Of course, the distinction between {\em physical}\/
$L$- and $R$-movers is carried by the index $\sigma = L,R$, 
and  \label{p:mathLR} $L$-$R$ backscattering
is described by the $\sigma \neq \sigma'$
terms in $V_{\eta \eta'}$.

% \bea
% \label{tr.Greeno}
%   G_{\eta \eta'} ( u- u')
%        &\equiv & - \langle T {\psi}_\eta (u)
%        {\psi}_{\eta'}^\dagger (u') \rangle \\
%   & = &
% \label{tr.GFoo}
%        {- \delta_{\eta \eta'} e^{- \mu_{\eta'} (u - u')/ \hbar  }
%        \over {\hbar \vf \beta \over \pi} 
%       \sin {\pi \over \beta \hbar}  (u-u' ) } 
%       \quad \stackrel{T \to 0}{\longrightarrow}  \quad
%       {- \delta_{\eta \eta'} \over \vf (u - u') }  
%       e^{- \mu_{\eta'} (u - u')/ \hbar }
%       \; .
% \eea

\subsection{Transformation to even-odd basis}
\label{transcribee-o}

As mentioned in section~\ref{nanointeraction},
the relation between the NTKM and the standard
2CK model is best understood in the 
even-odd basis (denoted by bars) of operators
$\bar c_{o \ve \bar \eta} =  N_{\bar \eta \eta} c_{o \ve \eta}$
[see  \Eq{cl.N}]. 
Therefore, we  define even-odd fields
\be
\label{psievenodd}
\lpsi_{\bar \eta} (ix) = N_{\bar \eta \eta} \psi_\eta (ix) \; ,
\ee
normalized as in \Eq{psinormalization}.
In terms of these, $H_o$ and $H_{int}$  of \Eqs{eq4.Hox} and
(\ref{eq4.Hee}) are:
\bea
\label{finalHo}
        H_o      &=&
        \sum_{\bar \eta} \! \int_{- \infty}^{\infty} \! \! 
        {\textstyle {dx \over 2 \pi}}
        \lpsi^{\dagger}_{\bar \eta} (i x)
        i \partial_x 
        \lpsi_{\bar \eta} (i x) \; , 
\\
\label{finalHscat}
        H_{int} &=&
        \sum_{\bar \sigma  \bar \alpha \bar \alpha' \bar i}
        \lpsi^\dagger_{\bar \sigma \bar \alpha \bar i} (0)
        \left( v_{\sss K} \delta_{\bar \sigma e}
        \case{1}{ 2} \vec{\sigma}_{\bar \alpha \bar \alpha '} 
                \cdot \vec S \right)
        \lpsi_{\bar \sigma \bar \alpha' \bar i} (0) \; .
\eea
 The odd channel ($\bar \sigma
= o$) decouples from $H_{int}$. In the even channel 
($\bar \sigma = e$), $H_o + H_{int}$ 
 is precisely the ``bare'' Hamiltonian of the equilibrium 2CK model solved
exactly by AL [see e.g.\ \cite{Lud94a}, Eq.~(2.17)].
Therefore, the even channels will display \nfl\ behavior
for $T,V \ll \Tk$.

\subsection{Definition of scattering amplitude 
$\tilde U_{\eta  \eta'} (\varepsilon') $}
\label{defscatamp}

Having rewritten the model in field theory language,
we can define the equilibrium  Green's function 
(in the original $L$-$R$ basis) that is to
be the link to AL's CFT results:\footnote{
In paper~III, this Green's function is denoted by
$ G^\sRL_{\eta \eta'} (z^\ast ; z')$, following the notation
used AL.}
\be
\label{eq4.GFtau}
        G_{\eta \eta'} ( \tau, -ir ; \tau', ir')
        \equiv  - \langle T \psi_{\eta} (\tau,- ir) 
        \psi^{\dagger}_{\eta'} (\tau', ir') \rangle \;  ,
\ee
with $r,r' > 0$. Since its arguments correspond to taking
$x = -r <0$ and $x' = r' > 0$, it  gives the amplitude 
that an incident $\eta'$-electron will emerge from the scattering
process as outgoing $\eta$-electron.

In order to calculate $G_{\eta \eta'}$ in 
terms of scattering states, we rewrite 
 the fields $\psi_\eta (\tau, i x)$  in terms of the $c_{\ve \eta}$'s.
\label{sec:SSwf}
Inserting the inverse of \Eq{eq4.utc} into \Eq{eq4.psiex} and defining
\be
\label{eq4.newphi}
        \phi_{\varepsilon' \eta'} (ix, \eta) \equiv
        \int \!\!  d \varepsilon
        e^{-i \varepsilon x }
        U_{\eta \eta'} (\varepsilon, \varepsilon' ) \; ,
\ee
we find
\be
\label{eq4.psiphi}
        \psi_{\eta} (\tau, ix) =
        \sum_{\eta'} \! \int \!\!  d \varepsilon'
        \phi_{\varepsilon' \eta'} (ix, \eta)
        c_{\varepsilon' \eta'} (\tau) \; , 
%\\
%       c_{\varepsilon' \eta'} (\tau) &=&
%        \case{1 }{\sqrt{ \hbar \vf}}
%        \sum_{\eta} \int \! {\textstyle {dx \over 2 \pi}}
%        \phi^{\ast}_{\varepsilon' \eta'} (ix, \eta)
%        \psi_{\eta} (\tau, ix) \; .
\ee
which implies that 
$
         \phi_{\varepsilon' \eta'} (ix, \eta)
        = \langle \psi_{\eta} (\tau, ix) 
                c_{\varepsilon' \eta'}^\dagger (\tau)
        \rangle
$.
Since by its definition (\ref{eq4.psiex}) 
$\psi^\dagger_{\eta} (\tau, ix) $ has the interpretation
of creating an electron with quantum numbers $\eta$ at $x$,
this shows that  $\phi_{\varepsilon' \eta'} (ix, \eta)$
may be thought of as the ``wave-function'' for the
scattering states $| \ve' \eta' \rangle$:\footnote{This
interpretation of $\phi_{\varepsilon' \eta'} (ix, \eta)$
as a wave-function
is meant as a mnemonic and  should not be taken  literally; as
mentioned in section~\ref{defnonequibmodel}, the
actual physical wave-functions are intractably complicated.}
it gives the amplitude for an electron
in state $| \ve' \eta' \rangle$  to
be found at $x$ with quantum number $\eta$.
The orthonormality and completeness of these
wave-functions is guaranteed by the  unitarity (\ref{eq4.Sunit})  
of $U_{\eta \eta'} (\varepsilon, \varepsilon' )$:
\bea
\label{eq4.ortho}
        \sum_{\tilde \eta} \int \! {\textstyle {d \tilde x \over 2 \pi}}
        \; \phi^{\ast}_{\varepsilon \eta} (i \tilde x, \tilde \eta)
        \phi_{\varepsilon' \eta'} (i\tilde x, \tilde \eta)
        & = &  \delta_{\eta \eta'} \delta (\ve - \ve') \; , 
\\
        \sum_{\tilde \eta} \! \int \!\!  d \tilde \varepsilon \;
        \phi^{\ast}_{\tilde \varepsilon \tilde \eta} (i x,  \eta)
        \phi_{\tilde \varepsilon \tilde \eta} (ix', \eta')
        &=&  2 \pi \; \delta_{\eta \eta'} \delta (x-x') \; .
\eea

%The advantage of the representation (\ref{eq4.psiphi})
%of $\psi_{\eta} (\tau,ix)$
%is that $c_{\varepsilon \eta} (\tau) = c_{\varepsilon \eta}
%e^{- \tau \ve / \hbar} $, because $c_{\varepsilon \eta}$  diagonalizes $H$
%[\Eq{eq4.Hdiag}]. Therefore
%the Heisenberg equation (\ref{eq4.Schroedinger})
%reduces to an eigenvalue equation for $\phi_{\varepsilon' \eta'}
%(ix,\eta)$:
%\be
%\label{eq4.ESchroedinger}
%       H_{\eta \tilde \eta} (x) \phi_{\varepsilon' \eta'} (ix, \tilde \eta)
%       = \ve' \phi_{\varepsilon' \eta'} (ix, \eta) \; .
%\ee

Now, because scattering takes place only at $x=0$,
for $x>0$ (i.e.\  before the scatterer is encountered)
the wave-function $\phi_{\varepsilon' \eta'} (ix, \eta) $
must correspond to the
free wave-function 
$e^{-i \ve' x}$ of the state $|\varepsilon' \eta'
\rangle_o$.
Thus, we make the following Ansatz:\footnote{
In writing \protect\Eq{eq4.scatwave}\protect,
we have assumed elastic scattering 
($\ve_{in}  = \ve_{out}$). For a 2CK model, 
this holds only if the impurity 
 energy splitting $\Delta = 0$, as assumed in this paper,
so that electrons cannot exchange energy
with the impurity.}
\be
\label{eq4.scatwave}
        \phi_{\varepsilon' \eta'} (ix, \eta) \equiv 
        e^{-i \varepsilon' x}
        \left[ \vphantom{|^Q}
        \tilde U_{\eta  \eta'} (\varepsilon') \theta (-x) +
        \delta_{\eta \eta'}  \theta (x) \right] \; .
\ee
This relation defines the matrix $\tilde U_{\eta  \eta'} (\varepsilon') $,
which clearly can be interpreted as a scattering amplitude,
since it specifies the amplitude for an electron
incident with quantum numbers $(\ve' \eta')$ to 
emerge with quantum numbers $(\ve' \eta)$.

The relation between the scattering amplitude
$ \tilde U_{\eta  \eta'} (\varepsilon')$
and the matrix $U_{\eta  \eta'} (\varepsilon , \ve')$ can be
found by inserting \Eq{eq4.scatwave} into the inverse
of \Eq{eq4.newphi}:
\bea
\label{eq4.Se}
        U_{\eta \eta'} (\varepsilon , \varepsilon' ) &=&
        \int \! {\textstyle {dx \over 2 \pi }}
        e^{i \varepsilon x}
         \phi_{\varepsilon' \eta'} (ix, \eta)
\\
\label{eq4.Sexplicit}
        &=& {1 \over 2 \pi i} \left[
        {\tilde U_{\eta  \eta'} (\varepsilon' ) 
        \over \ve - \ve' - i \epsilon} \:
        - \:  {\delta_{\eta \eta'} 
        \over \ve - \ve' + i \epsilon} \right] \, 
\eea
($\epsilon >0$ is infinitessimally small).
This  shows that $U_{\eta \eta'} (\varepsilon , \varepsilon' )$
is completely known once
 $\tilde U_{\eta  \eta'} (\varepsilon' )$ is known.
The unitarity condition \Eq{eq4.Sunit} on $U_{\eta \eta'}
(\varepsilon , \varepsilon' )$ then immediately implies
unitarity for $\tilde U_{\eta  \eta'} (\varepsilon' )$
(the $\int \! d \tilde \ve $ integral can trivially
be done by contour methods):
\be
\label{eq4.Seeunit}
        \sum_{\tilde \eta}
        \tilde U_{\eta \tilde \eta} (\varepsilon') 
        \tilde U^{\dagger}_{\tilde \eta \eta'} 
        (\varepsilon')  \equiv \delta_{\eta \eta'}\; .
\ee
The unitarity of $\tilde U_{\eta \tilde \eta} (\varepsilon')$
could of course also have been anticipated from
\Eq{eq4.scatwave}: it ensures that scattering conserves probability,
i.e.\  that $\sum_{\eta}  |\phi_{\varepsilon' \eta'} (ix, \eta)|^2$
is the same for $x > 0$ and $x<0$.

The current can be rewritten as follows by 
inserting \Eq{eq4.Sexplicit} into  \Eq{transformedcurrent}:
\widetext
\bea
\label{eq4.Ia}
        I & = &
        {|e| \over N_o h }
       \sum_{\eta \eta' \eta'' } 
         \! \int \!  d \ve'  \! \int \! d \ve''
        \mbox{Re} \left[
        {\sigma \over 2 \pi i} \left( 
        {\tilde U^\dagger_{\eta' \eta} (\varepsilon') 
        \tilde U_{\eta \eta''} ( \varepsilon'' )
        \over \ve' - \ve'' - 2 i \epsilon }
        - { \delta_{\eta' \eta} \delta_{\eta \eta''} 
        \over \ve' - \ve'' + 2 i \epsilon } 
        \right)
        \; \langle
        c_{\varepsilon' \eta'}^{\dagger} c_{ \varepsilon'' \eta''} 
        \rangle \right] \; 
\\
\label{eq4.Ib}
        &=&
        {|e| \over N_o h }
        \sum_{\eta \eta' \eta''}
        \!  \int \!\!  d \tilde \ve'
        \!  \int \!\!  d \tilde \ve''
         \sigma 
        \half   \left[ 
        \tilde U^\dagger_{\eta' \eta} (\varepsilon') 
        \tilde U_{\eta \eta''} (   \varepsilon'' )
        +  \delta_{\eta' \eta} \delta_{\eta \eta''} 
        \right] \delta (\ve - \ve'')  
         \; \langle
        c_{\varepsilon' \eta'}^{\dagger} c_{\varepsilon'' \eta''} 
        \rangle
\\
        &=&
\label{eq4.I}
        \sfrac{|e| }{h}
        \sum_{ \eta' \eta } \!  \int \!\!  d \ve'
         \sigma 
        \half   \left[ \tilde U^{\dagger}_{ \eta' \eta} (\varepsilon' )
       \tilde U_{\eta  \eta'} (\varepsilon') 
        \, + \, \delta_{ \eta' \eta} 
        \right]
        f (\ve' , \eta' ) \; .
\eea
\narrowtext
To obtain \Eq{eq4.Ia}, the $\int \! d \ve$ integral in 
 \Eq{transformedcurrent} was done using contour methods.
\Eq{eq4.Ib} follows since the diagonal nature of
$\langle 
        c_{o \varepsilon' \eta'}^{\dagger} c_{o \varepsilon'' \eta''} 
\rangle $
ensures that 
$
\tilde U^\dagger_{\eta' \eta} (\varepsilon') 
       \tilde  U_{\eta \eta''} (\varepsilon'' )
$
is real, so that we may use 
$
\mbox{Re} \left[ (2 \pi i) (\ve' - \ve'' \mp 2 i \epsilon ) \right]^{-1}
= \pm \half \delta(\ve' - \ve'') $. Finally, to obtain
\Eq{eq4.I}, we used footnote~\ref{f:diagonalm}.
The problem of calculating the current  has thus been
reduced to that of finding
the scattering amplitude  $ \tilde U_{\eta \eta'} (\varepsilon)$.

\subsection{Extracting $\tilde U_{\eta \eta'}$ from 
the Green's Function $G_{\eta \eta}$}
\label{sec:UfromGF}

 Using \Eqs{eq4.psiphi}, (\ref{eq4.scatwave}) and
(\ref{eq4.ceVav}) (with $\mu' = 0$), $G_{\eta \eta'}$ 
of \Eq{eq4.GFtau} can be reduced 
to the form
\widetext
\bea
\label{eq4.GFtauA}
         G_{\eta \eta'} ( \tau, -ir ; \tau', ir')
        &=&
        - \sum_{\tilde \eta \tilde \eta'} \! 
        \int \! d \tilde \ve \, d \tilde \ve'\, 
        \tilde U_{\eta \tilde \eta} (\tilde \varepsilon ) 
        \delta_{\tilde \eta' \eta'}
        \, \langle c_{\tilde \ve \tilde \eta} (\tau)
         c^\dagger_{\tilde \ve' \tilde \eta'} (\tau' ) \rangle
        \,  e^{- i(- \tilde \ve r - \tilde \ve' r') } 
\\
\label{eq4.Gexplicit}
        &=& -  \int \!\!  d \tilde \varepsilon \,
        \tilde U_{\eta \eta'} (\tilde \varepsilon ) \:
        { e^{- \tilde \ve  ( \tau -ir - \tau' - ir')}
        \over e^{- \beta \tilde \ve  } + 1 } \; .
\eea
Its Matsubara-transform is readily found to be
\be
\label{GSpectralform}
        G_{\eta \eta'} (i \omega_n; r, r')
        = \int \! d \ve 
        { \tilde U_{\eta \eta'} (\varepsilon ) e^{i \ve (r + r')}
        \over i \omega_n - \ve } \; .
\ee
This is the central result of this section:
$ G_{\eta \eta'} $ is completely determined by 
$\tilde U_{\eta \eta'} (\varepsilon )$, which is proportional
to the spectral function of $ G_{\eta \eta'} $, as is evident
from the form of \Eq{GSpectralform}. Conversely, 
if $ G_{\eta \eta'} $ is known, $\tilde U_{\eta \eta'} (\varepsilon )$ can be
extracted from it using 
\be
\label{spectralfunction}
\label{eq4.Saae}
         \tilde U_{\eta \eta'} (\varepsilon ) =
        {\textstyle {i \over 2 \pi}}  
        e^{-i \ve (r + r') } 
         \left[ 
        G_{\eta \eta'} (\ve  + i 0^+ ; r, r')
      - G_{\eta \eta'} (\ve  - i 0^+ ; r, r')
        \right]
        \; .
\ee
\narrowtext
{\em Thus, by equating $ G_{\eta \eta'} $ to the corresponding
exact CFT result,  $ \tilde U_{\eta \eta'} (\varepsilon )$
can be extracted from the latter using \Eq{eq4.Saae}.}\/ 
In the next section section, we shall cite the CFT results for
$\bar G_{\bar \eta \bar \eta'} $ and 
$\tilde {\overline U}_{\bar \eta \bar \eta'} (\ve)$
in the $e/o$ basis;
from these,  the above
 $G_{ \eta \eta'} $ and $\tilde U_{ \eta \eta'} (\ve)$,
which were defined in the $L/R$-basis, can be obtained by 
\be
\label{GbarG}
        G_{ \eta \eta'} 
        = N^{\dagger}_{ \eta \bar \eta}
        \bar G_{\bar \eta \bar \eta'} 
         N_{\bar \eta' \eta'} \; ,
\qquad
        \label{cl.UN}
        \tilde U_{ \eta \eta'} (\ve)
        = N^{\dagger}_{ \eta \bar \eta}
        \tilde {\overline U}_{\bar \eta \bar \eta'} (\ve)
         N_{\bar \eta' \eta'} \; .
\ee

In Appendix~\ref{app:simpleexamples} 
the above formalism is illustrated by a simple example, namely
 potential scattering of  two species
of fermions (i.e.\  $\eta = 1,2$).

\subsection{Result of CFT calculation for 
$\bar G_{\bar \eta \bar \eta'} $
and $\tilde {\overline   U}_{\bar \eta \bar \eta'} (\ve)$}

The CFT calculation of  $ \bar G_{\bar \eta \bar \eta'}$
and $\tilde {\overline U}_{\bar \eta \bar \eta'} (\ve) $
in the $e/o$ basis, which follows closely the work of AL,
 is outlined in %%IIIsection~\ref{sec:cftU} of 
paper III.  For present purposes, it suffices to consider CFT
as a ``black box'' that, starting from \Eqs{finalHo}
and (\ref{finalHscat}), allows one to calculate 
the Green's function $ \bar G_{\bar \eta   \bar \eta'} $ of
\Eq{GbarG}, and   produces the following results for
$\tilde U_{ \eta \eta'} (\ve)$ [extracted from 
$ \bar G_{\bar \eta   \bar \eta'} $  using 
\Eqs{eq4.Saae} and (\ref{GbarG})]: 

$\tilde {\overline U}_{\bar \eta \bar \eta'}
(\ve)$ has the form
\be
\label{Ueo}
        \tilde {\overline U}_{\bar \eta \bar \eta'} (\ve) 
        =
        \delta_{\bar \alpha \bar \alpha'}
        \delta_{\bar i \bar i'}
        \left( \begin{array}{cc}
                \Ue &  0 \\
                0 & \quad \Uo
                \end{array}  \right)_{\bar \sigma \bar \sigma'} \; ,
\ee             
where $\Ue$ and $\Uo$ can be interpreted as
the magnitudes of the  scattering
amplitudes in the even and odd channels, respectively.
Since the odd channels decouple, $\Uo = 1$. For the even channels,
 $\Ue$ has the following scaling
form: %%III (see \Eq{cl.GscaleUfinal} in paper~III):
\be
\label{cl.Ueoexp}
%        \tilde {\overline U}_{e \bar \alpha \bar i,
%       e \bar \alpha \bar i'} (\ve) =
%       \delta_{\bar \alpha \bar \alpha'}
%       \delta_{\bar i \bar i'} \Ue (\ve,T) \; , 
%\qquad \mbox{where} \qquad
        \Ue (\ve,T) = \lambda T^{1/2}
        \tilde \Gamma (\ve/T) e ^{i \phi_e} \; .
\ee
Here $e^{i \phi_e}$ is a trivial phase shift\footnote{
We shall assume that the phase shift $\phi_e$ is energy-independent.
In general, it can have an energy-dependence,
$\phi_e = \phi^{(0)}_e + {\ve \over \ve_F}  \phi^{(1)}_e  + \dots$,
but this will be very weak (since $\ve / \ve_{\sss F}$),
and only give rise to subleading
corrections in the conductance, i.e.\ terms of the form 
$(T^{3/2} / \ve_{\sss F}) \Gamma_{(1)} (V/T)$. } 
that can occur in the Kondo channel if 
particle-hole symmetry is broken 
(see \cite[section IV]{AL93}), and $\lambda$ 
is a non-universal constant  (called $\lambda_7$ in paper~III).
 $\tilde \Gamma (x)$ is a universal scaling function, whose 
explicit form was calculated by AL \cite{AL93}:
\bea
\label{ft.tildeGx}
{\tilde \Gamma}(x) 
&=& \left\{ \sfrac{3}{2 \sqrt{2}}
(2\pi)^{1/2} 2 \sin(\pi/2) \int_0^1 du  \right.
\\ \nonumber
& & \times
\left[ u^{(-i x)/(2 \pi)} u^{-1/2} (1-u)^{1/2} F(u)
\vphantom{\frac{\Gamma(2)}{\Gamma^2(3/2)}} \right.  \\
\nonumber
& & \qquad \qquad 
\left. \left. - \frac{\Gamma(2)}{\Gamma^2(3/2)} u^{-1/2} 
(1-u)^{-3/2} \right] \right\} \; .
\eea
$F(u) \equiv F(3/2, 3/2, 1; u)$
is a hypergeometric function.
The $\int \! du$ integral can be done numerically for any value
of $x$, thus giving us an explicit expression for
the scaling function $\tilde \Gamma (x)$. The real and imaginary parts of 
$
        \tilde \Gamma (x) \equiv  
        \tilde \Gamma_e (x) + i \tilde \Gamma_o (x) 
$
have the properties 
\be
\label{propoftildegamma}
        \tilde \Gamma_{e/o} (x) = \pm \tilde \Gamma_{e/o} (- x ) \; ,
        \qquad \mbox{and} \qquad \tilde \Gamma_e (x) < 0 \; .
\ee

\section{Calculation of the Current and Scaling Function}
\label{sec:Gscaling}
\label{ch:calc}
\label{sec:actualKondo}
\label{currentcalc}

We now have all the ingredients for step (S6) of section~\ref{sec:poorVV},
the actual calculation of the current, from which we shall 
extract the desired scaling function $\Gamma (x)$.

\subsection{Calculation of the current}

Using \Eq{cl.UN} to express the current $I$ of \Eq{eq4.I} in
 terms of the $e/o$ scattering
amplitude $\tilde {\overline U}_{\bar \eta \bar \eta'} (\ve)$
of \Eq{Ueo}, we find:
\bea
\label{cl.Scurrent}  
        I &=&  \sfrac{|e| }{h}
        \sum_{ \eta' } \!  \int \!\!  d \ve'
        \half \left[ P_{\eta'} (\ve')
        \: + \:  \sigma' T_{\eta'} \; \right]
        f (\ve' , \eta' ) \; ,
\\
\label{cl.P}
        P_{\eta'} (\ve')  &\equiv &
        \sum_{\eta} 
        \left( N^{\dagger} \tilde{\overline U}^{\dagger} 
        N \right)_{\eta' \eta}
         \sigma T_{\eta}
        \left( N^{\dagger} \tilde{\overline U} N \right)_{\eta  \eta'}
        \; .
\eea

Let us now analyze the matrix product of \Eq{cl.P} index by index. 
All matrices are diagonal in $\alpha, i$, hence the sums
$\sum_{\alpha i}$ in $P_{\eta'}$ are trivial. Next, 
consider matrix multiplication in the index $\sigma$.
Using \Eq{cl.Nexp} for $N_{\bar \sigma \sigma}$, we find
%\be 
%\label{cl.NUeo}
%       \left( N^{\dagger} 
%       \tilde{\overline U} N \right)_{\tilde \sigma  \sigma}
%       = \half \left( \begin{array}{cc}
%              \mbox{$[\Ue + \Uo ]\phantom{xxxx}$}  & 
%               \mbox{$[\Ue - \Uo ] $}\\
%              \mbox{$[\Ue - \Uo ]\phantom{xxxx}$}  & 
%               \mbox{$[\Ue + \Uo ]$}
%                \end{array}  \right)_{\tilde \sigma  \sigma} \; ,
%\ee
\be 
\label{cl.PUeo} P_{\eta'} (\ve') = \sigma'   \mbox{Re}
        \left( {\overline U}^{\dagger {\scriptscriptstyle (o)}} (\ve')
        \Ue (\ve') \right)      \; .
\ee
Note that in spite of the fact that the
current operator is diagonal in $\eta$ 
[see \Eq{geometrical basis statescurrent}], 
$P_{\eta}$
turns out to have $o/e$ cross terms,
$\tilde{\overline U}^{\dagger {\scriptscriptstyle (o)}} \Ue$.
This is a direct consequence of the $L$-$R$ scattering
matrix $v_{\sigma \sigma'}$ introduced in \Eq{bk.HeffV}:
it necessitated the $L/R$-to-$e/o$ basis transformation
$N_{\bar \sigma \sigma}$,
and this  produced a current operator that is off-diagonal
in the $e/o$ basis. The presence of $o/e$ cross terms in
$P_{\eta'} (\ve')$ is extremely important, since 
 $\Ue$, describing Kondo scattering in the even channel, 
 has a $T^{1/2}$ contribution, but
 $\Uo$, describing 
no scattering at all in the odd channel, does not. Thus we see
that our model  contains  a $T^{1/2}$ contribution to the 
current, as observed in experiment (compare property (Cu.6)
in paper~I).

On the other hand, had we attempted to use a model
without $L$-$R$ scattering, i.e.\ with 
$v_{\sigma \sigma'} = \delta_{\sigma \sigma'} $
(such as the model studied by Schiller and Hershfield \cite{SH94}),
no $L/R$-to-$e/o$ basis transformation would have been needed;
then $P_{\eta'} (\ve')$ would be proportional to 
$\tilde U^\dagger \tilde U $, i.e.\ to
\label{sec:Schiller}$(T^{1/2})^2$, 
not $T^{1/2}$. Thus, the inclusion of  $L$-$R$ scattering
into the model is absolutely essential  to obtain the
$T^{1/2}$ dependence. 

In the above presentation, we glossed over one important
subtlety: the scattering matrix $\tilde {\overline U}_{\eta \eta'}$
must be unitary [see \Eq{eq4.Seeunit}], but the form
given in \Eq{cl.Ueoexp} manifestly is not
(since, e.g.\ $\Ue = 0$ for $T=0$). This
reflects the so-called ``unitarity paradox'' \cite{AL93},
according to which the scattering matrix for free fermions
off a 2-channel Kondo impurity into free fermions is
not unitary, which seems to violate the conservation
of probability during a scattering process.
 The resolution of this paradox \cite{ML95,vDZ,Ye}
is that the ``missing probability''
is scattered into a sector of Hilbert
space that cannot be described in terms of linear combinations
of {\em single-particle}\/ fermionic excitations
(compare footnote~\ref{f:Malda})
but has a simple representation when the theory is bosonized.
In %%%%section~\ref{sec:unitpar} of 
paper~III we shall discuss this
issue in more detail, and %%%%in section~\ref{spinorsinnano} of paper III
show how to incorporate
the resulting complications into the present framework. 
The upshot is that the expression (\ref{cl.PUeo}) remains
valid.%%%%, see \Eq{cl.PUeospin} of paper~III. 

\subsection{Calculation of scaling function $\Gamma (v)$}

We now have gathered all the ingredients to derive
the sought-after scaling form for the current and conductance.
Inserting \Eq{cl.PUeo} into \Eq{cl.Scurrent} gives
\widetext
\be
\label{cl.Iexp}
        I \; = \;  \sfrac{|e| }{h} 4 
          \int \!\!  d \ve' 
        \half \left\{ 
        \mbox{Re} \left[ \Ue (\ve') \right]
        + 1  \right\}
        \left[ \vphantom{Q^2}
        f_o (\ve' \! - \! eV /2) - f_o (\ve' \! + \! eV/2 ) \right] \;
,
\ee
where the factor 4 comes from $\sum_{\alpha' i'}$ and 
the sum $\sum_{\sigma'}$, written out
explicitly, gives the two terms in the last factor.  
Now, the conductance can be written in the form\footnote{
To see this, use $ ( \partial_{\ve'} f_o) (\ve' + eV/2)
=  ( \partial_{\ve'} f_o) (- \ve' - eV/2)$ and then change 
integration variables, $\ve' \to - \ve'$ in the second term of
\protect\Eq{cl.Iexp}\protect.
Also recall the sign conventions of footnote~\ref{f:curpos}.}
\be
\label{cl.G}
        G = \left| {\partial I \over \partial V} \right|
        = {2 e^2  \over h}
        \int \!\!  d \ve'
        \left\{ 
        \half \mbox{Re} \left[ \Ue (\ve') \right] 
        + \half \mbox{Re} \left[ \Ue (- \ve') \right]
        + 1 \right\} 
        (- \partial_{\ve'} f_o) (\ve' - eV/2) \; .
\ee
\narrowtext
Thus, using \Eq{propoftildegamma} and (\ref{cl.Ueoexp}), $G$ reduces to
\be       
\label{cl.Ugamma}
        G = 2  \sfrac{e^2 }{h}  \left[ 1
        - \lambda \gamma_o \cos\phi_e
        T^{1/2} \Gamma (\gamma_1 eV /T)  \right] \; .
\ee
Here we introduced a universal scaling function $\Gamma (v)$,
which is defined as follows in terms of
the even part  $\tilde \Gamma_e $
of the exactly known function  $\tilde \Gamma$ of \Eq{ft.tildeGx}:
\be
\label{cl.Gamma}
        \gamma_o\Gamma (\gamma_1 v) \equiv -
        \int \!\!  d x \tilde \Gamma_e 
        \left(  x+ v/2  \right)
        \left[ - \partial_x f_o (x) \right] \, ,
\ee
where $v \equiv eV/T$, $x = \ve' / T$, $f_o (x) = 1 
/ ( e^x + 1) $. 
The positive constants 
 $\gamma_o$, $\gamma_1$ 
are by definition to be chosen such that $\Gamma (v)$ obeys 
the normalization conditions [compare Eq.~(12) of paper~I]: 
\be
\label{calc.Gammanorm}
\Gamma(0) \equiv 1 \; , \qquad
\Gamma(v) \; \mbox{vs.} \; v^{1 \over 2} \; \mbox{has \,\, 
slope \, = \, 1\,\,
as}\; v^{1 \over 2} \to \infty \; ,
\ee
and a minus sign has been included 
in the definition (\ref{cl.Gamma}) of $\Gamma$,  since
$\tilde \Gamma$ is negative definite [see \Eq{propoftildegamma}].

Thus, we have shown that within the present
model, the conductance obeys\footnote{
Note that consistency with the sign of the experimental
zero-bias anomaly  requires that
$B = - 2 e^2/ h  \lambda \gamma_o \cos \phi_e $
must be $> 0$, i.e.\  $\lambda \cos \phi_e < 0$.
This is in  agreement with AL \protect\cite[p.~7309]{AL93},
who concluded (for the case $\phi_e = 0$) that $\lambda < 0$ 
in the regime where the Kondo coupling constant is below
its critical value, $\lk < \lk^\ast$, i.e.\ 
if one flows towards $\lk^\ast$ from the weak-coupling
regime.}  the
scaling relation
\be
\label{cl.condGscale}
        G (V,T) = G_o + B T^{1/2} \Gamma   (\gamma_1 v) \; ,
\ee
with the universal scaling function $\Gamma (v)$, given by
\Eq{cl.Gamma}, known exactly. It is plotted as curve~6
 in Fig.~\ref{fig:scalingcurve}.

Note that this function is the same as that found in 
Eq.~(19) of paper~I (for $m=1$ there)  
by a back-of-the-envelope calculation. The reason
for this agreement is that  $\tilde \Gamma_e (x)$
also turns out to determine the
bulk scattering rate $\tau^{-1} (\ve,T)$ through 
the relation
\bea
\label{ft.selfenergy}
\lambda T^{1/2} \tilde \Gamma_e (\ve, T) &\propto&
2 \left(\mbox{Im}  \Sigma^R (\ve,T) - \mbox{Im} \Sigma^R (\ve,0) \right) 
\\ \nonumber
&=& -  \left( \tau^{-1} (\ve,T ) -  \tau^{-1} (\ve,0 ) \right)  \; ,
\eea
where $ \Sigma^R (\ve,T) $ is 
the retarded bulk electron self-energy
 calculated by AL \cite[eq. (3.50)]{AL93}.
This {\em a posteriori}\/
justifies the assumption made in section VI.A.2 of paper~I,
namely \label{p:boldjust} that the nanoconstriction conductance will be 
governed by $\tau^{-1} (\ve, T)$.

Note that according to the above calculation and \Eq{cl.condGscale}, 
the slope of the scaling curve $[G(V,T) - G(0,T)]/ BT^{1/2}$
seems to be universal, whereas in experiment it is
not [\label{p:universalslope}see Fig.~11(a) of paper~I]. 
The reason is
that in our calculation we assumed that the impurity
sits exactly at the center of the nanoconstriction,
where the non-equilibrium between $L$- and $R$-movers is
strongest,  and hence feels the full effect of the applied voltage.
However, as was explained in Section~III
of paper~I, an impurity not sitting exactly at the center of the
constriction experiences an effective voltage $a_i V$,
where the geometrical constant  $a_i$ (of order unity)
depends on the position of the $i$-th impurity.
When summing over all contributing impurities,
one thus finds expression (21) of paper~I,
which is simply a sum of terms of the form (\ref{cl.condGscale}), 
evaluated at slightly different voltages,
corresponding to different impurity positions in the nanoconstriction.
Our lack of knowledge about the  $a_i$'s forces us to 
introduce another non-universal scaling factor $A$,
and use the scaling form  
\be
\label{IIhowtoplot}
        G (V,T) = G_o + B T^{1/2} \Gamma   ( A v)       \; ,
\ee
when comparing theory with experiment below (see 
Eq.~(13) of paper~I). When checking in the next section
 whether epxerimental (or numerical) data for $G(V,T)$ obeys this 
relation, we shall  plot it in {\em maximally
normalized}\/ form (see Section~VI~B~3 of paper~I),
i.e.\ we shall  plot 
\be
\label{maxnorm}
        { G(V,T) - G(0,T) \over B T^{1/2}} \quad 
        \mbox{vs.} \quad
        (A v)^{1/2}\; ,
\ee
with $A$ determined by the requirement that the 
asymptotic slope of the resulting function be equal to 1
[compare \Eq{calc.Gammanorm}].  According to \Eq{IIhowtoplot},
curves with different $T$ should all collapse onto each 
other when plotted in this way, and the resulting curve should be 
identical to the universal curve $\Gamma (v) - 1$ vs.\ $v^{1/2}$.

\subsection{Deviations from Scaling}
\label{devfromscaling}

It should be emphasized that the scaling relation
found above is only expected to hold for  
$T/\Tk \ll 1$, because it is based on keeping
just the leading term in an  expansion of $G_{\eta \eta'}$
in powers of $T/ \Tk$. 
If $T/\Tk $  is not $\ll 1$, subleading  correction terms proportional 
to  $(T/\Tk)^{\alpha_n} $,
that have been neglected in the calculation
of $G_{\eta \eta'}$, will become important. They will give 
contributions of the form $(T/ \Tk)^{\alpha_n} \Gamma_n (v)$,
which will cause deviations from scaling.
In principle, it would be possible to calculate
the functions $\Gamma_n (v)$ within our CFT approach,
to obtain
\be
\label{lambdan}
{G(V,T) - G(0,T)\over B T^{1/2}}
=\sum_{n=0}^{\infty}
\lambda_n \  (T/ \Tk)^{\alpha_n } \Gamma_n (v) \; .
\ee

Here, all $\alpha_n$ are positive and thought to
be arranged in increasing order.
These terms arise from all the possible irrelevant operators,
generated upon the renormalization group flow
from the weak coupling to the strong coupling fixed point.

If one takes as starting point for the calculation some realistic general
anisotropic Kondo model which depends on a number of different couplings
(such as that emerging from Zawadowski and coworkers' RG analysis
of the original TLS-electron interaction, see Appendix~\ref{app:bk}), 
the amplitudes $\lambda_n$ of the subleading terms will depend on these
couplings and hence be non-universal. 
For the purposes of comparing theory with experiment,
these constants would have to be treated
as fitting parameters, leading to more freedom than one would want for
a meaningful\label{p:fitting} comparision of theory and experiment. 
Therefore
we have focussed in this paper only on the $n=0$ term, with
$\alpha_0=0$, $\lambda_0=1$ (by choice of normalization)
 and   $ \Gamma_o (v) =\Gamma (v)$.

On the other hand, for the sake of comparing our CFT 
calculation with other theoretical methods such
as the non-crossing approximation discussed below,
it is useful to consider the ``idealized'' situation that the starting
Hamiltonian is the isotropic 2CK model of Eq.~(\ref{cl.NHk}).
Then there really is only one coupling constant, $v_K$,
which determines $\Tk$, and all the amplitudes $\lambda_n$ are 
expected to be universal. This implies that the (maximally
normalized) differential conductance of  (\ref{lambdan})
would be  a universal function of two scaling variables,
\be
\label{universalF} 
{G(V,T) - G(0,T)\over B T^{1/2}}
=  \Gamma (T/ \Tk, V/T) \, ,
\ee
which could in principle be meaningfully compared to other calculations.

\section{Final Result for Scaling Curve}
\label{sec:expcompare}
\label{ch:comp}

In this section we compare the CFT prediction (\ref{cl.Gamma}) for the
universal scaling curve $\Gamma (v)$ to the experimental scaling curve
of Fig.~11(b) of paper~I. We also compare it to the results of
Hettler, Kroha and Hershfield (HKH) \cite{HKH94,HKH95}, who computed
the non-linear conductance within the non-crossing-approximation
(NCA).

\subsection{A Few Words on the NCA Method}
\label{sec:HKH94}
 
In order to understand what HKH did, a few
introductory remarks about the NCA method
and a summary of HKH's results are in order here.
Some more details  may be found in Appendix~\ref{NCA}. 

HKH adopt an $SU(2)$, 2-channel Anderson Hamiltonian 
in the slave-particle representation [see
Eq.~(\ref{calc.Anderson})] that in the infinite-$U$ limit that can be mapped
by a Schrieffer-Wolff transformation onto the NTKM of \Eq{bk.HeffV}
and~(\ref{cl.VKroha}).  The two models are therefore in the same universality
class and describe the same low-energy physics.  HKH treat their model with
the NCA technique \cite{Bic87}, which they generalize to $V \neq 0$ using
Keldysh techniques.  

The NCA approach is a self-consistent summation of an infinite set of selected
diagrams, and hence was conventionally viewed
as an uncontrolled approximation in the sense that in
general there is no small parameter.  However, when 
applied to the $N \to \infty$, $k \to \infty$ limit, with $k/N$ fixed, of the
$SU(N)$, $k$-channel Anderson model (Eq.~(\ref{calc.Anderson}) with $k$
channels of electrons $i = 1, \dots, k$, each with $N$ possible pseudo-spin
values $\alpha = 1, \dots, N$) [which maps under a Schrieffer-Wolff
transformation onto the $U(1) \times SU(N)_s \times SU(k)_f$ Kondo model],
the NCA  was recently shown \cite{CR93} to yield exact results
for some quantities (as discussed in more detail in
Appendix~\ref{NCA}).   In particular, 
for {\em all}\/ $N$ and $k$ (with $k > 2$) the NCA approach 
 gives  \cite{CR93} the {\em same}\/ leading critical exponents as
conformal field theory (i.e.\ the exact ones) for all physical properties
involving the 4-point slave particle correlation functions, including the
impurity spectral function $A_d(\omega)$ which in HKH's calculation governs
the point-contact conductance.  (For example, for general $N,k$ both CFT and
the NCA yield $\alpha = {N \over k + N}$ for the exponent in the scaling
relation (\ref{IIscaling}), which reduces to $\frac12$ for $k = N = 2$.)  The
NCA method can be therefore be regarded as a useful interpolation between the
regime $T/\Tk \ll 1$, where it gives the correct exact critical exponents, and
the regime $T/\Tk \gg 1$, where any perturbative scheme works.  Moreover, when
combined with the Keldysh technique, it deals with the non-equilibrium aspects
of the problem in a more direct way than our CFT approach (it can be regarded
as a self-consistent determination of the scattering amplitudes), and is able
to go beyond the weakly non-equilibrium regime ($V \ll \Tk$).

Therefore, it is certainly meaningful to compare the NCA results of HKH to
ours.  CFT serves as a check on how well the NCA does at $V=0$ and very low
temperatures, where CFT is exact and NCA only an uncontrolled approximation.
As we shall see, this check confirms the reliability of the NCA method in the
regime of very low energies.  Conversely, the NCA can then be used as a check
on our use of CFT for $V\neq 0 $ situations, for which the NCA presumably is
the more suitable method.

Nevertheless, a word of caution is warranted when comparing NCA and CFT
results: they cannot be expected to agree exactly, even deep in the scaling
regime $T\ll \Tk$, since the Anderson model on which the NCA is based breaks
particle-hole symmetry, whereas the 2CK model and its CFT solution does not.
Moreover, the NCA's deviations from scaling occur when $T/\Tk$ is no longer
$\ll 1$, but this is a regime in which the quality of the NCA can not be
checked using CFT, and where its results for $N=k=2$ can well differ somewhat
from those for the large-$N,k$ limit where it can be trusted better.
In Appendix~\ref{Parcollet}, we suggest
a way around these problems, based on a new approximation
proposed by Parcollet, Georges, Kotliar and Sengupta (PGKS) \cite{PGKS},
which is reminiscent of the NCA but preserves particle-hole
symmetry, and moreover becomes rigorously exact in the limit
$N \to \infty$, $k \to \infty$ at fixed $k/N$.

On the other hand, when comparing theory to experiment, the above
caveats  are probably less important than uncertainties of another
kind: as mentioned after Eq.~(\ref{lambdan}), the fact that
deviation-from-scaling contributions are {\em non-universal}\/ means that they
will {\em always}\/ depend to some extent on unknown quantities such as the
initial values of irrelevant couplings.  In this light, the NCA probably
should be viewed simply as a reasonable (if non-rigorous) interpolation
between to well-checked limits, as mentioned above.

\label{sec:NCAcond}

\subsection{NCA Results of Hettler, Kroha and Hershfield}

HKH calculated the conductance $G(V,T)$ for a series
of temperatures, measured in units of $\Tk$, ranging
from $T/\Tk = 0.003$ to 0.5. \fig{fig:NCAcurves}(a) shows
their results for $G(V,T)$, plotted according to \Eq{maxnorm}
with $A=1$ (i.e.\ without any adjustable parameters).
 The experimental
data for one of the Cu samples of Ralph and Buhrman
(called sample \#1 in paper I), for which $\Tk \simeq 8 K$,
are shown for comparison in \fig{fig:NCAcurves}(b).

The lowest $T/\Tk$ values in \fig{fig:NCAcurves}(a) show good scaling, in
accord with the CFT prediction. However, for larger $T$-values, marked
deviations from scaling occur, just as seen in the experimental curves of
\fig{fig:NCAcurves}(b).  It is one of the strengths of the NCA method that
these deviations from scaling are automatically obtained, without the need for
making a systematic expansion in powers of $T/ \Tk$ and $V /\Tk$, as would be
necessary in the CFT approach.

The striking qualitative
similarity between the two sets of curves in
Fig.~\ref{fig:NCAcurves} can be made quantitative by 
using $\Tk$ as a fitting parameter:
the choice of $\Tk$ determines 
which curves in Fig.~\ref{fig:NCAcurves}(a) and (b)
are to be associated with each other.
Choosing $\Tk = 8K $ for sample 1,
HKH were able to get ``quite good'' \cite{HKH94}
 simultaneous agreement between a significant number 
of the individual experimental data curves and
their NCA curves of corresponding temperature.
This is illustrated
in \fig{fig:NCAindiv} \cite{HKH95} for 3 curves from sample \#~1.
 In other words,
{\em by using a single fitting parameter, $\Tk$, 
HKH  obtained good quantitative agreement
between the NCA and experimental conductance
curves for a whole set of curves.}

\subsection{Comparison of CFT and NCA 
Results with Experimental Scaling Curve}
\label{subsec:expcompare}

Let us denote the result of plotting a given NCA numerical
$G(V,T)$ curve in the maximally normalized form of \Eq{maxnorm}
by $\Gamma (v,T) - 1$. Fig.~\ref{fig:NCAcurves}(a) shows that
for sufficiently small
$T$, the $\Gamma (v,T) -1$ curves for different $T$
all overlap, i.e.\ the NCA results show good scaling as $T \to 0$,
in agreement with the CFT prediction. The $\Gamma (v,T)$ curve with
the smallest $T$ calculated by HKH, namely $T/ \Tk = 0.003$,
is the most likely to agree with the CFT result for $\Gamma (v)$;
the reason is that for this curve the  $T/ \Tk$ deviations from
perfect scaling, which are neglected in the CFT
calculation [i.e.\ $\Gamma(v) = \Gamma(v,0)$], are smallest.

In \fig{fig:scalingcurve}
 we show  the three experimental scaling curves
of Fig.~11 of paper~I (curves 1-3), the CFT prediction
for $\Gamma (v) -1$ from \Eq{cl.Gamma} (curve 4),
and the NCA result for $\Gamma (v,T) -1$, for $T / \Tk = 0.003$
 (curve 5) and $T/ \Tk = 0.08$ (curve 6).
All these curves have been rescaled into the ``maximally
normalized form'' of \protect\Eq{calc.Gammanorm}\protect. 
We see that there
is rather good agreement between the CFT curve and the
$T / \Tk = 0.003$ NCA  result. 
The experimental scaling curves agree with 
neither of these, but agree remarkably well
with  the $T/ \Tk = 0.08$ NCA curve.

To make these statements quantitative,
we compare the values for the universal constant
$\Gamma_1$, defined  as follows from
the asymptotic large-$v$ expansion of $\Gamma (v)-1$
[compare Eq.~(26) of paper~I]:
\begin{equation}
\label{calc.etafunction}
\Gamma(v) -1 \equiv v^{1/2} + \Gamma_1+ O(v^{-1/2}) \, .
\end{equation}
$\Gamma_1$ is the $y$-intercept of the asymptotic slope
of the curve $\Gamma(v) -1$ vs. $v^{1/2}$,
extrapolated back to $v=0$. It measures
``how soon the scaling curve bends up'' towards
linear behavior, and is the single 
parameter that most strongly characterizes the
scaling function (which is otherwise rather featureless).
We find the following values for $\Gamma_1$:
\begin{mathletters}
\label{cl.Gamma1}
\bea
%\begin{array}{rclrcl}
\Gamma_1^{\sss CFT} &=&  - 1.14 \pm 0.10 \, , \\ % \quad &
\Gamma_1^{\sss NCA} (T/ \Tk = 0.003) &=&  - 1.12 \pm 0.10 \, ,
\\
\Gamma_1^{\sss NCA} (T/ \Tk = 0.08) &=&  - 0.74 \pm 0.10 \, , \\
\Gamma_1^{\sss EXP} &=&  - 0.75 \pm 0.16 \, . %    \quad &
%\end{array}
\eea
\end{mathletters}
Hence, {\em the CFT and NCA calculations for $T/ \Tk = 0.003$
agree rather well, which  inspires confidence
in the general reliability of the NCA method at
very low energies}\/. 

The agreement between the experimental curves and the 
$T/ \Tk = 0.08$ NCA curve  could actually have been anticipated,
for the following reason: HKH determined their (only) fitting
parameter $\Tk$ by choosing the value (namely $\Tk = 8 $~K)
that produces the best
agreement between the few {\em lowest}\/-$T$ curves in 
their set of calculated $G(V,T)$ curves and the corresponding
experimental ones. Thus, the very lowest $T$-curve in the experiment
(with   $T = 0.6 $K) is well-reproduced by the corresponding
NCA curve (with $T/ \Tk = 0.08$) because $\Tk$ was specifically
chosen to produce this agreement.\footnote{
The NCA calculations achieved more, though, than merely
fitting one curve with one parameter, because they succeeded
in reproducing quite well a whole set of curves
(see end of section~\ref{sec:HKH94}).}

It  is somewhat surprising, though, that
the difference between the $T/\Tk = 0.003$ and $T/\Tk = 0.08$ NCA
curves  is
so large. Perfect scaling would require all the various $\Gamma (v,T)$ 
curves for different $T$ to overlap, and the fact that they do not shows
that  the deviations from perfect scaling
which are expected to develop as $T/\Tk$ grows are already
significant at values as small as $T/\Tk = 0.08$.

Thus, as first pointed out by HKH,
the NCA results imply that {\em the $T/\Tk$ corrections
to the universal scaling curve
that were neglected in the CFT calculation}\/ 
(see section~\ref{devfromscaling}) {\em are
in fact not negligible in the present experiment:}\/
$T$ is still large enough that they matter, and the
experimental scaling curve is {\em not}\/ the truly universal one.
This conclusion explains why the CFT and experimental
scaling  curves don't agree; it also suggests that
if the experiments were repeated at lower temperatures,
better agreement might be achievable.

One might ask whether our conclusion that deviations
from scaling are important are not in conflict with
the claims in paper~I [property (Cu.6)] that the experimental curves show 
good scaling. The answer is that while the experimental
curves do scale well, they do not  scale quite well enough
to reproduce ``perfect'' scaling. Perfect scaling
requires that the curves overlap completely
{\em when plotted in maximally normalized form}\/
(as in figure~\ref{fig:scalingcurve}), a procedure
that involves rescaling the $x$-axis by a constant
$A$ to make the slope = 1. This procedure is
clearly very sensitive:  even curves that seem
to collapse well onto the same scaling curve
when not maximally normalized [as those in
Fig.~\ref{fig:NCAcurves}(b), or Fig.~8(b) of paper~I],
can show slight differences in slope
in the regime of largish $v$ when they begin
to bend away from the ideal scaling curve (note that some uncertainty is
involved in determining this slope, since the curves
are not perfectly linear in this regime).
 When being brought
into maximally normalized form,
these curves will have their $x$-axes rescaled
by different amounts to make all the slopes equal to 1 
(the exact amount of rescaling needed being subject to
the same uncertainty as the slope), and can  by this rescaling be
sufficiently deformed that they do not collapse
onto each other any more. This is vividly illustrated
by the observation that the $T/\Tk = 0.003$ and 0.08 NCA curves,
that in fact seem to overlap
rather well in the non-maximally normalized form
of Fig.~\ref{fig:NCAcurves}(a), 
differ so markedly in the maximally normalized
form of Fig.~\ref{fig:scalingcurve}.

In short, maximal normalization is very efficient in revealing small
deviations from perfect scaling, which is why the experimental data,
which scales well when not maximally normalized, does not scale so
well under maximal normalization.

One might be tempted to compare the CFT curve with experiment
in non-maximally normalized form, where deviations from
scaling do not reveal themselves so glaringly. However,
this would not be meaningful, because the slope of the CFT scaling
$\Gamma (v)$ curve is universal, whereas
those of the experimental scaling curves are not
(see Fig.~11(a) and
the last paragraph of section~\ref{p:universalslope}). The 
only meaningful comparison between CFT and experiment
is in a form in which the non-universality of the
experimental slopes has been rescaled  away,
i.e.\ the maximally normalized form.

>From a theorist's point of
view, the conclusion that the experimental
scaling curve is not the universal one
and that non-universal $T/\Tk$ corrections do play a role
is somewhat disappointing, 
since for a system about whose microscopic
nature so little is known, the  quantities that 
allow the most compelling comparison between theory
and experiment are universal quantities, which
are independent of the unknown details. However,
disappointing or not, this is the message
of \fig{fig:scalingcurve}.

Nevertheless, the good agreement between the CFT and NCA scaling curves, which
confirms the NCA method to be reasonably reliable [without fully eliminating
the caveats mentioned in the previous section, though], combined with the good
quantitative agreement between the NCA and the experimental conductance curves
when $\Tk$ is used as fitting parameter, allows the main conclusion of this
paper: \\ {\em The 2-channel Kondo model is in quantitative agreement with the
  experimental scaling $G(V,T)$ data.}

\section{Summary and Conclusions}
\label{sec:concc}

The calculation of this paper
was inspired by experiments of Ralph and Buhrman
on ZBAs in quenched Cu nanoconstrictions
(reviewed in paper~I). These
are qualitatively in accord with the assumption
that the anomalies are caused
by two-level systems in the constriction that
interact with electrons according to
Zawadowski's non-magnetic Kondo model,
which is believed to renormalize, at sufficiently
low temperatures, to the 2CK model. 

To obtain a quantitative test of this
interpretation of the experiment, 
 we performed a calculation of the 
non-linear conductance $G(V,T)$ of a nanoconstriction
containing 2-channel Kondo impurities, 
in the weakly non-equilibrium regime (weakly non-equilibrium regime)
of $V, T \ll \Tk$, and extracted
from it a certain universal scaling function $\Gamma (v)$,
which we compare with experimental scaling function.

To model the experimental situation, we introduced
a generalization of the bulk 2CK model, namely
the naconstriction 2-channel Kondo model (NTKM),
which keeps track of which lead (left or right)
an electron comes from and is scattered into.

The main conceptual challenge in the calculation
of $G(V,T)$ was how to deal with the non-equilibrium
aspects of the problem. On the one hand,  standard perturbative
Keldysh approaches do not work for $T \ll \Tk$, where
perturbation theory breaks down for the Kondo problem. On the other
hand, Affleck and Ludwig's conformal field theory
solution (CFT) of the 2CK problem was worked out only for an
equilibrium electron system. 

Therefore we proposed a conceptually new  strategy
(outlined in section~\ref{sec:Y-Kondo}, the heart of this paper)
which combines ideas from CFT with the Hershfield's $Y$-operator
formalution of non-equilibrium problems:
Hershfield showed that the calculation
of non-equilbirium expectation values
becomes simple when they are expressed in terms
of the scattering states of the problem.
We expressed these in terms of certain
scattering amplitudes 
$ \tilde U_{\eta \eta'}$, which we extracted 
from an equilibrium two-point function
$G_{\eta \eta'} = - \langle  T \psi_\eta \psi^\dagger_{\eta'} \rangle$
that is exactly known from CFT.
(This procedure only gives their $V=0$ values,
but we proposed that in the \nflr\ the
corrections of order  $V/ \Tk$ are negligible.)
Once the $ \tilde U_{\eta \eta'}$ were known, the calculation
of the current was straightforward. 

In the present paper, we implemented all parts of 
this strategy, except that which  requires a detailed
knowledge of CFT, namely the calculation of
$G_{\eta \eta'}$. This is discussed in detail in paper~III.

Our result for the scaling curve  $\Gamma (v)$ does not agree with
the experimentally measured scaling function, because
terms of order $T/ \Tk$ that are neglected in our calculation
are apparently not sufficiently small in the experiment;
however,  when our results are combined with the numerical
results of Hettler, Kroha and Hershfield \cite{HKH94}
(which implicitly do include the neglected terms),
{\em quantitative agreement of the 2CK calculations
with the experimental scaling results 
is achieved}\/  (see section~\ref{subsec:expcompare}).

Thus we conclude that the NTKM
is in quantitative agreement with
the experimental scaling data. This lends further
support to the 2CK interpretation of RB's experiments,
and the associated conclusion that they have indeed observed
 non-Fermi-liquid behavior.

However, the theoretical justification for 
assuming that the non-magnetic Kondo model will
under renormalization flow into
sufficiently close proximity of  the non-Fermi-liquid fixed
point of the 2CK model has recently been called into question.
There are presently unresolved theoretical concerns \cite{WAM95,MF95},
summarized  in Appendix~\ref{sec:criticism},
whether a realistic TLS-electron system will
ever flow sufficiently close to
this fixed point
to exhibit the associated non-Fermi-liquid behavior,
because of the inevitable presence of various relevant
perturbations that can prevent the flow towards
this fixed point.

Therefore, not all questions regarding the Ralph-Buhrman experiments 
have been resolved to everyone's satisfaction.
In our opinion, the outstanding question that remains is:
why does the 2CK interpretation of 
this experiment seem to work so well despite
the concerns about  the theoretical justification
for assuming proximity to the 2CK model's non-Fermi-liquid
fixed point? In view of the fact that
at present no alternative explanation for the experiment
is known that is in agreement with all
experimental facts, we believe that the question of the 
flow towards and stability
of the non-Fermi-liquid fixed point of the non-magnetic
Kondo problem is worthy of further theoretical investigation.

{\em Acknowledgements:}
It is a pleasure to thank D. Ralph and R. Buhrman
for an extremely stimulating collaboration, and 
D. L. Cox, D. Fisher, S. Hershield,  M. Hettler, Y. Kondev, J. Kroha,
 M. Moustakas,  A. Schiller, N. Wingreen, G. Zar\'and and 
in particular A. Zawadowski for discussions. This work was 
partially supported  by the MRL Program of the National Science
Foundation, Award No. DMR-9121654, and Award No. DMR-9407245 
of the National Science Foundation.

\newpage

\widetext
\appendix
\narrowtext

\section{Semiclassical Description of Non-Equilibrium Transport}
\label{sec:sharvin}
\label{app:Vtransport}

In order to motivate the form
of the free density matrix $\rho_o$ introduced in 
section~\ref{sec:freedensity},
we recall in this appendix some standard results from the
semi-classical theory of non-equilibrium transport
through a ballistic nanoconstriction. 
Usually, this is described using a semi-classical
 Boltzmann formalism to calculate the semi-classical
electron distribution function $f_{\vec k} (\vec r)$ 
and the electrostatic potential energy $ e \phi (\vec r)$.
This was first worked out in \cite{KSO77,OKS77}; a very  
careful treatment may  be found in  \cite{KOS77}, which is well-reviewed
 in \cite{JvGW80}. A more up-to-date review is \cite{DJW89}. 

In the semi-classical 
 strategy, one first calculates $f^{(0)}_{\vec k} (\vec r)$ 
and $ e \phi^{(0)} (\vec r)$, the distribution function
and electrostatic potential in the absence of any electron
scattering mechanism, and thereupon uses these functions
to calculate the backscattering current due to 
electrons that are backscattered while attempting to traverse the hole.
The results for  $f^{(0)}_{\vec k} (\vec r)$ and $ e \phi^{(0)} (\vec r)$
are standard and shown in Figs.~\ref{fig:fkr} and 
\ref{fig:electrstat}.\footnote{Our figures and
 arguments are given for the case
$eV > 0$. We take  $e = - | e |$ and hence $V= -|V|$.
With $\mu \pm e V/2$ for $R/L$ leads, 
there then is a net flow of electrons from
right to left, \label{f:curposa}and the current to the right is 
positive.}
Fig.~\ref{fig:fkr} is a position-momentum
space hybrid, showing $f^{(0)}_{\vec k} (\vec r)$ at $T=0$, 
with its $\vec k$-space origin drawn at the position $\vec r$
to which it corresponds. One can understand Fig.~\ref{fig:fkr}(a)
 almost without calculation, simply
 by realizing that in the absence of collisions,
electrons will maintain a constant total energy $E_{\vec k}$.
Thus, an electron that is injected from $z = \pm \infty$ in
the $R/L$ lead with total energy
$E_{\vec k} (z = \pm \infty)  = \ve_{\vec k} \pm e V/2$ 
and traverses the hole, will experience a change in 
its potential energy from $e \phi (\pm \infty) = \pm e V/2$ 
to $e \phi (\mp \infty) = \mp eV/2 $ and hence
accelerate or decelerate     in such a way that 
$E_{\vec k} ( \vec r) = \varepsilon_{\vec k} + e \phi (\vec r)  $
remains  constant.

The key feature of Fig.~\ref{fig:fkr} is that the
distribution of occupied electron states in momentum space, at any
point $\vec r$ in the vicinity of a ballistic constriction, 
is highly anisotropic and consists of {\em two}\/
sectors, to be denoted by $L$ and $R$.
The $L/R$ sector contains the momenta
of all electrons that are {\em incident as $L/R$-movers,}\/
i.e.\  originate from the $\pm V/2$ or 
$R/L$ side of the device, {\em and}\/  have reached $\vec r$ along ballistic
straight-line paths, including paths that traverse the hole
(the bending of paths due to the electric field is of order 
$eV /\varepsilon_F$ and hence negligible). 
At a given point $\vec r$, 
the momentum states in the $L/R$  sectors 
are filled up to a maximum energy of $\left(E_{\vec k} (\vec r)
\right)_{\sss F} $ which,
because of energy conservation along
trajectories, is equal to $ \mu \pm e V/2$, 
the Fermi energy at $z = \pm \infty$
from where the electrons where injected.
 Thus, for $\vec k$ in the $L/R$ sector, one finds
\bea
\nonumber
        f^{(0)}_{\vec k \in L/R } (\vec r) &= &
        f_o \! \left[ E_{\vec k} (\vec r) - \left(E_{\vec k} (\vec r)
        \right)_{\sss F}  \right]
\\ \label{eq:2.fr}
        &=& f_o \! \left[ \ve_{\vec k} - \left(\mu \pm eV/2- e \phi^{(0)}
        (\vec r) \right)  \right] \; .
\eea
Fig.~\ref{fig:electrstat} 
shows that 
$e \phi^{(0)} (\vec r)$ changes smoothly from $- e V/2$ 
to $+e V/2$ (the change occurs  within a few 
constriction radii $a$ from the hole).
It is worth emphasizing, though,  that the electrostatic potential
energy $e \phi (\vec r)$  plays only an indirect role when
it comes to calculating low-energy (i.e.\ $T/\varepsilon_F, 
V/\varepsilon_F \ll 1$)
transport properties.  The reason is simply that
the only role of $e \phi (\vec r)$ is to define the {\em bottom}
of the conduction band, hence causing acceleration and
deceleration of electrons to maintain $E_{\vec k}(\vec r) = constant$.
 Low-energy transport
properties, however, are determined by what happens at the
{\em top}\/ of the conduction band, in particular 
by the sharply anisotropic features characterizing 
Fig.~\ref{fig:fkr} and \Eq{eq:2.fr}.\footnote{
This is illustrated, for example, in the calculation
of the Sharvin formula for the conductance $G_o$ of the
a circular constriction (radius $a$) in the absence of scattering
\cite{JvGW80}:
\be
\label{eq:2.current}
        I_o = \int_{\sss hole} \! \! \! \! \! \! dx dy \,
        { 2 e \over \mbox{Vol}}
         \sum_{\vec p} ({v}_{\vec p})_z
        f_{\vec p} (x,y,z=0) = 
         a^2 e^2 m \ve_{\sss F} / (2 \pi \hbar^3) |V| \; .
\ee
It depends on the electrostatic potential only through
 $ e \phi (x,y,z=0) = 0$, and it is easy to 
verify that the $V$-dependence arises
solely from the $L/R$ anisotropy of $f_{\vec p} (x,y,z=0)$.}

The above considerations suggest that 
the essence of the non-equilibrium nature
of the problem will be captured correctly if
we adopt the following  simplified picture: ignore
the spatial variation of the electrostatic
potential $ e \phi (\vec r)$ altogether,
and  simply consider two leads ($R/L$) with
chemical potentials (measured relative to the 
equilibrium $\mu$) $\mu_\eta = \sigma \half e V$,
which inject $L/R$-moving ballistic electrons into each other
(recall that $\sigma = (+,-)$ for $(L,R)$-movers).
The two leads are assumed infinitely large and
hence ``independent and unperturbed'', in the sense 
 that their thermal distribution properties 
are not perturbed when a small number of electrons 
are transferred from one to the other. 
This simplified picture is the basis for the Ansatz~(\ref{eq4.rhoo})
for the free density matrix $\rho_o$ in section~\ref{sec:freedensity}.

\section{The Bulk Non-Magnetic Kondo Model}
\label{ch:bk}
\label{sec:VZ}
\label{sec:bk}

In this appendix we recall some basic properties of Zawadowksi's
non-magnetic (or orbital) Kondo model for 
the interaction of a TLS with conduction electrons in a bulk metal.
Thus, this appendix
provides the background material that was assumed known
when we introduced the nanoconstriction
two-channel Kondo model in section~\ref{sec:pcmodel}.
Zawadowski proposed his model in  Ref.~\cite{Zaw80}, subsequently
developed it with his coworkers in Refs.~\cite{BVZ82,VZ83,VZ85,VZZ86,VZZ88},
and rather recently, together with   Zar\'and, introduced 
some important refinements \cite{Zar93,ZZ94a,ZZ94b,Zar95}.
Brief, lengthy and exhaustive reviews may be found
in \cite{ZZ96a}, \cite{ZV92,ZV97} and \cite{CZ95}, respectively.

\subsection{Zawadowski's Bulk Bare Model}
\label{sec:defmodel}

\label{sec:genconsiderations}

\begin{figure}
\vphantom{.}
\vspace{1cm}
\centerline{\psfig{figure=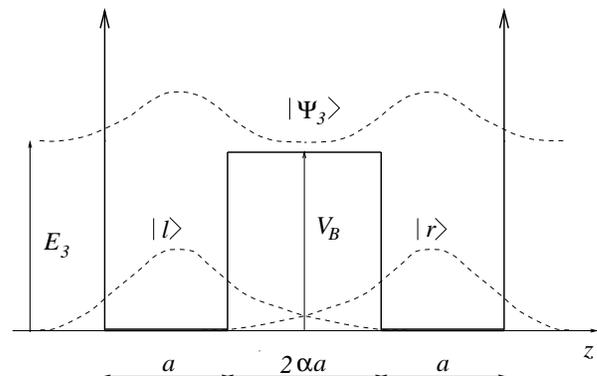,width=0.9\linewidth}}
\vphantom{.}
\vspace{1cm}
\caption[A symmetrical square double well potential
and its left, right and first excited eigenstates.]{\label{fig:excitedstates}
%[[4.3]]
A symmetrical square double well potential (heavy line),
such as that used
by Zar\'and and Zawadowski for their model calculations,
and the wave-functions for the states $| r \rangle$,
$| l \rangle$ and the first excited state $| \Psi_{3} \rangle$.
% We show a square well 
%to illustrate the parameters chosen
%by Zar\'and and Zawadowski for their model calculations 
%\protect\cite{ZZ94b}\protect.
%The choices most favorable for obtaining a large 
%Kondo temperature were \protect\cite[Table I]{ZZ94b}\protect:
%$a = 0.1 \AA$, $\alpha = 2.5$; the choices
%$V_{\sss B} = 494$~K or 740~K then gave $E_3 = 245$~K or 456~K,
%and $\Tk = 2.76$~K  or 3.19~K, respectively.
%The effective couplings $v^x$, $v^y$ and $v^z$,
%all equal to $v_{\sss K}$ at the 2-channel fixed point
%[see \protect\Eq{bk.Hinteff}\protect], are then on
%the order of $v_{\sss K} \simeq 0.1 -  0.2$ .
%The parameter that most strongly influences
%$\Tk$ is $\alpha$, because it affects the
%overlap between the states $| r, l \rangle$, and
%$|\Psi_3\rangle$. Even though the barrier is
%very high relative to the other energy scales, 
%Kondo temperatures in experimentally accessible ranges
%result.}
}
\end{figure}

\begin{figure}
\vphantom{.}
\vspace{1cm}
%\centerline{\psfig{figure=2.3.ps,width=0.9\linewidth}}
\centerline{\psfig{figure=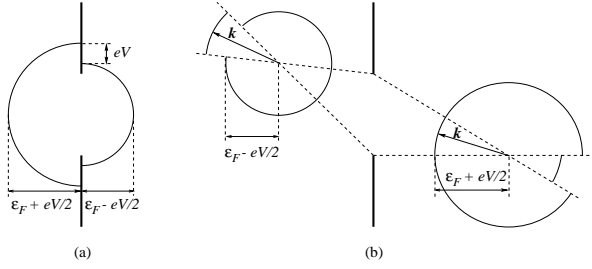,width=0.9\linewidth}}
\vphantom{.}
\vspace{1cm}
\caption[The electron distribution function
in a nanoconstriction.]{
%[[2.2]]
\label{fig:fkr} (taken from \protect\cite[Fig.~7]{JvGW80}\protect):
The $T=0$ electron distribution function $f^{(0)}_{\vec k} (\vec r)$
shown (a) {\em at}\/ the
hole and (b) at two points near the hole. The picture is a 
position-momentum
space hybrid, showing the momentum-space distribution
function $f^{(0)}_{\vec k}$ with its origin drawn at the position $\vec r$
to which it corresponds. A finite temperature simply smears out
the edges of the two ($R/L$) Fermi seas.}
\end{figure}

\begin{figure}
\vphantom{.}
\vspace{1cm}
%\centerline{\psfig{figure=2.4.ps,height=3cm}}
\centerline{\psfig{figure=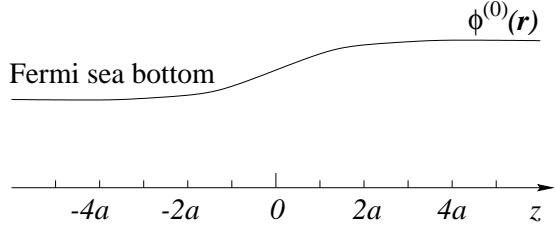,height=3cm}}
\vphantom{.}
\vspace{1cm}
\caption[The electrostatic potential energy $e \phi^{(0)}$
in a nanoconstriction.]{
%[[2.3]]
\label{fig:electrstat} (taken from \protect\cite[Fig.~6]{JvGW80}\protect):
The electrostatic potential energy $e \phi^{(0)} (\vec r)$,
which defines the {\em bottom}\/
of the conduction band, 
near a point contact with radius $a$, shown along the $z$-axis
for the case $eV > 0$. 
Within a few radii $a$ from the hole,
$e \phi^{(0)} (\vec r)$ changes smoothly from $-eV/2 $
on the left to $+ eV/2$ on the right.}
\end{figure}

\begin{figure}
\vphantom{.}
\vspace{1cm}
\centerline{\psfig{figure=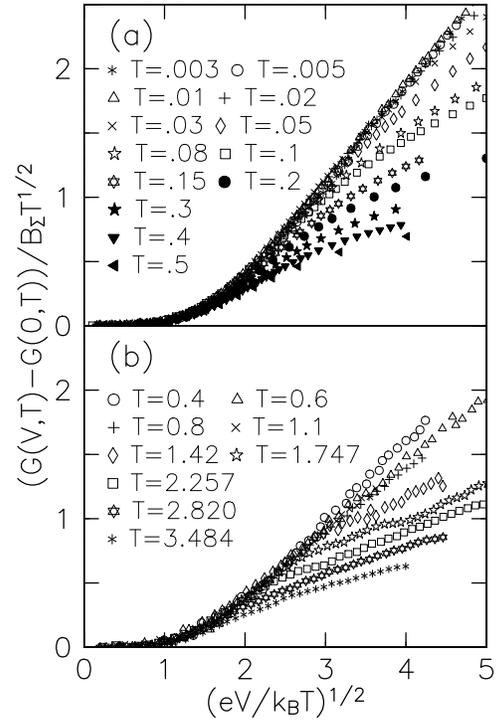,width=1.1\linewidth}}
\vphantom{.}
\vspace{4cm}
\caption[Comparison of NCA scaling plots
and experiment for the rescaled conductance curves.]{
\label{fig:NCAcurves}
%[[9.4]] 
Scaling plots of the conductance for (a)
the NCA calculations of Hettler {\em et al.}\/
\protect\cite{HKH94}\protect\ and
(b) experiment (sample \#1). With
$B_{\sss \Sigma}$ determined from the
zero-bias conductance,   
 $ G(0,T) = G(0,0) + B_{\Sigma} T^{1/2}$
[Eq.~(60)],
 there are no adjustable parameters.
The temperatures in the NCA- and experimental 
plots are in units of $\Tk$ and Kelvin, respectively.
}
\end{figure}

\begin{figure}[t]  %[p]
\vphantom{.}
\vspace{4cm}
\centerline{\psfig{figure=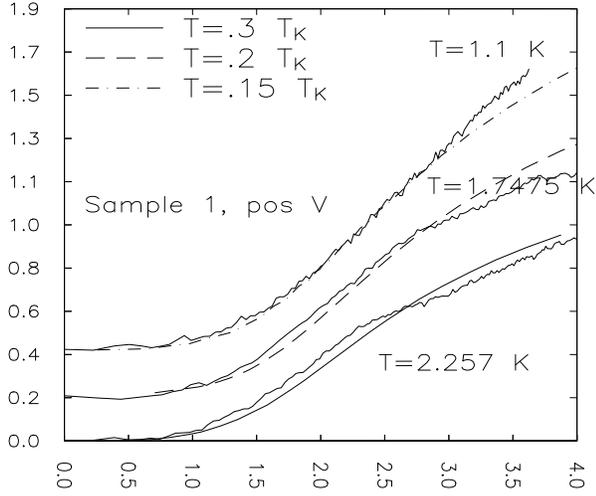,width=0.9\linewidth}}
\vphantom{.}
\vspace{1cm}
\caption[Comparison between
NCA theory and experiment for three individual
conductance curves from sample \#~1.]{\label{fig:NCAindiv}
%[[9.5]] 
Comparison between NCA theory and experiment for three individual
conductance curves from sample \#~1. By using
$\Tk$ as a single fitting parameter
and choosing $\Tk = 8 K$ for sample~1, this
kind of agreement is achieved simultaneously for
a significant number of individual curves
[Hettler, private communication], \protect\cite{HKH95}.
The NCA curves shown here correspond to 
$T = 0.3 \Tk = 2.4 $K, $0.2 \Tk = 1.6 $K and
$0.15 \Tk = 1.2 $K (NCA curves for the actual
experimental temperatures of $T = 2.257$K,
$1.745$K and $1.1$K were not calculated.)
}
\end{figure}

\begin{figure}\hspace{-1.8cm}
\vphantom{.}
\vspace{1cm}
\centerline{\psfig{figure=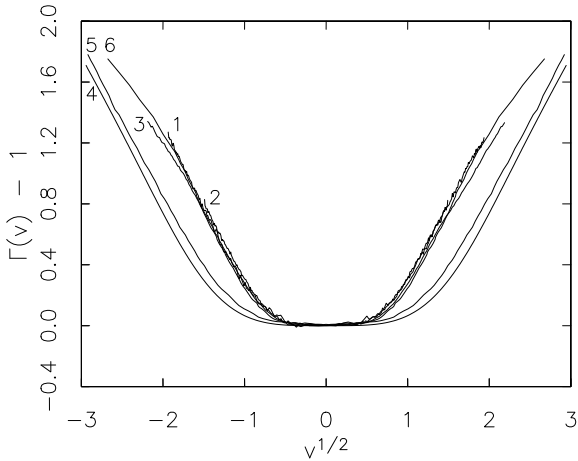,width=0.7\linewidth}}\vspace{3cm}
\vphantom{.}
\vspace{1cm}
\caption[Comparison of the universal 
conductance scaling function $\Gamma (v)$
from experiment, CFT and the NCA.]
{\label{fig:scalingcurve}
%[[9.6]]
The  conductance scaling function $\Gamma (v)$.
Curves 1,2,3 are the experimental curves of
Fig.~11(b) of paper~I. Curve 4 is the CFT prediction
from \protect\Eq{cl.Gamma}\protect. Curves 5 and 6
are the NCA results of HKH, with 
$T/\Tk = 0.003$ and $0.08$, respectively. All curves have
been rescaled in accordance with \protect\Eq{calc.Gammanorm}.}
\end{figure}

Consider a tunneling center (TC)  in a bulk smetal, i.e.\  an atom or
group of atoms that can hop between two different
positions inside the metal, modelled by a double-well potential
[see \fig{fig:excitedstates}, and Fig.~6 of paper~I].
At low enough temperatures and if the barrier is sufficiently
high, hopping over the barrier through thermal activation
becomes negligible. However, if the separation between the 
wells is sufficiently small, the atom can still move between
them by tunneling.

If the tunneling is {\em slow}\/
(hopping rates \cite{CZ95} $\tau^{-1} <10^8 s^{-1} $), 
the atom is coupled
only to the density fluctuations of the electron sea,
which can be described by a bosonic heat bath \cite{YA84,HMG84}.
The tunneling is then mainly incoherent, and
the only effect of the electron bath is  to 
``screen'' the tunneling center: an electron
screening cloud builds up around the center and
moves adiabatically with it, which leads to 
a reduced tunneling rate due to the non-perfect
overlap of the two screening clouds corresponding
to the two positions of the tunneling center.

In this paper we are interested only in the case where the tunneling is
{\em fast}\/ (at rates  \cite{CZ95}
$10^8 s^{-1} < \tau^{-1} < 10^{12} s^{-1}$),
in which case the tunneling center is usually called a two-level system (TLS)
[though in this appendix and the next we shall continue to call
it a tunneling center, because in general more than two states
can be associated with it, see \Eq{bk.TC}].
Then the energy corresponding to 
the tunneling rate, determined by the uncertainty principle,
is in the range 1~mK to  10~K.
(If the tunneling is ``ultra-fast'' ($\tau^{-1} >  10^{12}
s^{-1}$), the energy splitting $E_2 - E_1$
 between the lowest two
eigenstates due to tunneling becomes too large ($> 10 K$)
and the interesting dynamics is frozen out.)
Moreover, the TLS-electron
coupling is assumed strong enough that in addition to 
screening,  an electron scattering off the tunneling center can
{\em directly induce}\/ transitions
between the wells: it can either induce direct tunneling
 through the barrier ({\em electron-assisted tunneling}\/), or
excite the atom to an excited state in one well, from where
it can decay across to the other well ({\em electron-assisted hopping}\/
over the barrier). 

To describe such a system, Zawadowski introduced the following model.
The Hamiltonian is the sum of three terms:
\be
\label{bk.fullH}
        H = H_{\sss TC} + H_{\sss el} + H_{\sss int} \; .
\ee
The first term describes the motion of
the tunneling center the double well,
in the absence of electrons
[see \fig{fig:excitedstates}, and Fig.~6 of paper~I]:
\be
\label{bk.TC}
\label{bk.HTLS}
        H_{\sss TC} = \sum_a E_a b^\dagger_a b_a \; .
\ee
 This problem is 
considered to be already solved: the energies
$E_a$ ($E_1 < E_2 < \dots$) are the 
exact eigenenergies of the   exact eigenstates
$|\Psi_ a \rangle = b^\dagger_a |0 \rangle$ of the tunneling center,
with corresponding wave-functions $\varphi_a (\vec R)$.
The spectrum will contain two nearly-degenerate 
energies $E_1$ and $E_2$, split by an amount $
E_2 - E_1$, corresponding to even and odd
linear combinations of the lowest-lying eigenstates
of each separate well; the remaining energies,
collectively denoted by $E_{\sss ex}$, correspond
to more highly excited states in the well, with 
$E_{\sss ex} - E_2$ typically on the order of the Debye temperature of the
metal, i.e.\  several hundred Kelvin.

The tunneling-center-electron interaction  is described by
a pseudo-potential $V(\vec R- \vec r)$,
which describes the change in energy of 
the tunneling center at position $\vec R$ due to the presence
of an electron at position $\vec r$, 
and is assumed to depend only on the relative 
coordinate $\vec r - \vec R$:
\bea
\nonumber
         H_{int} &=& 
        \sum_i \int \!\! d^3 \vec r \, \Psi_i^\dagger (\vec r) \Psi_i (\vec r)
        V(\vec r - \vec R) \\
&& \label{bk.Hint}
\times         \int \!\! d \vec R \, \sum_{a a'}
        b^\dagger_a \varphi_a^\ast (\vec R)
        b_{a'} \varphi_{a'} (\vec R) \; ,
\eea
where 
$
        \Psi_i ( \vec x) = (\mbox{Vol})^{-1/2} 
        \sum_{\vec p} e^{ i  \vec p \cdot \vec x } 
        c_{o \vec p   i} \; .
$
Here $c^\dagger_{o \vec p   i}$ creates a {\em free}\/ electron
(hence the subscript $o$) with momentum $\vec p$ $(= \! p \hat p)$,
energy $\ve_p$ (assumed independent of the direction
$\hat p$ or $\Omega_{\hat p}$ of $\vec p$) and 
Pauli spin $i= \uparrow, \downarrow$ 
(we use the index $i$ because
this will turn out to be the channel index).
Terms in \Eq{bk.Hint} with $a \neq a'$ correspond to transitions between
eigenstates of the tunneling center induced by the scattering of an electron.

Now, let $\{ F_\alpha ( \hat p) \} $
be any complete set of orthogonal functions of $\hat p$ 
(e.g.\ $F_\alpha ( \hat p) = \sqrt{4 \pi} Y_{lm}  ( \hat p)$,
but in principle
 any  set of orthogonal angular  functions can be used), labelled by
a discrete index $\alpha$ and satisfying 
\bea
\nonumber        \sum_\alpha F^\ast_\alpha ( \hat p) F_\alpha (\hat p')
        &=& 4 \pi \delta (\hat p - \hat p') \; , 
\\ \label{bk.orthogonal}
         \int \! {d \Omega_{\hat p} \over 4 \pi} \,
         F^\ast_\alpha ( \hat p) F_{\alpha'} (\hat p)
       & = & \delta_{\alpha \alpha'} \; .
\eea
Then the electrons' continuous direction index $\hat p $ 
can be traded for the discrete index $\alpha$ by making
a unitary transformation
($N_o$ is the density of states per spin at $\ve_{\sss F}$):
 \bea
\nonumber
         c_{o \ve \alpha i} &=& N_o^{1/2}
         \!  \int \!  {d \Omega_{\hat p} \over 4 \pi} \, 
         F^\ast_\alpha ( \hat p) 
        c_{o \vec p   i} \; , 
\\ \label{bk.transforms}
        c_{o \vec  p   i}
        &=&  N_o^{-1/2}
        \sum_\alpha F_\alpha (\hat p)  c_{o \ve \alpha i} \; ,
\eea
The new set of operators  $\{ c_{o \ve \alpha i} \}$
are labelled by the continuous energy index $\ve $ ($= \ve_p$)
and the discrete index $\alpha$, 
to be called the conduction electron {\em pseudo-spin 
index}, for reasons that will become clear below. 

In the new basis, the electrons' kinetic energy and interaction
with the tunneling center can be written in the following form:
\bea
\label{bk.Helnew}
        H_o &=&  \int_{- D}^D \! \!
         d \varepsilon \sum_{\alpha i}
        \varepsilon \, c^\dagger_{o \ve \alpha i} c_{o \ve \alpha i}
        \; ,
\\
\label{bk.Hintnew}
        H_{int} &=& \sum_{i}
        \sum_{\alpha \alpha'} \sum_{a a'}
        \int_{-D}^D \! \!  d \varepsilon \! \int_{-D}^D 
         \! \! d \varepsilon' \,
        v^{a a'}_{\alpha \alpha'}
        c^\dagger_{o \ve \alpha i} c_{o \ve' \alpha' i}
        b^\dagger_a b_{a'} \; \; .
\eea
For simplicity, the standard assumptions were made that
electron energies
lie within a band of width $2D$, symmetric
about $\varepsilon_{\sss F}$, with constant density of states
$N_o $ per $\alpha,i$ species, and that the energy dependence of
the coupling constants $v^{a a'}_{ \alpha \alpha'}$ can be neglected.
(These assumptions are  justified by the fact that
the Kondo physics to be studied below is dominated by excitations
close to the Fermi surface.)
The $v^{aa'}_{\alpha \alpha'}$ are 
volume-independent, dimensionless  constants (typically
of order 0.1 or smaller),
whose actual values are determined by the potential $V(\vec r- \vec R)$ and
the tunneling center eigenstates $\varphi_a (\vec R)$.

Written in this form, the interaction has the form of a
generalized, anisotropic Kondo interaction: 
$a$ and $\alpha$ can be regarded as impurity- and electron 
{\em pseudo-spin}\/ indices 
(since $\alpha$ takes on infinitely many values,
the electrons have an infinitely large pseudo-spin), 
and the interaction describes 
electron-induced ``spin-flip'' transitions of the impurity.
Note, however, that because the nature 
of the interaction is non-magnetic (to which fact the model owes its name),
the interaction is diagonal in the Pauli spin index 
$i= (\uparrow, \downarrow)$.
Thus we have two identical {\em channels}\/ of conduction electrons,
the $\{ c^\dagger_{o \ve \alpha \uparrow} \} $-
  and the $\{ c^\dagger_{o \ve \alpha \downarrow} \} $ operators, 
and accordingly $i$  is  called  the {\em channel}\/ index.

\subsection{The Renormalized Bulk Model}
\label{sec.renormbulk}

The formal similarity of the interaction of Eq.~(\ref{bk.Hintnew})
with the usual (magnetic) Kondo interaction implies that here too
perturbation theory will fail at temperatures below a characteristic
Kondo temperature $\Tk$, leading to complicated many-body physics as
$T \to 0$ and a strongly correlated ground state.  Perturbation theory
fails for $T< \Tk$ because the effective ($T$-dependent) coupling
constants $v^{a a'}_{ \alpha \alpha'}$ grow as $T$ decreases, and
eventually become too large (see Fig.~\ref{fig:isotropicscaling} in
Appendix~\ref{app:bk}).  The way in which this happens was studied in
great detail by Zawadowski and co-workers. Using Anderson's poor man's
scaling technique to analyse the renormalization group evolution of
the bare model, they concluded that the renormalized model to which it
flows as the temperature is lowered is \cite{MG86} the {\em isotropic
  two-channel Kondo model} (see Eq.~(\ref{bk.Hinteff}) below). Below
we briefly give the starting point and final result of their poor
man's scaling analysis. A summary of the intermediate steps and main
assumptions made along the way can be found in Appendix~\ref{app:bk}.

The interaction vertex, calculated to second order in perturbation
theory, is given by the following expression:
\widetext
\bea
\label{bk.2ndvertex}
 \Gamma^{a a'}_{\ve \alpha \ve' \alpha'} &=& 
        v^{a a'}_{\alpha \alpha'}
        \; + \; 
        \sum_{b \beta} 
         \! \int^D_{-D} \! \! d \bar \varepsilon \; 
        \left[ v^{a b}_{\alpha  \beta} 
        v^{b a'}_{\beta  \alpha'} 
        {1 - f_{ \beta}(\bar \ve)  \over 
        \varepsilon' \!+\! E_{a'} - (\bar \varepsilon \!+\! E_b)}
        \; - \; 
        v^{a b}_{\beta  \alpha'} 
         v^{b a'}_{\alpha  \beta} 
        { f_{\beta} (\bar \ve) \over 
        \varepsilon'  \!+\! E_{a'} - (-  \bar \varepsilon
        \!+\! \varepsilon' \!+\! \varepsilon
        \!+\! E_b)} 
        \right]
\\
\label{bk.2ndvertexbb}
        & \simeq &
        v^{a a'}_{\alpha \alpha'}
        \; + \; 
        \sum_{b \beta} \ln 
        \left[ \mbox{max} \{E_{a'},  E_b, T,  \varepsilon ,
        \varepsilon' \} / D \right]
        \; \left[  v^{a b}_{\alpha  \beta}  v^{b a'}_{\beta \alpha'} 
        -  v^{b a'}_{\alpha  \beta}  v^{a b}_{\beta  \alpha'} \right]
\; ,
\eea
\narrowtext
To obtain the second line, only the logarithmic terms were retained.

Note the occurence of the ``commutator''
$ \left[  v^{a b}_{\alpha  \beta}  v^{b a'}_{\beta \alpha'} 
        -  v^{b a'}_{\alpha  \beta}  v^{a b}_{\beta  \alpha'}
\right]$; the fact that this is in general non-zero, due
to the non-trivial angular dependence of the
coupling constants, is crucial for the presence
of logarithmic corrections (and is the reason
why this model is sometimes called a {\em non-commutative}\/
model). 

Now Anderson's poor man's scaling RG \cite{And70} is
implemented: one changes the bandwidth from $D$ to a slightly smaller
$D'$,  and compensates this change by introducing new coupling
constants that depend on $x = \ln D/D'$, namely
$ v = v(x)$, with the $x$-dependence 
chosen such that $\Gamma^{a a'}_{\alpha \alpha'}$
remains invariant. The procedure is repeated successively until
$D'$ reaches  $ E_{max} = \mbox{max} \{E_{a'},  E_b, T,  \varepsilon ,
        \varepsilon' \}$, at which point the RG flow
is cut off, and the resulting renormalized model,
with coupling constants $v(\ln(D/ E_{max}))$,  has
to be analyzed anew.

The upshot of a lengthy analysis (summarized in Appendix~\ref{app:bk})
is the following: All but the lowest two of
the excited states of the tunneling center  decouple from 
the interaction, which eventually involves an impurity 
with effectively only two states,\footnote{The
two states are considered here in the energy representation, i.e.
their wavefunctions are $\varphi_{1,2} = {1 \over \sqrt{2}} 
(\varphi_{r} \pm \varphi_l)$ in terms of the wavefunctions
$\varphi_{r}$ or $\varphi_l$ describing the tunneling center localized mainly
 in the $r$ or $l$ wells.} $a = 1,2$
(i.e.\  a TLS  with  an effective pseudo-spin $S_{imp} = 1/2$),
with a renormalized splitting $E_2 - E_1  \equiv \Delta$ and Hamiltonian
$H_{\sss TC} = \sum_{a,a'= 1,2}  
b_a^\dagger (\case{1}{2} \Delta \sigma^z_{a a'} ) b_{a'}$.
 Likewise, for the conduction electrons
all but two of the pseudo-spin degrees of freedom,
which we label by $\alpha = 1,2$, 
decouple from the interaction.  
These two ``surviving'' channels, $c_{o \ve 1 i}^\dagger$ and
$c_{o \ve 2 i}^\dagger$, are in general two complicated linear
combinations of the initial $c_{o \ve \alpha i}^\dagger$'s.
They represent
those two angular degrees of freedom that initially were coupled most strongly
to the impurity and for which the couplings hence grow faster under
the renormalization group than those of all other channels
(which hence effectively decouple).
Furthermore, the resulting effective 
interaction is spin-isotropic (spin-anisotropy can be
shown to be an irrelevant perturbation \cite{AL92b}), 
so that the effective
interaction can be written in the form [see \Eq{ap.Hinteff}]:
\bea
\nonumber
        H_{\sss int} &=& 
        \int \!\! d \varepsilon \! \int \! \! d \varepsilon' \! \!
          \sum_{\alpha \alpha' = 1,2} \sum_{a a' = 1,2 }
        \sum_{i = \uparrow, \downarrow}
\\ & & \label{bk.Hinteff}
\times                 v_{\sss K}  \, \left( c_{ o \ve \alpha i}^\dagger
        \half  \vec \sigma_{\alpha \alpha'}
        c_{o \ve' \alpha' i}  \right) \cdot
         \left( b^\dagger_a \half   \vec \sigma_{a a'} b_{a'}
        \right) \; .
\eea
Here  $v_{\sss K}$ is the magnitude of
the effective tunneling-center-electron coupling (and estimated  \cite[Table
1]{ZZ94b} to be of order $v_{\sss K} \simeq 0.1 - 0.2$).
Thus, {\em the effective Hamiltonian}\/\footnote{
Of course, the flow toward the isotropic 2CK model
 only happens provided that all relevant
perturbations that would drive the system
away from this fixed point are sufficiently small --
this implicit assumption 
 will be critically discussed in 
Appendix~\protect\ref{sec:criticism}.}
{\em 
has exactly the form of the isotropic,
magnetic 2CK problem,
with impurity pseudo-spin $S_{imp} = 1/2$ $(a= 1,2)$, 
electron pseudo-spin $ s_{el} = 1/2$
($\alpha = 1,2$), and the Pauli spin $i = \uparrow, \downarrow$
as  channel index.}

When the temperature is lowered even further, then,
provided that $\Delta = 0$,  this model
flows towards a non-trivial, {\em non-Fermi-liquid fixed point}\/
 at $T=0$, at which
the system shows non-Fermi-liquid behavior \cite{AL93,Lud94a}.
However,  $\Delta$ is a relevant perturbation
(with scaling dimension $- \case{1}{2}$,
see section~VI~C of I). This means that if
 $\Delta \neq 0$, the flow towards the
non-Fermi-liquid fixed point will be cut off
when $T $ becomes smaller than $\Delta^2 / \Tk$,
after which the flow will be towards 
a different, Fermi-liquid fixed
point that corresponds to potential scattering off a static
impurity.
In subsequent sections we shall always
adopt assumption (A2) of paper~I
(for reasons explained in Section~VI~C of paper~I)
namely that $\Delta$ is sufficiently small relative to $T$
($ \Delta \ll \sqrt{ T \Tk}$) that the physics {\em is}\/ governed 
by the non-Fermi-liquid fixed point, and that the departure
of the flow from the latter
 towards the Fermi-liquid fixed point has not yet started.

It is tempting \label{MFinterpretation}to propose for
 the effective Hamiltonian of \Eq{bk.Hinteff}  the following physical 
interpretation (which is given in this form by 
Moustakas and Fisher \cite{MF95}, and can be viewed as complimentary to
Zawadowski's picture of electron-induced tunneling).
A charged impurity in a metal
will be screened by a screening cloud of electrons,
which can be thought of as part of the ``dressed'' impurity.
If the impurity is a two-state system, it
will drag along its tightly bound screening
cloud as it tunnels between the wells. In doing so,
it will redistribute the low-energy excitations
near the Fermi surface. In particular,
it will likely interact most strongly 
with two spherical waves of low-energy electrons, ``centered''
on the two impurity positions in the left and right wells
\cite[p.1575]{VZ83},  
with which one can associate a pseudo-spin index $\alpha = L,R$. 
Now, when the impurity and its screening cloud tunnels from the left
 to  the right well, low-energy electrons around 
the right well will move in the opposite direction to the left well,
to compensate the movement of electronic charge bound
up in the screening cloud, and thereby to decrease the orthogonality
between the pre- and post-hop configurations. 
Thus, a flip in the  impurity pseudo-spin is always
accompanied by a flip in electron pseudo-spin,
as in \Eq{bk.Hinteff}. In two very recent papers \cite{MF95},
Moustakas and Fisher have used this interpretation as a starting point
for a related but not quite equivalent  description
of the tunneling center-electron system \cite{MF95}.

\section{Poor Man's Scaling Analysis of Bulk Non-Magnetic Kondo Model}

\label{app:bk}

In this appendix, 
we summarize, following  the recent papers by  
Zar\'and and Zawadowski \cite{ZZ94a,ZZ94b}
and Zar\'and \cite{Zar95,ZV96},  the poor man's scaling
arguments that suggest that the bare, equilibrium  non-magnetic Kondo
model of \Eq{bk.Hint}  renormalizes to the
isotropic 2-channel Kondo model \Eq{bk.Hinteff}.
It should be mentioned at the outset, though, 
that the ensuing analysis has a somewhat heuristic character,  since it
employs scaling equations  derived
in the weak-coupling limit, based on perturbation theory
in the coupling constants. Since such scaling equations
 cease to be strictly valid as soon as one scales
into strong-coupling regions of parameter space,
by  such an analysis the conclusion that the bare model
flows towards the 2CK model
can at best be made plausible, and never be proven conclusively. 
In fact, this conclusion
has recently been called into question \cite{WAM95,MF95},
on the basis of theoretical considerations (controversial themselves
\cite{ZZ96b}), that are discussed in Appendix~\ref{sec:criticism}.

\subsection{Hamiltonian and Initial Parameters}

The starting point is the Hamiltonian introduced and motivated
 in Appendix~\ref{ch:bk}, 
written in the form of \Eqs{bk.HTLS},
 (\ref{bk.Helnew}) and (\ref{bk.Hintnew}).
%\bea
%\label{ap.HTLS}
%        H_{\sss TC} &=& \sum_a E_a b^\dagger_a b_a \; ,
%\\
%\label{ap..Helnew}
%        H_{el} &=&  \int_{- D}^D \! \!
%         d \varepsilon \sum_\alpha
%        \varepsilon \, c^\dagger_{o \ve \alpha i} c_{o \ve \alpha i}
%        \; ,
%\\
%\label{ap.Hintnew}
%        H_{int} &=& 
%        \sum_{\alpha \alpha'} \sum_{a a'}
%        \int\! \!  d \varepsilon \! \int \! \! d \varepsilon' \,
%        v^{a a'}_{\alpha \alpha'}
%        c^\dagger_{o \ve \alpha i} c_{o \ve' \alpha' i}
%        b^\dagger_a b_{a'} \; \; .
%\eea
Let $\Delta_b = E_2 - E_1$ be the bare energy difference
between the two lowest-lying, nearly degenerate
eigenstates states of the well. The remaining energies,
$E_a$, $a = 3, 4, \dots$, 
collectively denoted by $E_{\sss ex}$, correspond
to more highly excited states in the well.

We are interested in the regime where $\Delta_b \ll T \ll E_{\sss ex}
\ll D$. Hence we take $\Delta_b \simeq 0$,
i.e.\  consider a symmetrical double well 
with a two-fold degenerate ground state. 
It is then convenient
to make a change of basis from the exact
symmetrical and anti-symmetrical ground states
$|\Psi_1 \rangle$ and $|\Psi_2 \rangle$
to the right and left states
$|r \rangle$ and $|l \rangle$  $= {1 \over \sqrt{2}} 
\left( | \Psi_1 \rangle \pm | \Psi_2 \rangle \right)
$.

Note that, since a non-zero bare tunneling matrix element ($\Delta_0$)
between the wells always leads to a splitting
$E_1 - E_2 \simeq \Delta_o$, by taking $\Delta_b \simeq 0$ 
we are also  implicitly assuming
that $\Delta_o \ll T$. This means that direct tunneling
events are very unlikely,
raising the question of whether Kondo-physics
will occur at all.\footnote{This was a serious limitation
of  Zawadowski's original model, which did not include
excited states: to give non-trivial
many-body physic (i.e.\  a sufficiently large Kondo energy $\Tk$), 
the bare  tunneling rate $\Delta_o$ could not be too small;
yet at the same time, the model only flows
to the interesting non-Fermi liquid fixed point
if  $E_1 - E_2 \ll T$. This would have required a rather
delicate and perhaps questionable fine-tuning of parameters.
This problem has been overcome by including
excited states in the model \protect\cite{ZZ94a,ZZ94b}\protect,
as explained above.}
However, the inclusion of excited states in
the model overcomes this potential problem as follows \cite{ZZ94a,ZZ94b}:
a careful \label{roleofex}
estimate of the coupling constants \cite{Zar93}
in terms of the overlap integrals (\ref{bk.Hint})
has shown that
\begin{mathletters}
\label{bk.couplings}
\bea
        |v^{r,l}|   &\simeq &10^{-3}  |v^{l,l} - v^{r,r} | \; , 
\\ 
        |v^{l,ex}| &\simeq &
|v^{r,ex}| \simeq |v^{l,l} - v^{r,r} | \; .
\eea
\end{mathletters}
The first relation reflects the fact that direct 
electron-assisted tunneling, para\-me\-te\-rized by
$|v^{r,l}|$, is proportional to the bare tunneling
rate $\Delta_o$ and hence 
very small. However, the second relation
shows that the matrix elements for electron-assisted transitions
to excited states, parametrized by $|v^{l,ex}|$ and
$|v^{r,ex}|$, are of the same order of magnitude
as for the usual ``screening term'' $ |v^{l,l} - v^{r,r} |$
[this is because the overlap integrals in \Eq{bk.Hint}
are larger for $\varphi^\ast_{\sss ex} \varphi_{r,(l)}$
than for $  \varphi^\ast_{r}  \varphi_{l}$,
since the excited state wave-function spreads
over both wells (see \fig{fig:excitedstates})].
Although the amplitudes for such processes are
proportional to  the factor $1/E_{\sss ex}$ (which is small,
since $E_{\sss ex}$ is large), Zar\'and and Zawadowski
showed that such terms also grow under scaling [see \Eq{bk.scalingeq}
below], and eventually lead to a renormalized model which has
sufficiently large effective tunneling amplitudes to
display Kondo physics.

\subsection{Poor Man's Scaling RG}
\label{sec:poorman}

The interaction vertex, calculated to second order in perturbation
theory from  the diagrams
in \fig{fig:diagrams}(a), is given by \Eq{bk.2ndvertexbb}:
\bea
\nonumber
        \Gamma^{a a'}_{\ve \alpha \ve' \alpha'}
        &=&  v^{a a'}_{\alpha \alpha'}
        \; + \; 
        \sum_{b \beta} \ln 
        \left[ \mbox{max} \{E_{a'},  E_b, T,  \varepsilon ,
        \varepsilon' \} / D \right]
\\ \label{bk.Gammaalpha}
& & \times \left[  v^{a b}_{\alpha  \beta}  v^{b a'}_{\beta \alpha'} 
        -  v^{b a'}_{\alpha  \beta}  v^{a b}_{\beta  \alpha'} \right]
\; .
\eea

Now Anderson's poor man's scaling RG \cite{And70} is
implemented (very nicely explained in
\cite[sections 3.2.2]{CZ95}): electron or hole excitations with large
energy values do not directly participate in {\em real} 
physical processes; their only effect occurs through
virtual excitions of the low-energy states to
intermediate high-energy states. Hence such processes
may be taken into account by introducing renormalized 
coupling parameters, which sum up all the virtual processes
between a new, slightly
 smaller cut-off $D'$ and the original D. In other words,
all virtual processes between the energies $D'$ and $D$ 
are integrated out and their contributions
are incorportated in new, $D'$-dependent coupling constants.
This procedure is repeated for smaller and smaller $D'$,
until $D'$ becomes on the order of
 $\mbox{max} \{ E_c, T, \varepsilon_{ p'} \} $.

Concretely, this is done by writing $v^{ab}_{\alpha \alpha'}
= v^{ab}_{\alpha \alpha'} (x)$, where $x = \ln (D'/D)$
and  the $x$-dependence of the coupling constants
is determined by the requirement that
the interaction vertex be invariant under poor man's scaling, i.e.\ 
$       \partial_x \Gamma^{ab}_{\alpha \alpha'}  = 0$.
By \Eq{bk.Gammaalpha}, this leads to the following {\em leading-order
scaling equation}\/:
\be
\label{bk.scalingeq}
        \partial_x \uv^{ab} (x)  =     
        \sum_c \theta (D' - E_c)
        [ \uv^{ac} (x) , \uv^{cb} (x)  ] \; ,
\ee
where we have adopted the matrix notation
$v^{ab}_{\alpha \alpha'} \equiv \uv^{ab}$.
[The significance of $\theta (D' - E_c)$ is explained
in \sect{sec:excited}.]
This equation, to be solved with 
 the boundary condition $ \uv^{ab} (0) = (\uv^{ab})_{\sss
bare}$, determines the nature of the RG flow 
away from the weak-coupling limit.

In the following two sections we outline the results
obtained by Zawadowski and co-workers concerning
the nature of the fixed point that the Hamiltonian
flows towards as it scales out of the weak-coupling
region. However, the arguments that are to follow
all have a somewhat heuristic character: since they
are based on scaling equations that were derived
in the weak-coupling limit, based on perturbation theory
in the coupling constants,  in principle 
they cease to be strictly valid as soon as one scales
into strong-coupling regions of parameter space.
(The only method that gives quantitatively reliable
results for the cross-over region is Wilson's
numerical NRG \cite{Wil75,KWW80,CLN80,AL92b,PC91}.)
Many of the results obtained below are therefore of mainly qualitative
value, and not expected to be quantitatively accurate.

\subsection{Scaling to 2-D Subspace}
\label{sec:2Dsubspace}

Let us for the moment consider
the model without any  excited tunneling center states,
i.e.\ with $\sum_c = \sum_{r,l}$ (as was done in 
the first papers \cite{Zaw80,BVZ82,VZ83}),
postponing  the more general case to \sect{sec:excited}.
In this case, the coupling constants $\uv^{ab} (x)$
can be expanded in terms of Pauli matrices in
the 2-dimensional space of the tunneling center
(in the $l$-$r$ basis),
\be
\label{bk.2statefactor}
        v^{ab}_{\alpha \alpha' } (x)
        = \sum_{A = 0}^3 \tilde v^A_{\alpha \alpha' } (x)       
         \sigma_{ab}^A  \; , \quad \qquad a,b = l,r \; ,
\ee
where $A= (0,1,2,3) = (0,x,y,z)$ and $\sigma^0_{ab} \equiv
\delta_{ab}$. The $v^z$ term is called the {\em screening}\/ term,
and characterizes the difference in scattering
amplitudes for processes in which an electron
scatters from the atom in the right or left
well {\em without}\/ inducing a transition
to the other well. The $v^x$ and $v^y$ terms
are called {\em electron-assisted tunneling}\/ terms,
and describe the amplitude for processes in which 
the scattering of an  electron induces the tunneling center to make a transition
to the other well. According to  \Eq{bk.couplings},
$ \tilde {\uv}^x (0) \simeq  \tilde {\uv}^y (0)
\ll \tilde {\uv}^z (0)$. If one chooses the wave-functions of
the tunneling center to be real,  time-reversal invariance
requires $ \tilde {\uv}^y (0) = 0$ (see \cite[(a), eq.(2.11)]{VZ83}).

The problem is now formally analogous
to a (very anisotropic)
 magnetic Kondo problem in which a spin-$\toh$
impurity is coupled to  conduction electrons
with very large pseudo-spin (since $\alpha$ takes
on a large number of values). 
However, Vlad\'ar and Zawadowski (VZ) have shown \cite[(a), section
III.C]{VZ83} that (with realistic
choices of the initial parameters) the problem always scales
to a 2-dimensional subspace in the electron's $\alpha$-index,
so that the electrons have pseudo-spin $S_e = \toh$
(this happens indepedent of the signs of the initial
couplings, see \sect{sec:isotropic}).
Their argument goes as follows:

In the notation of \Eq{bk.2statefactor}, 
the scaling equation (\ref{bk.scalingeq}) 
takes the form  \cite[p.1573, eq.(3.3)]{VZ83}
\be 
\label{bk.2by2caling}
        {\partial \uv^A \over \partial x}
        \; = \; - 2 i \sum_{BC} \varepsilon^{ABC}  \; \uv^B \uv^C \; .
\ee
Since $ \tilde {\uv}^x (0) \ll \tilde {\uv}^z (0)$ 
and  $ \tilde {\uv}^y (0) = 0$,
\Eq{bk.2by2caling}
can be linearized in $\uv^x$ and $\uv^y$. 
VZ solved the linearized equations
in a basis in $\alpha$-space in which
$ \tilde v^z_{\alpha \beta} (0) $ is diagonal
[$  \tilde v^z_{\alpha \beta} (0) 
= \delta_{\alpha \beta}  \tilde v^z_{\alpha} (0) $],
and obtained the following solution
\cite[p.1576, eq.3.17]{VZ83}:
\bea
\label{bk.expgrowth}
         \tilde v^z_{\alpha \beta} (x) &=&
        \delta_{\alpha \beta}  \tilde v^z_{\alpha} (0) 
\\ 
         \tilde v^x_{\alpha \beta} (x) 
        &=&
         \tilde v^x_{\alpha \beta} (0)  
        \cosh 2x\left[  
        \tilde v^z_{\beta} (0) -  \tilde v^z_{\alpha} (0) \right] \; ,
\\
         \tilde v^y_{\alpha \beta} (x) 
        &=&
         i \tilde v^x_{\alpha \beta} (0)        
        \sinh 2x \left[  
        \tilde v^z_{\beta} (0) -  \tilde v^z_{\alpha} (0) \right] \; .
\eea
Barring unforeseen degeneracies in the matrix
$\tilde  {\uv}^z$, this shows that the two 
elements  of $\tilde  {\uv}^z$ which produce the largest
difference 
$ | \tilde v^z_{\beta} (0) -  \tilde v^z_{\alpha} (0) |$
will generate the most rapid growth in the corresponding
couplings $\tilde v^x_{\alpha \beta} (x) $
and $\tilde v^y_{\alpha \beta} (x) $. In fact, since
this growth is exponentially fast, any couplings
with only slightly smaller 
$ | \tilde v^z_{\beta} (0) -  \tilde v^z_{\alpha} (0) |$
will grow much slower and hence decouple. Thus, we conclude
that according to the leading-order scaling equations,
the system {\em always renormalizes to a 2-D
subspace in which the electrons have pseudo-spin $S_e = \toh$.}\/

\begin{figure}
\vphantom{.}
\vspace{1cm}
\centerline{\psfig{figure=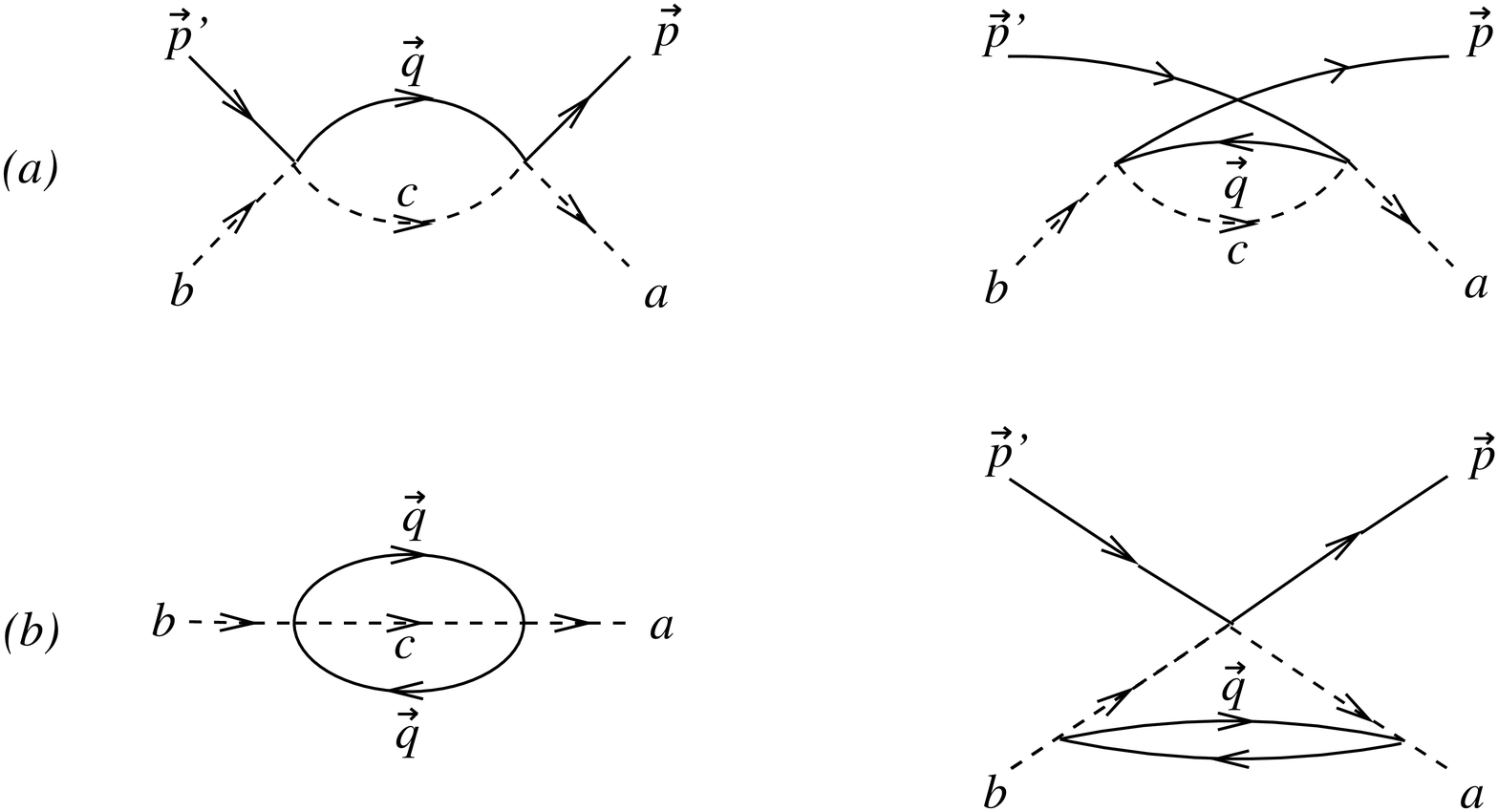,width=0.9\linewidth}}
\vphantom{.}
\vspace{1cm}
\caption[The vertex- and self-energy correction diagrams
that contribute to the poor man's scaling equations.]{
%[[4.4]] 
\label{fig:diagrams} (a) The second-order
vertex corrections that contribute to
\protect\Eq{bk.2ndvertex}\protect\
and  generate the leading order 
scaling equation (\protect\ref{bk.scalingeq}\protect).
(b) The impurity self-energy correction 
and the third-order next-to-leading-logarithmic
vertex correction that generate the
subleading terms  in the second-order
scaling equation (\protect\ref{bk.2ndorderscaling}\protect). 
(Note that subleading diagrams that are generated by the
leading-order scaling relation derived from the diagrams
in (a) have to be omitted.) Dashed and
solid lines denote impurity and electron Green's functions,
respectively.}
\end{figure}

The argument just presented is not quite waterproof, though.
Firstly, it depends on the assumption of extreme initial
anisotropy in the couplings, and secondly, it
is based only on the leading-order scaling equations.
As one scales towards larger couplings, sub-leading
terms in the scaling equations can conceivably become
important.  Zar\'and has investigated this
issue by including next-to-leading-order
logarithmic terms [generated by the diagrams in \fig{fig:diagrams}(b)]
in the scaling equations, which turn out to be
\cite[eq.(2.6)]{Zar95}: 
\bea
\nonumber
        \partial_x \uv^A 
        & = & - 2 i \sum_{BC} \varepsilon^{ABC} 
        \uv^B \uv^C 
\\
& & \label{bk.2ndorderscaling}
-  2 N_f \sum_{B \neq A}
        \left( \uv^A \mbox{Tr} [ (\uv^B)^2 ] 
        - \uv^B \mbox{Tr} [ \uv^A \uv^B  ] \right) \; . 
\eea
Note that the number of channels,  $N_f$ (equal to 2
for the case of interest), shows up here for the first time in
the  next-to-leading order, since each electron loop
[see \fig{fig:diagrams}(b)] carries a factor $N_f$.
Performing a careful analysis of the stability
of the various fixed points that occur, 
he concluded that the above-mentioned  $S_e = \toh$ fixed point
is the only {\em stable}\/ fixed point in of these equations.

Zar\'and and Vlad\'ar also investigated the effect of
the other channels, that don't couple as strongly
as the two dominant ones, near the fixed point
\cite{ZV96}.
They  produce irrelevant operators that eventually
scale to zero (which is why these channels decouple),
but that can nevertheless influence  the critical
behavior near the fixed point.
However,  Zar\'and and Vlad\'ar found that they have the {\em same}\/
critical exponent as the leading irrelevant operator
in the pure 2CK model, which means that these extra operators
don't change the universal critical behavior, merely some of the
corresponding amplitudes.

Since these results are independent of the value of $N_f$ and the number
of orbital channels considered,
and Zar\'and's   analysis  is exact in  the limit $N_f \to \infty$,
he expects his  results to also be valid for $N_f = 2$.
(However,  no completely rigorous proof exists yet for this
expectation; in particular, his analysis 
assumes $\Delta_b = 0$, and the case $\Delta_b \neq 0$
is substantially more complicated, see \cite[(b), section III]{VZ83}.
For another, symmetry-based argument in favor
of $S_e = \toh$, see \cite[section 3.3.2 (iii)]{CZ95}.)

\subsection{The fixed point is Pseudo-Spin Isotropic}
\label{sec:isotropic}

Next one shows, following \cite[section III]{Zar95},
that the  $S_e = \toh$ fixed point is actually {\em isotropic}\/
in pseudo-spin space. 

The last term in \Eq{bk.2ndorderscaling} can be eliminated
from the fixed-point analysis by making  a suitable
 orthogonal transformation
$\uv^A \to \sum_B O_{AB} \uv^B$. Therefore, it is sufficient to 
consider the first two terms on the right-hand side of
\Eq{bk.2ndorderscaling}. At the fixed point, where 
$\partial_x \uv^A   = 0$, we have
\be
\label{bk.Z3.1}
        \sum_{BC} \varepsilon^{ABC} \uv^B \uv^C \; = \; i N_f \uv_A \sum_{B \neq A}
         \mbox{Tr} [ (\uv^B)^2 ] \; .
\ee
Multiplying by $\uv^A$ and taking the trace, one obtains
the three relations
\bea
\label{bk.Z3.2}
& &       i N_f \balpha^A (\balpha^B + \balpha^C) = \bbeta \; , 
    \\     
&& \nonumber
\mbox{where}\quad
        \{      A,B,C \} = \{ x, y,z\}  \quad   (\mbox{cyclically}) \; ,
\eea
where we have defined $\balpha^A \equiv \mbox{Tr} [ (\uv^A)^2 ] $ and
$\bbeta \equiv  \mbox{Tr}( \uv^A \uv^B \uv^C - \uv^C \uv^B \uv^A) $.
This immediately implies one of two possibilities: either 
at least two of the $\balpha^A$'s are zero, which is the 
trivial (commutative)  case without
electron-assisted tunneling ($\uv^x = \uv^y = 0$); 
 or else they are all equal:
\be
\label{bk.isotropic}
        \balpha^A = \balpha^B = \balpha^C = \balpha \; .
\ee
The latter case  is the one of present interest. 
The conclusion that the couplings are all
equal (i.e.\  the effective Hamiltonian isotropic)
was checked numerically by Zar\'and \cite[Fig.4]{Zar95},
and is illustrated in \fig{fig:isotropicscaling}.

What is the matrix structure of the $ \uv^A$'s? Introducing 
the notation $J^A = {1 \over 2 N_f \balpha} \uv^A $,
\Eqs{bk.Z3.1} and (\ref{bk.isotropic}) imply that
the $J^A$ satisfy the $SU(2)$ Lie algebra,
\be
\label{bk.SU(2)}
        [J^A, J^B] = i \varepsilon^{ABC} J^C \; ,
\ee
which means that they must be a direct sum
of irreducible $SU(2)$ representations:
\be
\label{bk.irred}
        J^A = \sum_{\oplus k = 1}^n S_{(k)}^A \; .
\ee
According to the analysis of Zar\'and mentioned
in the previous section, only a single
subspace $S_e = \toh$ in this sum corresponds to 
a stable fixed point (all the others correspond
to unstable fixed points), in the vicinity of which we can therefore
write
 $J^A_{\alpha \alpha'}  = \toh \sigma^A_{\alpha \alpha'}
$.

After a rotation in $\alpha$-space to line up
the quantization axis of the pseudospins of the
impurity and the electrons, the effective Hamiltonian
to which (\ref{bk.Hint}) renormalizes can  be written as:
\bea
\nonumber
        H_{\sss int} &=& 
        \int \!\! d \varepsilon \! \int \! \! d \varepsilon' \! \!
          \sum_{\alpha \alpha' = 1,2} \sum_{a a' = 1,2 }
        \sum_{i = \uparrow, \downarrow}
\\
& & \label{ap.Hinteff}
\times   v_{\sss K} \left( c_{ o \ve \alpha i}^\dagger
        \half  \vec \sigma_{\alpha \alpha'}
        c_{o \ve' \alpha' i}  \right) \cdot
         \left( b^\dagger_a \half   \vec \sigma_{a a'} b_{a'}
        \right) \; .
\eea
Here  $v_{\sss K}$ is the magnitude of
the effective tunneling center-electron coupling (and estimated  \cite[Table
1]{ZZ94b} to be of order $v_{\sss K} \simeq 0.1 - 0.2$).
This is the main result of the RG analysis:
{\em The effective Hamiltonian
has exactly the form of the isotropic,
magnetic 2-channel Kondo problem;
the two surviving orbital indices $\alpha = 1,2$ 
play the role of pseudo-spin indices
and the Pauli spin indices $i = \uparrow, \downarrow$
the role of channel indices.}

An intuitively appealing interpretation of this model,
due to Moustakas and Fisher, is given in Appendix~\ref{ch:bk},
after \Eq{bk.Hinteff}.

We conclude this appendix with a number of  miscellaneous
comments:

The fact that one always scales towards
an {\em isotropic}\/ effective Hamiltonian
is rather remarkable 
(though in accord with the conformal field theory
results that show that anisotropy is an irrelevant
perturbation \cite[eq.(3.17)]{AL92b}): the initial extreme
anisotropy of the couplings is dynamically removed,
and a $SU(2)$ symmetry emerges that is not present in
the original problem!

Note that the initial signs of the
anisotropic coupling constants did not matter in the above arguments.
A more careful argument \cite[section 3.3.2 (ii)]{CZ95} shows that 
the flow toward this fixed point indeed occurs irrespective of
the initial signs of the coupling constants.

{\em Relevant perturbations:}\/
When the initial splitting $\Delta_b$ is non-zero,
the 2-nd order RG is considerably more complicated 
\cite[(b), section III]{VZ83}. The result is that
$\Delta_b$ gets normalized downward \label{p:Deltalower}
by about
two orders of magnitude \cite[(b), Fig.3]{VZ83}.
However, as emphasized in \cite[section 3.4.1 (c)]{CZ95},
{\em the splitting $\Delta_b$ is nevertheless a relevant
perturbation}\/: it can be shown to scale downward much slower than
the bandwidth $D'$, so that $\Delta_b (D') / D'$
{\em grows}\/ as $D'$ is lowered. 

By analyzing the stability of the fixed point equations
against a perturbation that breaks {\em channel}\/
symmetry, it can likewise be shown that
{\em channel anisotropy}\/ is a relevant perturbation
\cite[section 3.4.1 (c)]{CZ95}.

{\em Kondo temperature}\/:
The Kondo temperature is the cross-over temperature at which 
the couplings begin to grow rapidly. It can be estimated
from an approximate solution of the second order scaling equation
(\ref{bk.2ndorderscaling}).\footnote{Since $\Tk$
is only a statement about the {\em onset}\/
of rapid growth of coupling constants,
the value obtained from  scaling 
equations derived by perturbation theory  is expected 
to give  approximately the correct scale even though the scaling
equations themselves become invalid when the couplings become
too large \cite{Wil75}.}  The result found for $\Tk$
by VZ  \cite[p.1590, eq.(4.11)]{VZ83} is 
\be
\label{bk.TKondo}
        \Tk = D \left[ v^x (0) v^z (0) \right]^{1/2}
         \left( {v^x (0) \over 4 v^z(0) } \right)^{1 \over 4 v^z(0)} \; .
\ee
Note that the factor $\left[ v^x (0) v^z (0) \right]^{1/2}$
is absent\footnote{The presence of 
the  prefactor to the exponent in (\protect\ref{bk.TKondo}\protect)
is of course a well-known feature of second-order
scaling, see e.g.\ \cite[eq.(3.47)]{Hew93}.}
 if one estimates $\Tk$ only from the leading-order
scaling equation (\ref{bk.scalingeq}) \cite[p.1577, eq.(4.11)]{VZ83}.
Since the bare
$v^x (0) \ll 1$, this factor causes a substantial
suppression of $\Tk$ (by about two orders of magnitude),
if one simply inserts $v^x (0)$ into \Eq{bk.TKondo},
leading to pessimistically small values
of $\Tk \simeq 0.01-0.1$~K \cite{ZZ94a,ZZ94b}. However, the inclusion
of excited states remedy this problem, in
that excited states renormalize $v^x $ to larger values
by about two orders of magnitude (see below).

\subsection{The Role of  Excited States}
\label{sec:excited}
\label{sec:excitedstates}

Let us now return to the more general problem where the
excited states in \Eq{bk.TC} with energies $E_{\sss ex}$, are not 
neglected from the beginning. 

The first important consequence of including excited states in
the model has already been discussed in section \ref{roleofex}:
electron-assisted hopping transitions between the 
two wells {\em via excited states}\/ allow
Kondo physics to occur even if the barrier is so 
large that direct and electron-assisted
tunneling through the barrier is negligible (i.e.\  $\Delta_o \simeq
0$). This is good news, since the energy splitting
$\Delta_b = E_1 - E_2$ is limited from below by $\Delta_o$,
but simultaneously $\Delta_b$
(being a relevant perturbation)
needs to be very small if scaling to the 2-channel
fixed point is to take place.

Secondly, in the presence of excited states, 
poor man's scaling towards strong-coupling, based on \Eq{bk.scalingeq},
has to proceed in several steps: the  excited
state $| \Psi_c\rangle$ only contributes as long as the effective
bandwidth $D'$ is larger than $E_c$, as is 
made explicit by the $\theta (D' - E_c)$ in \Eq{bk.scalingeq}.
As soon as $D' < E_c$, the excited state decouples.

  Assuming that the presence of excited states does not affect
the result found in \sect{sec:2Dsubspace}, namely 
 that the effective Hamiltonian
scales towards a 2-D subspace in which the electrons
have pseudo-spin $S_e = \toh$,
Zar\'and and Zawadowski \cite{ZZ94a,ZZ94b} have
analyzed the successive  freezing out of excited
states. They concluded that when $D'$ becomes smaller
than the smallest excited-state energy $E_3$,
one ends up with a tunneling center of formally precisely  the same nature
as the one discussed in sections~\ref{sec:2Dsubspace}
and~\ref{sec:isotropic}, {\em but with renormalized
couplings}. 

The renormalized couplings turn out to be still small, which
means that the perturbative scaling analysis of sections~\ref{sec:2Dsubspace}
and~\ref{sec:isotropic} still applies; however,
$v^x$ and $v^y$ are renormalized upward 
by a factor of up to 50 from their bare values
(which were three orders of magnitude
smaller than $v^z$ see \Eq{bk.couplings}).
This has  very important consequences 
for the Kondo temperature \Eq{bk.TKondo}, which strongly
depends on $v^x$:
with realistic choices of parameters, 
the Kondo temperature turns out to be about 2 orders of magnitude larger
with than without excited states in the model,
and Kondo temperatures in the experimentally relevant range
of 1 to 3 K were obtained \cite[table II]{ZZ94b}.

To summarize: the inclusion of excited states in the model
leads to more favorable estimates of the important
parameters $\Delta_o$ (can be zero) and $\Tk$ (larger);
but since the excited states
eventually decouple for small enough effective
band-widths, they do not affect the flow toward 
the 2-channel Kondo fixed point in any essential way.

\section{Recent Criticism of the 2-Channel Kondo Scenario}
\label{sec:criticism}

Very recently, the claim that 
Zawadowksi's non-magnetic Kondo problem will renormalize to 
the \nfl\ fixed point of the
2-channel Kondo model at sufficiently low temperatures
has been called into question in two separate papers
\cite{WAM95,MF95}. We ignored the concerns stated there
when introducing our NTKM in  section~\ref{sec:pcmodel},
because there our attitude was phenomenological
and our aim merely to write down a phenomenological Hamiltonian that
accounts for the observed phenomena. 
However, the question as to whether or not
the bare non-magnetic
Kondo model does indeed renormalize toward the \nfl\ fixed point
2CK model is an interesting theoretical one in its own right,
which, in our view, has become all the more relevant
in the light of the apparent success of the NTKM
in accounting for all aspects of RB's experimental results. 
Therefore, we summarize the relevant issues in this appendix.

\subsection{Large $\Delta$ due to Static Impurities}

 Wingreen, Altshuler and Meir
have recently argued \cite{WAM95} that tunneling centers with
very small  splittings ($\Delta < 1 K$) can not
occur at all in a disordered material
if the TLS-electron coupling has the large values
that apply to the over-screened 2-channel Kondo fixed point.
Their argument goes as follows:

If $H_{\sss TC}$ of \Eq{bk.TC} is truncated to
the lowest two states and written in the 
left-right basis [see \Eq{bk.2statefactor}], it 
has the following general form: 
$H_{\sss TC} =  \half \sum_{A= x,y,z} \Delta_A \sigma^A$,
where $\Delta_z$ is the asymmetry energy and $\Delta_x$ the
spontaneous hopping rate (for the bare system,
time-reversal symmetry enables one to choose 
$\Delta_y = 0 $ by choosing real eigenfunctions,
but under renormalization $\Delta_y \neq  0 $ can be
generated, see below).
Hence $\Delta_A $ can be be interpreted as an effective local field
at the TLS site (in the language of the magnetic Kondo problem,
this would be called a local ``magnetic field''). 
The energy splitting is of course $\Delta = (\Delta_x^2 + 
\Delta_y^2 + \Delta_z^2)^{1/2}$.

Now, WAM pointed out that ordinary elastic scattering
of electrons off other static defects 
%(at positions $\vec R_i$) 
in the system,
which causes Friedel oscillations (wavelength $1/k_{\sss F}$)
in the electron density (see e.g.\ \cite{KV60}),
will make a random contribution to $ \Delta_A$ (not considered
in Zawadowski's theory). WAM suggested
that the magnitude of this effect can
be characterized by the  {\em typical splitting}\/ $\bar \Delta$
that arises for a typical TLS in the presence of static disorder. 

As estimate for $\bar \Delta$, WAM used $\langle \Delta^2 \rangle^{1/2}$, 
the average of $\Delta$ over all realizations of disorder.
Calculating this 
using simple 2nd-order perturbation theory in the 
coupling between the electrons and static impurities, they 
found\footnote{Cox 
has reproduced WAM's result \protect\cite{Cox95}\protect\
 by a simple calculation analogous to the one by which one obtains
the RKKY interaction between two magnetic impurities.}
\be
\label{bk.WAMestimate}
        \bar \Delta \simeq \varepsilon_{\sss F} v_{\sss K} 
        / \sqrt{ k_{\sss F} \ell} \, . 
\ee
Here
$\ell$ is the mean free path (a measure of the concentration of 
static impurities),  and  $v_{\sss K}$ the effective  TLS-electron
coupling strength in \Eq{bk.Hinteff}.
Moreover, WAM argue that because $\Delta_A$ has three
components, the probability distribution
$P (\Delta)$ goes to 0 at $\Delta = 0$, because
the probability to simultaneously find {\em all three}\/  components
$\Delta_A =  0$ is vanishingly small.

WAM estimated $v_{\sss K} \simeq 0.1$ at the 2CK fixed point, by 
using a Kondo temperature
of about 4K (as cited in \cite{RLvDB94}) in the standard  formula
$v_{\sss K} = \log( k_{\sss B}T_{\sss K}/\epsilon_{\sss F})$
obtained from the leading-order scaling equations.
(This value for $v_{\sss K}$ agrees with the values estimated
by Zar\'and and Zawadowski \cite{ZZ94b}
for the 2CK fixed point (see caption of
\fig{fig:excitedstates}).
Using $l \simeq 3$~nm, WAM then  found  a value of
$\bar \Delta \simeq 100 K$ (the result is so large because
according to \Eq{bk.WAMestimate} $\bar \Delta$ is 
proportional to $\varepsilon_{\sss F}$).

Since 100 K is a huge energy scale compared
to all other scales of interest, WAM argued
that the 2-channel Kondo physics evoked
in  paper~I to explain the Ralph-Buhrman experiment 
would never occur. Instead, they proposed an alternative
explanation of the  experiment 
based on disorder-enhanced electron interactions.
The latter suggestion, which we
believe contradicts several experimental facts \cite[(b)]{WAM95},
 is critically discussed  Appendix~A~1 of paper~I. 
Here we briefly comment 
 on their estimate of $\bar \Delta$, following \cite[(b)]{WAM95} 
and \cite{ZZ96a}.

We believe that WAM  are correct in pointing out 
that static disorder interactiond can act to
increase the energy splitting, $\Delta$, of  the TLS. 
However, we suggest that $\bar \Delta \sim 100$~K
 may be a considerable overestimate, for the following reason.

Firstly, it was pointed out by Smolyarenko \cite{Smolyarenko} that
the quantity $\langle \Delta^2 \rangle^{1/2}$ calculated
by WAM does {\em not}\/ correspond to the {\em typical}\/
splitting (i.e.\ the splitting at which the distribution $P(\Delta)$
of splittings peaks), but instead to the {\em average splitting}\/
 $ \int \! d \Delta P ( \Delta) \Delta$. 
This distinction is important,
since the  {\em average}\/ of $\Delta$ over all realizations
of disorder can be dominated
by a few rather rare realizations of disorder that give rise to 
very large splittings (e.g.\ where some static defect is 
very close to the TLS, so that the  Friedel oscillations have
very large amplitudes at the TLS), and hence can be much
larger than the typical splitting. However, for the experiments
at hand, clearly  the relevant quantity is the {\em typical}\/
 splitting,  since TLS that have a very large splitting
due to a very nearby static defect would simply not
exhibit Kondo physics. Smolyarenko was able to estimate
the typical splitting, and found a result of $\bar \Delta \simeq 15$K,
much smaller than WAM's value for the average splitting. 

Furthermore, according to Zar\'and and Zawadowski (ZZ) \cite{ZZ96a}, 
WAM's statements are equivalent to assuming 
that $\Delta_A$ is renormalized
by Hartree-type corrections to the TLS self-energy
(see Fig.~2 of \protect\cite{ZZ96a} for the Feynman diagram):
$\Sigma_A (\omega) = \int \! d \omega \rho_{\alpha \alpha'} (\omega)
\tilde v^A_{\alpha' \alpha} [\ln(\omega/D)]$. Here
$\rho_{\alpha \alpha'} (\omega)$ is the spectral function of
the conduction electrons in the presence of impurities
and $\tilde v^A$ the renormalized vertex function 
of \Eq{bk.2statefactor} [with $x = \ln(\omega/D)$ there;
 the corresponding bare vertex
function would be $ \tilde v^A (0)$].  WAM's (and
Smolyarenko's) estimate of $\bar \Delta$ is obtained if one
simply uses the unrenormalized diagonal part of the
spectral function. 
However, this is too simplistic, since if 
the renormalized spectral function is used,
$\Delta_A$ is {\em reduced significantly}\/ \cite{ZZ94a,Zar96} 
(despite the growth in the couplings $\tilde v^A$ under renormalization).
The spontaneous hopping rates $\Delta_x, \Delta_y$, in particular,
 {\em decrease}\/ by as much as three orders of magnitude under 
renormalization (ZZ estimate their final typical
value to be $\bar \Delta_x \simeq \bar \Delta_y \lsim 1$~K 
to 0.1~K \cite{ZZ96a}). This simply reflects the screening
of the TLS by conduction electrons: when tunneling between
the wells, the tunnling center  has to 
drag along its screening clowd, which becomes increasingly
difficult (due to the orthogonality catastrophy) at lower
temperatures. In contrast, the asymmetry term $\Delta_z$
is not reduced as much \cite{Zar96}
(ZZ estimate that after renormalization
$\bar \Delta_z \gsim 1$~K), because
of a much larger value of the bare coupling 
$ \tilde v^z (0)$ [$\simeq 10^3 \tilde v^x (0) \simeq 10^3 \tilde v^y (0)$,
see \Eq{bk.couplings}]. 

Thus, we believe that the reason for
the large estimates of $\bar \Delta$ is the neglect
of the reduction of $\Delta_A$ under scaling.
Though ZZ's studies of this reduction were performed without
considering static disorder, disorder should
not essentially change matters\footnote{To check this
statement, an extra term, including the effects of static disorder,
should be added to Zawadowski's Hamiltonian, and then a full
RG analysis should be performed  to determine {\em self-consistently}\/
how the couplings and the ``local field'' $\Delta_A$, 
evolve together under renormalization.}  (since this reduction simply
reflects the well-understood physics of screening). 
Moreover, because  $\Delta_x, \Delta_y$ end up
being so much smaller than $\Delta_z$, 
WAM's conclusion that $P(0) = 0$ for the distribution of splittings
is not persuasive,  because the distribtions $P_A (\Delta_A)$ of the 
individual $\Delta_A$ are not equivalent, as they assumed.

Note that although ZZ estimated that $\bar \Delta_z \gsim 1$~K, implying
that also $\bar \Delta \gsim 1$~K, this is only a statement
about the {\em typical}\/ splitting of a typical TLS.
In a disordered system, it seems very likely that {\em some}\/
TLSs will exist with a splitting $\Delta$ significantly
smaller than the typical $\bar \Delta$. In particular,
ZZ's estimate that typically $\Delta_x, \Delta_y $ are $\lsim 0.1$~K
implies that assumption (A2) of paper~I, Section~V~D,
namely that the  nanoconstriction does contain TLS with
$\Delta < 1$~K, does not seem unreasonable,
despite the fact that $\bar \Delta \gsim 1$~K. 

Finally, note that 
two-state systems  with small energy splittings ($\Delta < 1 K$) have
been directly observed in disordered metals in at least two
experiments: Graebner {\em et al.}\/ \cite{Graebner77}
found a linear  specific heat in
amorphous superconductors below $T_c$ in the
regime $ 0.1 < T 2$~K, which they attributed to two-state
systems; and Zimmerman {\em et al.}\/ directly
observed individual slow fluctuators 
in Bismuth wires \cite{GZC92,ZGH91}, with $\Delta$s as small as
0.04~K. Though the detailed properties of these two-state
systems may be  different than those of fast TLS, 
this does illustrate that even in systems where
the average splitting is expected to be large, the
physics {\em can}\/ be sometimes dominated by those two-state
systems that have smaller splittings.

The relevance of WAM's calculation 
 to the interpretation  of the experiments discussed in paper~I
are discussed in  Section~V~C~3 of paper~I.

\subsection{Another Relevant Operator}
\label{sec:MF}

The theoretical justification for the
non-magnetic Kondo model proposed by Zawadowski has recently 
also been questioned by
 Moustakas and Fisher (MF) \cite{MF95}. 
Reexamining a  degenerate two-level system
interacting with conduction electrons, they argued
that the model  of Eqs.~(\ref{bk.Helnew}) and (\ref{bk.Hintnew})
used by Zawadowski is incomplete, because it neglects certain
subleading terms in the TLS-electron
interaction that have the same symmetries as the leading terms.
%These subleading terms are initially small when the coupling is small, and
%moreover are RG irrelevant. Therefore, neglecting them is 
%perfectly valid for the purpose 
%for which Zawadoski constructed his model,
%namely estimating the Kondo temperature at which the leading terms in
%$H_{int}$ begin to grow. However, such subleading terms {\em can}\/
%become important when investigating the
% nature of the fixed point towards which
%the system flows under renormalization.
MF showed that when combined in certain ways, these subleading terms generate
an extra {\em relevant}\/ operator, not present in
Zawadowski's analysis, which in general prevents
the system from flowing to the $T=0$ fixed point. 
Therefore, unless a fine-tuning of parameters
miraculously causes this new relevant operator to vanish, it
 will eventually always become large, and the system
will never reach the $T=0$ fixed point.

Zawadowski {\em et al.}\/ \cite{ZZ96b} have recently reinvestigated
the nature of this new relevant term within the context of
a somewhat simpler (commutative) model than that of MF,
and suggested that it arises due to the
breaking of particle-hole symmetry. They estimate that 
before renormalization,
its prefactor in the bare model is smaller than the effective Kondo
coupling at the fixed point by a factor of $10^{-6}$,
and still by $10^{-3}$ after poor man's scaling  
renormalization to effective bandwiths of
order $D' = \Delta_o$ (the spontaneous tunneling rate).
Thus, they conclude that this effect can probably be
neglected in realistic systems.  
Moreover, they questioned the path-integral renormalization group scheme
employed by MF, arguing that it leads to misleading
conclusions about the renormalization of the physical
couplings. At present, this matter still seems controversial
and seems to deserve further investigation \cite{DvD96}.

\section{The Limit of Large Channel Number: $k \to \infty$}
\label{naivepert}

In this appendix, we perform a 
check on  the CFT calculation of the backscattering current
of section~\ref{sec:Gscaling}, by 
considering the (unphysical) limit of a large number of conduction
electron channels, $i = 1, \dots, k$, with $k \to \infty$.
In this limit,  the poor man's scaling approach becomes exact,
even though it is based on perturbation theory. The
reason for this is that (for the isotropic
model) the over-screened fixed point occurs
when the coupling constant has the special value $v^\ast = {2 \over 2
+ k}$,
%%III [see e.g.\ section~\ref{sec:sabs} of paper~III for the case $k=2$],
which $\rightarrow 0$ as $k \to \infty$.
Thus, in this limit one never scales into a ``strong-coupling''
regime, and the perturbative expressions from
which the scaling equations are derived retain
their validity throughout. Therefore, results from the
poor man's scaling approach should agree with exact
results from CFT in the limit
$k \to \infty$, which serves as a useful check on both
methods. 

Thus, in this appendix 
we consider the $k$-channel version of the
NTKM of section~\ref{sec:pcmodel}, 
in which the index $i = 1, \dots, k$, but
the interaction still has the form of
\Eq{effectiveVV}.

\subsection{Backscattering current}

To begin, we need  a perturbative expression for the 
backscattering current $\Delta I$ (i.e.\  the negative
contribution to $I$, which we defined such that $I>0$ if
it flows to the right) due
to the backscattering events of $ V_{\eta \eta'}^{a a'}$.
We use the most naive 
approach for treating the effects of the interaction
in a non-equilibrium situation: we simply
do perturbation theory in $H_{int}$ according
to the rules of $T=0$, $V=0$ quantum mechanics,
and insert by hand, into all sums over intermediate states,
appropriate non-equilibrium distribution functions that indicate
with what probabilitiy  the corresponding states are occupied or empty,
$\sum_\eta \int\! d \ve f_{\sigma} (\ve)$ or
$\sum_\eta \int\! d \ve [1-f_{\sigma} (\ve)]$,
as in \Eq{bk.2ndvertex}.
This is the method Kondo \cite{Kon64} used when
deriving his famous $\log T$ for the first time,
and when one is merely interested in the lowest few
orders of perturbation theory, it is certainly the 
simplest approach. 

In this approach, $\Delta I$ is given quite generally by\footnote{
Though the relation of the perturbative 
expression (\protect\ref{bk.backI}) to the 
non-perturbative ones of section~\ref{currentcalc}
is not readily apparent, note that \protect\Eq{FinCurrent}
has the same form as \protect\Eq{cl.Iexp}.}
\bea 
\label{bk.backI}
        \Delta I &=&  - {\tilde K N_o  \over |e| }
         \int \! d \varepsilon \! \int \! \! d \varepsilon'
         \left[ \vphantom{Q^2}
        b_{\eta \eta'}
        (1 - f_{\sigma} (\varepsilon))  
         f_{\sigma'} (\varepsilon') 
        \Gamma  (\varepsilon',\sigma' \to \varepsilon,
         {\sigma} )  
         \right.
        \\
        && - \left.  \vphantom{Q^2}
        b_{\eta' \eta}^\ast
        (1 - f_{\sigma'} (\varepsilon')) 
        f_{\sigma} (\varepsilon)        
        \Gamma(\varepsilon,{\sigma}
        \to \varepsilon',\sigma') \right] \; ,
\eea
where $\sigma = L$ and $\sigma' = R$,
and the backscattering rate from $\sigma'$-movers
to $\sigma$-movers is 
\be 
\label{bk.Gamma}
        \Gamma(\varepsilon',\sigma' \to \varepsilon ,{\sigma}) 
        = 2 \pi / \hbar \, \delta(\ve' - \ve) \, 
        N_o^{-2}  \, 
        \toh \sum_{\alpha i a, \alpha' i' a'} 
        |T^{a a'}_{\ve \eta \ve' \eta'}|^2 \; .
\ee
In writing down \Eq{bk.backI}, 
the fact that the Fermi functions do not depend on the
indices that appear in the sums $\sum_{\alpha i a, \alpha' i' a'} $
has been exploited to pull them out to the front of
these sums, which could thus be included in the definition 
(\ref{bk.Gamma}) of $\Gamma$. In \Eq{bk.Gamma}  the factor 
$\toh$ has been included so that $\toh \sum_{a'}$
represents an average over 
the initial states of the TLS.  $T^{a a'}_{\ve \eta \ve' \eta'}$
is the generalization of the interaction
vertex $\Gamma^{a a'}_{\eta \eta'}$ of \Eq{bk.2ndvertex}
to all orders of perturbation theory, and depends not only on 
the matrix elements $V^{a a'}_{\eta \eta'}$, but also on 
the distribution functions $f(\ve,\eta)$ (as \Eq{bk.2ndvertex}
illustrates explicitly). The factor
$\tilde K \equiv e^2 \sum_\eta \tau_\eta (0) / h$ 
(where $\tau_\eta (\ve)$ is defined below) is 
included  in \Eq{bk.backI} for dimensional reasons, and the dimensionless
constants $b_{\eta \eta'}$  characterize all those details of scattering 
by the impurity that are energy-independent and 
 of a sample-specific, geometrical nature, such as 
the position of the impurity relative to the constriction, etc.
(compare section~III of paper~I).

Now, since $V_{\eta \eta'}^{a a'}$ is independent
of the indices $\sigma, \sigma'$, the same is true
for $|T^{a a'}_{\ve \eta \ve' \eta'}|^2 $.
(Although  the ``internal sums'' $\sum_{\sigma''}$
involving intermediate states are highly non-trivial, because of the 
presence of $\sigma''$-dependent $f (\ve, \eta'')$ functions,
$T^{a a'}_{\ve \eta \ve' \eta'}$ of course does not depend on 
such $\sigma''$-indices, since they are summed over.)
Therefore it follows immediately that
\begin{equation}
\label{bk.symmetry}
        \Gamma(\varepsilon', \sigma' \to 
        \varepsilon,{\sigma} ) = 
        \Gamma(\varepsilon,{\sigma} \to 
        \varepsilon',  \sigma') \; .
\end{equation}
Exploiting eq.~(\ref{bk.symmetry}) and
the $\delta(\varepsilon' - \varepsilon)$
function in $\Gamma(\ve' , \sigma' \to \ve ,\sigma )$, the 
backscattering current can be brought into the suggestive form
[compare with \Eq{cl.Iexp}]
\be
\label{FinCurrent}
        \Delta I =  - {\tilde K b\over |e| }
        \int \!\! d \varepsilon' 
        \left[ f_R(\varepsilon') - f_L(\varepsilon') \right]
        \toh \sum_{\alpha', i'} 
        \frac{1}{\tau_{\eta'} (\varepsilon')}  \; .
\label{finalcurrent}
\ee
Here we have taken $b_{\eta \eta'} = b$ for simplicity, and have defined
the total scattering rate $\frac{1}{\tau_{\eta'} (\varepsilon')}$
for a electron with energy $\ve'$ and discrete quantum numbers $\eta'$
by\footnote{The $\half$ in \Eq{FinCurrent} is needed because 
the definition (\ref{lifetime}) of $\tau_{\eta'}^{-1}$ contains
a sum $\sum_{\sigma'}$ that does not occur in \Eq{bk.Gamma},
and is $=2$, since $T^{a a'}_{\eta \eta'}$ 
is independent of $\sigma'$.}
\begin{equation}
\label{lifetime}
        \frac{1}{\tau_{\eta'} (\varepsilon')} 
        \equiv N_o^{-1}
        \int \!\! d \ve \sum_{\eta}{ 2 \pi \over \hbar }
        \delta(\ve' - \ve)
        \toh \sum_{a a'} |T^{a a'}_{\ve \eta \ve \eta'}|^2 \; .
\end{equation}

How is $T^{a a'}_{\ve \eta \ve \eta'}$ to be calculated 
explicitly? For $T> \Tk$, the leading order logarithmic 
terms of the perturbation series in powers of $H_{int}$ can be
summed up using the poor man's scaling approach, discussed
in the next subsection. On the other hand, an analysis of
the regime around the $T=0$, $V=0$ fixed point requires the use of
CFT (see paper~III). A consistency check between the two 
methods can be performed by taking the limit $k\to \infty$
in which perturbation theory becomes exact, as discussed
above.

\subsection{Gan's Results for Large Channel Number}
\label{sec:Gan}

A calculation of $\tau^{-1}_\eta (\ve)$ for
the bulk isotropic $k$-channel Kondo model, in the limit 
$k \to \infty$, has been carried out by Gan \cite{Gan94}. 
More specifically, he calculated  the imaginary part
of the electron self-energy, 
$\Sigma^{\sss I} ( \omega, D, g)$, perturbatively\footnote{Since 
the coupling constant $ v \sim 1/k$, and
closed electron loops get a factor $k$, Gan had
to include up to 8-th order diagrams!} to order $k^{-4}$
(we cite only the lowest relevant terms below).
By then using  poor man's scaling methods,
he was able to obtain agreement to order $k^{-2}$
with the exact CFT results for $\Sigma^{\sss I}$. 

Since Gan considered
precisely the interaction Hamiltonian  of  \Eq{cl.NHk} that 
governs our even channels,
we can directly use his results.
He obtained the following expression at $T=0$, $V=0$:
\be
\label{bk.sigmafinal}
         \Sigma^{\sss I} (\omega, D, v_{\sss K}) 
        \propto
        \left[ 1 - c_1 \left( {\omega \over  \Tk
        }\right)^{\bar \alpha} \right] \; .
\ee
His perturbative expression for the exponent 
occuring here is $\bar \alpha = {2 \over k}(1 - {2 \over k})$, 
which agrees to order $k^{-2}$ with
the exact CFT result $\alpha = {2 \over 2 + k}$.

Since \Eq{bk.sigmafinal} was derived using poor man's scaling methods,
{\em it also holds in the non-equilibrium case,}\/ as long as $\omega
> V$ (see section~\ref{sec:poorV}).
This condition does not hold strictly in the integral (\ref{finalcurrent}).
Nevertheless,   if we use ${1 \over \tau(\omega)}
= - 2  \Sigma^{\sss I} (\omega)$ in 
Eq.~(\ref{finalcurrent}) with $T= 0$, $V \neq 0$, 
the resulting asymptotic  expression
for the backscattering conductance $\Delta G(V,0) = 
\partial_V \Delta I (V,0)$,
\be
\label{bk.DeltaG}
        \Delta G(V,0) \propto V^{\bar \alpha} \; ,
\ee
should still be approximately correct up
to logarithmic corrections that are typical of the poor man's
scaling approach. 
Indeed, the corresponding expression
 that we obtained  in section~\ref{ch:calc} from our
CFT approach has the same asymptotic form
[see \Eq{cl.condGscale}], 
but with $\bar \alpha$ replaced by the exact value for the exponent,
namely $\alpha = {2 \over 2 + k}$.
[Actually, in section~\ref{ch:calc} we always use
$k=2$,  and hence $\alpha = \toh$ in \Eq{cl.condGscale}.]
This agreement is a reassuring confirmation that the
methods used in sections~\ref{sec:Y-Kondo}
to \ref{currentcalc} agree with
the present perturbative results in the one limit ($k \to \infty$) where
perturbation theory can be trusted.

\section{Hershfield's $Y$-operator approach to Non-Equilibrium Problems}
\label{sec:non-eqapp}

In this appendix we summarize the main ideas of Hershfield's
$Y$-operator approach to non-equilibrium problems.

\subsection{The Kadanoff-Baym Ansatz for $V\neq 0$}
\label{KadanoffBaym}

The problem at hand is defined by the free Hamiltonian
$H_o$ of \Eq{eq4.Hoe}, the free density matrix $\rho_o$ of
\Eq{eq4.rhoo} and the interaction $H_{int}$ of \Eq{bk.HeffV}.
$H_o$ and $\rho_o$ describe free electrons that move between
two leads or baths ($R/L$), 
at different chemical potentials ($\mu_\pm$),
by passing ballistically through a nanoconstriction,
in which   $L$- and $R$-movers can be scattered into each
other by $H_{int}$.

How does one calculate statistical  averages for such a system,
in other words, how does one define the full
density matrix in the presence of $H_{int}$?
The main complication that has to be confronted is
that the number of electrons in each bath is not
conserved, in that $[N_{\sL, \sR} , H_{int}] \neq 0$
(compare footnote \ref{f:tunnel} on page~\pageref{f:tunnel}).
Therefore,  any attempt to naively  replace
$\rho_o (V)$ in \Eq{eq4.rhoo} by  $e^{-\beta (H - Y_o)}$ 
will (apart from lacking first-principles justification)
quickly run into problems: since $[H, Y_o ] \neq 0$,
many of the standard properties of equilibrium
Green's functions [e.g.\ $G(\tau  + \beta) = \pm G(\tau )$],
no longer hold.

Kadanoff and Baym have shown how such a general 
problem can be dealt with, by using the
notion of adiabatically switching on the interaction
 \cite[eq.(6.20)]{KB62}:
Thermal weighting has to be
done with the initial density matrix $\rho_o$ at some early
time $t_o \to - \infty$, at which all interactions
$H_{int}$ are switched off, and then
$H_{int}$ 
is adiabatically turned on 
[$H_{int}(t) \equiv H_{int} e^{\alpha t}$, with $\alpha \to 0^+$]
while the system
is time-evolved to the time $t$ of interest.
Concretely, to evaluate the thermal expectation
of an operator $O$,
one writes the operator in  the Schr\"{o}dinger picture,
and uses the thermal weighting factors 
$ e^{- \beta [E_{on} - {1 \over 2}
        eV (N_{\sL } - N_{\sR })_n]}$
appropriate to a trace $\sum_n  \! \langle n, t_o |
\quad | n, t_o \rangle $ of Schr\"{o}dinger states 
taken at some early time $t_o \to - \infty$
(where they are eigenstates of $H_o$ with eigenvalues $E_{on}$).
However, one then takes the actual
trace between the {\em time-evolved versions}\/ of these
states $|n, t \rangle  =  U(t, t_o) | n, t_o \rangle$,
where%\footnote{No time-ordered exponential  is needed
%here, because $H$ is assumed to be time-independent.}
 $U= {\cal T} e^{-i\int_{t_o}^t dt' H(t')}$ 
is the Heisenberg time-evolution operator:
\bea 
\nonumber
        \langle O (t)  \rangle_{\sss V}
        & \equiv &
        {  \sum_n  e^{- \beta [E_{o} - {1 \over 2}
        eV (N_{\sL } - N_{\sR })_n]} 
        \langle n , t| O | n , t\rangle \over 
        \sum_n  e^{- \beta [E_{on} - {1 \over 2}
        eV (N_{\sL } - N_{\sR })_n]} }  
\\ \label{tr.KadBaym}
&=& 
        {\mbox{Tr} \, \rho_o (V, t_o)\,  U^\dagger (t, t_o) O U(t, t_o) 
        \over \mbox{Tr}  \, \rho_o (V, t_o) } \; ,
\eea
where in the second equality
the trace is taken between the states $|n, t_o \rangle$.
(Since steady-state expectation values of a single operator
are time-independent, $t$ is here just a dummy variable,
and is often taken to be 0.)

\Eq{tr.KadBaym} is the defining prescription for taking  non-equilibrium
expectation values in the presence of interactions.
\label{sec:v=0standard}
For $V = 0$, it 
reduces to the standard equilibrium prescription,\footnote{The
second argument $t_o$ in $\rho (0,t_o )$ is superfluous;
it is retained here only for the sake of notational
consistency with the $V \neq 0 $ case.}
\be
\label{tr.V=oexp}
   \langle O \rangle_{\sss V=0} \equiv  {\mbox{Tr} \rho (0,t_o) O
         \over \mbox{Tr} \rho (0,t_o ) } \; ,
        \qquad \mbox{where} \; \qquad \rho (0,t_o  )
        \equiv  e^{-\beta {H}}  \; ,
\ee
as shown, e.g., by Hershfield in \cite{Hers93}. \Eq{tr.V=oexp}
is of course the starting point for
familiar equilibrium statistical mechanics. 
One of its most useful features is that 
the  thermal weighting factor $e^{-\beta H}$ and
the dynamical time-evolution factor $U (t, t_o) = e^{-i H(t-t_o)}$
commute;  Green's functions therefore
have the periodicity property $G(\tau + \beta) = \pm G(\tau )$,
which makes it convenient to formulate perturbation
expansions in $H_{int}$ along the negative imaginary axis,
$t = -i \tau \in [0, -i \beta]$.

\subsection{Hershfield's Formulation of the case $V\neq 0$}
\label{sec:hersVneq0}

If $V \neq 0$ so that \Eq{tr.KadBaym} and not \Eq{tr.V=oexp} 
is the starting point, there are no obvious periodicity
properties along the imaginary time axis, and the
conventional approach, due to Keldysh, is to
formulate perturbation expansions in $H_{int}$ along the
real axis \cite{Kel64,RS86}. The various non-equilibrium
diagrammatic techniques 
that have been devised are simply ways of doing the real-time
integrals  $\int_{t_o}^t dt'$ that result from  the
expansion of $U(t, t_o)$. However, for our purposes such
expansions are inconvenient:
firstly, perturbation expansions have limited use in the
Kondo problem, and secondly,
 we would in the end like to apply Affleck and Ludwig's non-perturbative
CFT results.

Hershfield has recently shown that  \Eq{tr.KadBaym} can
be rewritten in a way that exactly meets our needs.
The first step is trivial: 
using the cyclical property of the trace to move $U(t, t_o)$
to the front, \Eq{tr.KadBaym} can be written as 
\be 
\label{eq4.rhoV}
  \langle O (t) \rangle_{\sss V}
         \equiv {\mbox{Tr} \rho (V, t) O
         \over \mbox{Tr} \rho (V, t ) } \; ,
\ee
where
\be
\label{tr.rhoVdef}
        {\rho (V,t) \over \mbox{Tr} \rho (V,t) }
         \equiv  { U(t, t_o) \rho_o (V, t_o)  U^\dagger (t, t_o)
        \over \mbox{Tr} \rho_o (V,t_o) } \; .
\ee
The formal definition (\ref{tr.rhoVdef}) makes it
clear that $\rho (V,t)$ is the density operator
that $\rho_o (V, t_o)$ develops into as 
the interaction is switched on and the system time-evolves
from $t_o$ to $t$, with appropriately changing 
normalization.  Thus, all complications introduced through 
the adiabatic switch-on procedure are lumped into the time-evolved
density operator $\rho (V,t)$.

Next, Hershfield transfers these complications to a new
operator, $Y$, which he defines by writing  $\rho (V,t)$ in
the form
\be
\label{eq4.rho}
        \rho (V,t) \equiv 
 e^{-\beta [ H - Y(V,t) ]}  \; ,
\ee
purposefully constructed to resemble the definition
of $\rho_o (V, t_o)$ in \Eq{eq4.rhoo}.
Then he was able to show\footnote{
Hershfield's proof is perturbative:
using \Eq{eq4.Yevolve} he showed explicitly
that \Eq{eq4.rho}, expanded in powers of
$H_{int}$, reproduces the Keldysh perturbation expansion obtained
from the Kadanoff-Baym Ansatz (\ref{tr.KadBaym}).}
 (and herein lies the hard work)
that the operator $Y$ thus defined can be characterized as follows:
\begin{enumerate}
\item[(P)]
${Y}$ is the operator into which ${Y}_o$ evolves 
as the interactions are turned on 
[as is suggested by a comparison of \Eqs{tr.rhoVdef}
and (\ref{eq4.rhoo})]. It satisfies the relation
\be
\label{eq4.Yevolve}
  [ {Y}, {H} ] = i \alpha ( {Y}_o - 
    {Y}) \; , \qquad \mbox{where} \quad  \alpha \to 0^+ \; ,
\ee
which  implies that {\em $Y$ is  a conserved quantity.}
\end{enumerate}

{\em The fact that  the $Y$-operator is a conserved
quantity is the great advantage of the $Y$-operator approach.}\/
It implies that the problem is now formally equivalent
to an equilibrium one (for which one has $\mu  N$ ($N$= total
electron number) instead of $Y$, and $[H,  N] = 0$).

If $Y$ is known,  one can therefore apply the usual methods of {\em 
equilibrium}\/ statistical mechanics,\footnote{From
\protect\Eq{eq4.Y}\protect\
it is clear that $Y$ can actually be shifted
away in $\rho = e^{- \beta (H - Y)}$ by defining
new energies $\varepsilon' \equiv  \ve - \mu_\eta$
associated with $c_{\ve \eta}$, 
i.e.\ measuring the energy of an excitation
relative to the Fermi surface of the bath
from which it originates.
 The   weighting factor then completely resembles its 
 equilibrium form, but because
$c_{\ve \eta} (\tau) = c_{\ve \eta} e^{- \tau ( \ve' + \mu_\eta)}$,
 extra factors of $e^{\pm \tau 2\mu_\eta}$ appear
on some  operators that are not diagonal in $\sigma$,
such as $H_{int}$. We shall not follow this approach here.}
 using the density matrix $\rho \equiv
e^{-\beta({H} - {{Y}})}$ and Heisenberg
time-development $\hat O (\tau) = e ^{ {H} \tau}
\hat O e ^{-  {H} \tau }$, to calculate physical quantities.

In general, finding $Y$ explicitly is just as difficult
as solving the full non-equilibrium problem. However,
for scattering problems, it is sometimes possible to
write down $Y$ explicitly in terms of the problem's scattering
states \cite{Hers93}. 
Explicit expressions for $H$ and $Y$ in a typical scattering
problem are given in section~\ref{sec:non-eq}.

\section{Example: 2-species Potential Scattering}
\label{sec:simpex}
\label{app:simpleexamples}

In this appendix we illustrate the formalism developed in
section~\ref{sec:UfromGF} by applying it to a very simple
\label{sec:2-species}
scattering problem, namely the scattering of only two species
 of (spinless) electrons off a static scattering potential.
We take $\eta$ equal
to the species index, $\eta \equiv \sigma =  (L,R)
= (+,-)$ (i.e.\  $\eta$ contains no extra channel indices $i$,
and $L/R$ denotes physical $L/R$ movers).
As Hamiltonian we take 
[compare \Eqs{eq4.Hox} and (\ref{eq4.Hee})]:
\bea
\nonumber
        H  &=& H_o + H_{int} 
\\  \label{eq4.Hax}
&=&
        \sum_{\sigma} \! \int \! {\textstyle {dx \over 2 \pi}}
        \psi^{\dagger}_{\sigma} (ix)
        \left( 
        \delta_{\sigma \sigma'} i \partial_x 
        + {2 \pi} \delta(x)  V_{\sigma \sigma'}
        \right)
        \psi_{\sigma'} (ix') \; .
\eea
Here $V_{\sigma \sigma'}$ is simply a hermitian
$2 \times 2$ matrix representing potential scattering
of the two species  into each other (i.e.\ the impurity
is not a dynamical object with internal degrees of freedom).
Since $V_{\sigma \sigma'}$ 
is Hermitian, we can make a unitary transformation of the form
\be
\label{eq4.lpsi}
        \lpsi_{\sigma}  \equiv M_{\sigma \sigma'} \psi_{\sigma'} \; ,
\ee
with $M$ chosen such that it diagonalizes
\bea
\nonumber
        H_{\rm scat} & = & \sum_{\sigma \sigma'}
        \lpsi_{\sigma}^{ \dagger} (0) 
        \left( MVM^{-1} \right)_{\sigma \sigma'} 
        \lpsi_{\sigma'} (0)
\\ \label{eq4.cs.diagint}
       & \equiv & 
         \lpsi_{\sigma}^{ \dagger} (0)
        \left(  v_o \half \delta_{\sigma \sigma'}
             +        v_3 \half \sigma^3_{\sigma \sigma'} \right) 
        \lpsi_{\sigma'} (0) \; .
\eea
Since the scattering term is now diagonal, its only effect
on the $\lpsi_{\sigma}$-fields can be to cause
a phase shift of the outgoing fields relative to the
incident ones:
\bea
\label{eq4.cs.pshift}
        \lpsi_{\sR \sigma } (ix) &= & P_{\sigma \sigma'} \lpsi_{\sL \sigma'
        } (ix)
                \qquad \mbox{for} \; x < 0 \; ,
\\ \nonumber 
        \mbox{where} \qquad 
        P_{\sigma \sigma'} &=& \delta_{\sigma \sigma'}
        e^{-i \left( \phi_o + \sigma \phi_3 \right) } \; .
\eea
and the phase shifts $\phi_0$ and $\phi_3$ are
functions of $v_o$ and $v_3$.
Rotated back into the $\psi_{\sigma}$-basis, this
phase shift of course becomes an actual [$SU(2)$] rotation
of the two species into each other:
\be
\label{eq4.twist}
        \psi_{\sR \sigma } (ix) = 
        \tilde U_{\sigma \sigma'} \psi_{\sL \sigma' } (ix)
        \vspace{-3mm} \; , 
\ee
 where $\tilde U_{\sigma \sigma'}$ is a unitary matrix of the form:
\be
\label{eq4.Uconst}
\tilde U_{\sigma \sigma'}
        \equiv \left( M^{-1} P M \right)_{\sigma \sigma'} \; 
        \equiv 
        \left( \begin{array}{cc}
                 {\cal T} & {\cal R} \\
                - {\cal R}^{\ast}  {\cal T} / {\cal T}^{\ast}  &
         {\cal T}          
        \end{array} \right) \; ,
\ee
$      [ |{\cal T}|^2 + |{\cal R}|^2 \equiv 1 ] $.
Comparing \Eq{eq4.twist} with
\Eq{eq4.scatwave} and \Eq{eq4.psiphi}, 
we see that $\tilde U_{\sigma \sigma'} (\ve')
= \tilde U_{\sigma \sigma'} $, i.e.\  in
this simple case $\tilde U$ is $\ve'$-independent.
Physically, this rotation of physical $L$- and $R$-movers
into each other simply reflects the fact that
$H_{int}$ causes backscattering: an incoming $L$-mover
has amplitude ${\cal T}$ to undergo forward scattering
and emerge as a $L$-mover,
and ${\cal R}$ to be backscattered into a $R$-mover.
This illustrates how our formalism is able to deal with backscattering
despite the fact that we  expressed both $\sigma= L$ and
$\sigma = R$ as mathematical $L$-movers in \Eq{eq4.psiex}, for which both the
transmitted (${\cal T}$) and reflected (${\cal R}$)
 parts of $\psi_{\sigma}$ live at $x < 0$.

%As a concrete example, we give the matrices $M$ and
%$ \tilde U$ corresponding to a very simple case.
%If  $V_{\sigma \sigma'} \equiv \half ( v_o \delta_{\sigma \sigma'}
%        + v_1 \sigma^x_{\sigma \sigma'} )$, then
%\be
%\label{eq4.example}
%        M = \sfrac{1}{\sqrt{2}}
%        \left( \begin{array}{cc}
%        1 & 1 \\
%        1 & -1 
%        \end{array} \right) \quad \mbox{and}\quad
%        \tilde U = 
%        e^{-i  \phi_o}
%        \left( \begin{array}{cc}
%        \cos \phi_3 & -i \sin \phi_3 \\
%        -i \sin \phi_3    &  \cos \phi_3 
%%        \end{array} \right).
%\ee

To calculate the current, insert \Eq{eq4.Uconst}
into \Eq{eq4.I}.
One readily finds 
\be
\label{eq4.Iscalar}
        I =  \sfrac{|e| }{h}
        \sum_{\bar \sigma} \!  \int \!\!  d \bar \ve
        \; |{\cal T} |^2 \; \bar \sigma \; f (\bar \ve , \bar \sigma ) 
        =         \sfrac{e^2 }{h}  |{\cal T} |^2 |V| \; .
\ee
As expected, the conductance $G \equiv \partial_V I = \sfrac{e^2
}{h} |{\cal T} |^2$ is reduced from its customary
value  for a single channel in the absence of
scattering, namely $\sfrac{e^2}{h}$, by
the transmission coefficient squared, $|{\cal T} |^2$.

\Eq{eq4.Iscalar} can also be used to illustrate
that  the conductance assumes a $V/T$ {\em scaling}\/  form
if the transmission coefficient ${\cal T}$ is {\em energy dependent.}\/ 
Assume that for some reason the ${\cal T}$ 
 in \Eq{eq4.Iscalar} depends on
the energy distance from the Fermi surface,
and can be expanded as 
$|{\cal T} |^2 \equiv A_o + (\ve/ \ve_{\sss F}) A_1 +
( \ve / \ve_{\sss F})^2 A_2 + \dots $
Then the conductance $G = \partial_V I$
is readily found to be
\be
\nonumber
        G (V,T)  =
%&=&
%        \sfrac{e^2}{h} \! \int \!\! d \ve
%        \left( A_o + (\ve/ \ve_{\sss F}) A_1 +
%        ( \ve / \ve_{\sss F})^2 A_2 + \dots \right)
%\\ & &
%\label{eq4.GT2}
%         \times
%        (- \half) \left[ \partial_\ve f_o (\ve - eV/2 ) +
%         \partial_\ve f_o (\ve - eV/2 ) \right]
%\\
%        &=&
        \sfrac{e^2}{h} \left[
        A_o \: + \: A_2 \sfrac{\pi^2}{3} 
        \left( {T \over \ve_{\sss F}} \right)^2
        \left( 1 \, + \, \sfrac{3}{4 \pi^2} 
        \left( {e V \over T } \right)^2 \, \right) \right] \; .
\ee
This has the scaling form $G(V,T) = G (0,0) + B T^2
\Gamma(v)$, where  $\Gamma (v) = \left( 1 + \sfrac{3}{4 \pi^2} v^2 \right)$
is a universal function, and $v \equiv eV/T $.

In the 2CK case, 
a scaling form for the conductance  arises
in a similar fashion, namely from an energy-dependence
in the transmission coefficient.  The non-trivial
difference is  that there we have
$|{\cal T} (\ve)|^2 = A_o + A_1 T^{1/2} \tilde \Gamma
(\ve / T)\; $, see  section~\ref{sec:Gscaling}.

\section{The NCA approach}
\label{NCA}

In  section~\ref{sec:expcompare} we compared our
results to  recent numerical calculations by Hettler,
Kroha and Hershfield (HKH) \cite{HKH94}, who used
the non-crossing-approximation (NCA) approach
to the Kondo problem.
Therefore, a few words about their work are in order here.

\subsection{Anderson model used for NCA}

HKH represent the system by the following 
infinite-$U$, $SU(2)$, 2-channel Anderson Hamiltonian
in a slave boson representation:
\bea
\nonumber
H_1 &=& \sum_{p, \sigma, \alpha, i} (\ve_p - \mu_\sigma) c^\dagger_{p \sigma
  \alpha i} c_{p \sigma \alpha i} + \ve_d \sum_\alpha f_\alpha^\dagger
f_\alpha 
\\ \label{calc.Anderson}
& & + \sum_{p, \sigma, \alpha, i} {\cal V}_\sigma \left( f^\dagger_\alpha
  \, b_i c_{p \sigma \alpha i} + \mbox{H.c.} \right) \; .  
\eea
The first term
describes conduction electrons in two leads, $\sigma = (L,R)= (+,-)$,
separated by a barrier and at chemical potentials $\mu_\sigma = \mu + \sigma
\toh eV$.  The electrons are labeled by a momentum $p$, the lead index
$\sigma$, a pseudospin index $\alpha = (1,2)$, and their Pauli spin $i =
(\uparrow, \downarrow) $, which plays the role of channel or flavor index.
[Note that for $V=0$ there is no need to distinguish between $L$ and $R$
electrons, and the $\sigma$-index can be absorbed into the $p$-index.]  The
barrier is assumed to contain an impurity level $\ve_d$ far below the Fermi
surface, hybridizing (with matrix elements ${\cal V}_\sigma$, with ${\cal
  V}_\sL = {\cal V}_\sR$ for our purposes) with the conduction electrons,
which can get from one lead to the other only by hopping via the impurity
level.  $f$ and $b$ are slave fermion and slave boson operators, and the
physical electron operator on the impurity is represented by
$d^\dagger_{\alpha i}= f^\dagger_\alpha b_i$, supplemented by the constraint
$\sum_{\alpha} f^\dagger_\alpha f_\alpha + \sum_i b^\dagger_i b_i = 1$.

Although this picture of two disconnected
leads communicating only via hopping through an impurity level
does not directly describe the physical 
situation of ballistic transport through
a hole accompanied by scattering off two-level systems,
the Hamiltonian (\ref{calc.Anderson})
can be mapped by a Schrieffer-Wolff transformation onto
the more physical  one [\Eq{bk.HeffV}] introduced 
in section~\ref{defnonequibmodel}.
It is therefore in the same universality class and describes
the same low-energy physics, provided that one
identifies the impurity-induced ``tunneling current'' $I_{tun}$ in the
HKH model with the impurity-induced backscattering current $\Delta I$ in 
the actual nanoconstriction.

HKH calculate the tunneling current,
\be
\label{NCAcurrent}
I_{tun} (V,T) = \int d \omega A_d (\omega) [ f_o (\omega - eV/2) - f_o (\omega
+ eV/2) ] \; , \ee where $ f_o(\omega) = 1 /( e^{\beta \omega} + 1)$, by
calculating the impurity spectral function $A_d(\omega)$ using the NCA
approximation \cite{Bic87}, 
generalized to $V \neq 0$ using Keldysh techniques.

\subsection{NCA and the limit $N,k=\infty$}
\label{NCAlargeN}

The NCA technique was historically developed as a self-consistent summation of
an infinite set of selected diagrams, yielding a set of coupled integral
equations (the so-called NCA equations) for the self-energies of the slave
fermions and slave bosons, which have to be solved numerically.  Hence the NCA
has conventionally been viewed as an uncontrolled approximation, with no small
parameter.  However, Cox and Ruckenstein \cite{CR93} have recently clarified
the nature of the approximation by 
recognizing that the NCA equations can be obtained
by taking a certain large-$N$, large $k$-limit for a generalized $SU(N)_s
\times SU(k)_f$ Anderson model. This model has precisely the same form as
the equilibrium version of 
Eq.~(\ref{calc.Anderson}) (with $V=0$ and no $\sigma$-index), 
but the spin and flavor indices take the values
$\alpha = 1, \dots, N$ and $i = 1, \dots k$, and the constraint becomes
\begin{equation}
  \label{eq:generalconstraint}
  \sum_{\alpha = 1}^N f_\alpha^\dagger f_\alpha +
\sum_{i = 1}^k b^\dagger_i b_i = 1 \; .
\end{equation}
(In the limit $\varepsilon_f < 0$, ${\cal V}_\sigma/|\varepsilon_f|
\ll 1$, this becomes equivalent to the $SU(N)_s \times SU (k)_f$
Coqblin-Schrieffer model.) They calculated the impurity contribution
to the partition function, $Z_{imp}$, as a functional integral using
an approximation reminiscent of a saddle point approximation,   and showed
that the limit $N \to \infty$, $k \to \infty$ at fixed $\gamma = k/N$
yields precisely the NCA equations for this generalized model.
[It was pointed out in Ref.~\onlinecite{PGKS}), however, 
that the model does not have a true large-$N$ saddle point, so
that $1/N$ corrections cannot be computed systematically.]
Moreover, the scaling dimensions which the ``saddle point'' yields for the
leading spin and flavor operators, $\Delta_s = 1/ (\gamma + 1)$ and 
$\Delta_f = \gamma/(\gamma + 1)$, 
turn out to agree precisely, for arbitrary $N$
and $k$ (with $ k \ge 2$), with the exact results derived by Affleck and
Ludwig \cite{AL94} for the $U(1)_c \times SU(N)_s \times SU(k)_f$
Kondo model, with the impurity spin transforming in an arbitrary
$SU(N)$ representation. This is useful, since $\Delta_s$, for example,
determines the leading critical exponent of the conduction electron
self-energy [and hence of the conductance in the RB experiments], 
as can be seen directly  in the CFT approach by doing perturbation theory
in the leading irrelevant operator. Moreover, for $N=k$, these
exponents are actually {\em independent} of $N$. 

However, it should be 
noted that many other properties
are not the same for $N=k=2$ as for $N,k = \infty$ at fixed $k/N = \gamma$. 
These include for example the temperature dependence
of the entropy and  the detailed form of the frequency dependence of the 
$T$-matrix (which depends on the amplitudes of
subleading terms in an expansion in the leading
irrelevant operator), etc. 
In particular, the Anderson model and the NCA approximation  
break particle hole symmetry, whereas the original 2CK model and
its CFT solution do not.  

These caveats should be kept in mind when when comparing the NCA results of
HKH and our CFT results for the universal scaling function $F$ of
Eq.~(\ref{lambdan}); they imply that it would be unreasonable to expect
``perfect agreement''.  Nevertheless, the NCA does have the distinct advantage
that when combined with the Keldysh technique, it deals with the
non-equilibrium aspects of the problem in a more direct way than our CFT
approach, and is able to go beyound the weakly non-equilibrium regime $(V \ll
\Tk$) (i.e.\ it incorporates some of the corrections to scaling mentioned in
section \ref{devfromscaling}). For our purposes, the NCA is thus to be regarded
as an approximation which happens to give correctly the critical exponents
near the non-Fermi-liquid fixed point for some (but not all) quantities, and
which thus interpolates between the $T \ll \Tk$ and $T \gg \Tk$ regimes (in
the latter it always works well, due to its perturbative origin).

\subsection{Electron Self-Energy}
\label{sec:selfNCA}

One would expect that the most direct comparison between
CFT and the NCA could be obtained 
by comparing %[see \fig{fig:selfenergy}]
the retarded self-energy $\Sigma^{\sss R} (\omega)$
for conduction electrons at $V=0$, 
calculated from the NCA, with that from CFT 
[essentially the function $\tilde \Gamma (x)$ of \Eqs{ft.tildeGx}
and (\ref{ft.selfenergy})].
However, the usefulness of this comparison is somewhat 
diminished by the fact that the NCA self-energy
is not a symmetric function of frequency,
which is a result of using the asymmetric Anderson model. This 
asymmetry disappears when calculating the conductance,
because $I_{tun} (V) = I_{tun} (-V)$ in \protect\Eq{NCAcurrent} even if
$A_d (\omega) \neq A_d (\omega)$,
meaning that the zero-bias conductance is the more
meaningful quantity to compare (see next section). Nevertheless,
for $\omega < 0$, the CFT and NCA results agree very well,
%[\fig{fig:selfenergy}].
see Fig.~4 of Ref.~\onlinecite{HKH95}.

\subsection{Impurity Spectral Function $A_d (\omega)$}

The NCA result for the impurity spectral function $A_d (\omega )$ is shown in 
Figure~\ref{fig:NCAVT}, which  is very instructive,
in that it illustrates what when $eV \gg \Tk$
(a regime not accessible
to CFT), the Kondo resonance
 splits into two  (as also found in \cite{WM94}
for a related model).
($\Tk$  is defined as the width
at half maximum of the $V=0$ impurity spectral function
at the lowest calculated $T$.)
 However, note that even for $V \simeq \Tk$, this splitting
has not yet set in, illustrating that non-equilibrium
effects are not important for $eV < \Tk$.
 This is the main justification
for the the approach followed in section~\ref{sec:HeffII}
of  neglecting all $V/\Tk$ corrections to the scattering
amplitudes.

\subsection{Possible Improvements on the NCA}
\label{Parcollet}
Very recently, Parcollet, Georges, Kotliar and Sengupta (PGKS) \cite{PGKS}
have devised a novel approximation scheme which avoids the problems of the NCA
mentioned in Section~\ref{NCAlargeN} and is based on yet another
generalization of the multi-channel Kondo model:
\bea
\nonumber
  H_{PGKS} &=&  \sum_{p, \alpha, i} 
\ve_p  c^\dagger_{p \alpha i} c_{p \alpha i}
\\   \label{eq:Hamilton-PGKS}
& & + 
{\cal V}_K  \sum_{p, \alpha, \beta, i}  \left( f^\dagger_\alpha
f_\beta - \delta_{\alpha \beta} Q / N \right) 
c^\dagger_{p \beta i} c_{p \alpha i} \; 
\eea
Here $f_\alpha$, $\alpha = 1, \dots, N$ are a set
of Abrikosov fermions satisfying the constraint
\begin{equation}
  \label{eq:Abrikosovconstraint}
  \sum_{\alpha = 1}^{N} f^\dagger_\alpha f_\alpha = Q \; ,
\end{equation}
which describe an impurity spin transforming as an antisymmetric
representation of $SU(N)$ of rank $Q$; the conduction electrons with spin
index $\alpha = 1, \dots, N$ and flavor index $i = 1, \dots , k$ transform
under the fundamental representation of the $SU(N)_s$ and $SU(k)_f$ groups.
PGKS showed that in the limit $N \to \infty$, $k \to \infty$, $Q \to \infty$,
with $k/N = \gamma$ and $q_0 = Q/N$ held fixed, the model has
a {\em true}\/ saddle point,  and that its saddle point equations 
are again identical with the NCA equations (bosonic operators
$b_i$ corresponding to those of the NCA arise via a Hubbard-Stratonovitch
decoupling of the Kondo interaction). Nevertheless
the approach of PGKS, 
though reminiscent of the NCA, is not fully equivalent to
it, since the constraint
(\ref{eq:Abrikosovconstraint}) evidently
differs from the constraint (\ref{eq:generalconstraint}) of the NCA.
[For $Q=1$ the constraints would be equal if in the NCA
 the limit $\varepsilon_d \to - \infty$ is taken, since
then the boson states are energetically so unfavorable that
they are never occupied, implying that the second term in
(\ref{eq:generalconstraint}) can be dropped. In this limit, 
the NCA becomes  particle-hole symmetric.]

The scheme of PGKS shares with the NCA the property that in the limit $N \to
\infty$, $k \to \infty$, the scaling exponents $\Delta_s$ and
$\Delta_f$ for the spin and flavor fields
are equal to the exact values ${N \over N +k}$ and ${k \over N+k}$ of
CFT.  In addition, the PGKS scheme has several advantages over the NCA.
Firstly, for $q_0 = 1/2$, {\em particle-hole symmetry is maintained}\/ for all
$N$ and $\gamma$, including the saddle point at $N= \infty$ and the 2-channel
Kondo case of present interest, $N=k= 2$, $Q=1$, for which
(\ref{eq:Hamilton-PGKS}) is just is the standard 2-channel Kondo Hamiltonian
in the Abrikosov-fermion representation.  Secondly, since it is based on a
{\em true}\/ saddle point, $1/N$ corrections can systematically be
incorporated by calculating fluctuations about the saddle.  Thirdly, if
generalized to the non-equilibrium case, it could be used to obtain a $1/N$
expansion also for the non-equilibrium properties of interest in the
Ralph-Buhrman nanoconstriction experiments, since the non-equilibrium
conductance can be computed within a Lagrangian (Keldysh) framework.  More
precisely, by including fluctuations about the saddle, it should in principle
be possible to obtain an expansion in powers of $1/N$ of the universal scaling
function $\Gamma$ in \Eq{universalF} [this expansion should not be confused,
though, with that of Eq.~(\ref{lambdan})]: \be
\label{largeNF}
{G(V,T) - G(0,T)\over B T^{1/2}} =\sum_{j=0}^{\infty} \ ({1\over N})^j \, 
\bar \Gamma_j [T/\Tk, V/T] \; .
\ee
 The advantage of such an  $1/N$ expansion is
that each term would go beyond the scaling regime, i.e.\ 
cover the entire crossover of the Kondo model, from weak all the way to strong
coupling, or correspondingly from $T \gg T_K$ to $T \ll T_K$.
Moreover, in the extreme  scaling regime, i.e.\ in the limit  $T \ll T_K$,
we expect that  $\bar \Gamma_0$ should coincide exactly with
the CFT result for $F$. The reason is that the latter
is in this limit determined {\em completely}\/ by
the scaling dimension $\Delta_s$, and that in PGKS's approach this dimension
is reproduced  {\em at}\/ the saddle point, at which 
all $1/N$ corrections vanish and all their results become exact.

It would thus be very interesting to repeat HKH's calculation
using PGKS's approach.

\section{$V$-dependent corrections to $ \tilde U_{\eta \eta'}$}
\label{app:Vcorr}

In this appendix, $V$-dependent corrections to the
scattering amplitudes $ \tilde U_{\eta \eta'}$ are discussed.

A key assumption made throughout this paper was that
the scattering amplitudes  that describe scattering  in 
the \nflr\
are $V$-{\em independent,}\/ for reasons given in section~\ref{sec:HeffII}.
However, a simple poor man's scaling
argument shows that this assumption can not be correct in general:
If $V>T$, then the RG flow will eventually be cut off at an energy scale
of order $V$. In the poor man's RG approach, this is implemented
by replacing the renormalized bandwidth 
$D'$ by $V$ in the effective interaction vertex.
This means that the effective renormalized Hamiltonian now
is explicitly $V$-dependent, implying that
the same will be true for its scattering amplitudes.

Intuitively, the  $V$-dependence 
arises because when $V \neq 0$, the difference in Fermi energies
of the $L$- and $R$ leads causes the Kondo peak in the density of states to 
split \cite{MWL93,WM94} into two
separate peaks (at energies $\mu \pm \half e V$, 
(see \fig{fig:NCAVT} of Appendix~\ref{NCA}, taken from \cite{HKH95}).
However,  in the \nflr, this $V$-dependence
can nevertheless be neglected, because 
when $V/\Tk \ll 1$, the  splitting of the Kondo peak by $eV$ 
is negligible compared to its width, which is $\propto \Tk$
(said differently, then $V \neq 0$ cuts off the RG
 flow  at a point sufficiently close to the \nfl\ fixed
point that the latter still governs the physics). 

To investigate the onset of Kondo 
peak-splitting effects with increasing  $V$ but still in the \nflr,
we use the same kind of
arguments as the ones used by AL to find the leading $T/\Tk$ term
in $G_{\eta \eta'}$ (see 
%%IIIsections~\ref{perturbationTTk} and~\ref{identifyPhi} of 
paper~III): $V \neq 0$ breaks a symmetry
of the system (namely $\sigma = L \leftrightarrow R$),
and the breaking
of a symmetry allows boundary operators
to appear in the action describing
the neighborhood of the fixed point that had been previously
forbidden (for extensive applications of this principle,
see \cite[section III.C]{AL92b}). 

To find the form of the leading $V\neq 0$ boundary operator,
we argue as follows: $V$ enters the formalism only via $Y_o$
[see \Eq{eq4.Yo}], which takes the following form when
written in terms of the fields $\psi_\eta$ of
\Eq{eq4.psiex} or $\lpsi_\eta$ of \Eq{psievenodd}:
\widetext
\bea
\label{cl.Yoapp}
        Y_o &=& \toh e V 
        \sum_{\eta}  \int_{-\infty}^{\infty} 
         \! {\textstyle {dx \over 2 \pi}}
        \psi_{\eta}^{\dagger} (ix)
        \sigma \psi_{\eta} (ix) \; , 
\\
\label{cl.Yoevenoddapp}
            &= & \toh e V 
        \sum_{\bar \alpha \bar i }  \int_{-\infty}^{\infty} 
         \! {\textstyle {dx \over 2 \pi}}
        \left[ \lpsi_{e \bar \alpha \bar i }^{\dagger} (ix)
         \lpsi_{o \bar \alpha \bar i } (ix) \; + \; 
         \lpsi_{o \bar \alpha \bar i }^{\dagger} (ix)
         \lpsi_{e \bar \alpha \bar i } (ix)  \right] \; .
\eea
\narrowtext
[For the second line we used
$(N \sigma^z N^{-1})_{\bar \sigma \bar \sigma'} 
= \sigma^x_{\bar \sigma \bar \sigma'} $, with 
$N$ given by \Eq{cl.N}.]
\Eq{cl.Yoevenoddapp} shows that $Y_o$ {\em mixes even and odd channels.}\/
Since the CFT solution was formulated only
in the even sector, the present model\footnote{However,
related models exist which can be treated exactly
by CFT even if $V \neq 0 $, for example
 the model used by Schiller and Hershfield
in \cite{SH95}. There, the pseudospin index is also the
$L$-$R$ index (i.e. the interaction 
matrix elements are ${1\over 2} \vec \sigma_{\sigma \sigma'}
\cdot \vec S$), which means that 
 $$Y_o 
= \toh e V 
        \sum_{\alpha i}  \int_{-\infty}^{\infty} 
         \! {\textstyle {dx \over 2 \pi}}
        \psi_{\alpha i }^{\dagger} (ix)
        \sigma^z_{\alpha \alpha'}  \psi_{\alpha' i} (ix) 
        = \;  e V  \int_{-\infty}^{\infty} 
         \! {\textstyle {dx \over 2 \pi}} J^z_{s } (i x)$$
where $\vec J_s  $  is the spin current.
 %%III (see \Eq{E.defs}  of paper~III).  
Now, in this case it is easy to find the exact $Y$-operator
in the presence of the Kondo interaction. 
$Y$ must both commute with $H$ and reduce to 
$Y_o$ when the interaction is switched off.
This is evidently satisfied by $Y = e V \int_{-\infty}^{\infty} 
         \! {\textstyle {dx \over 2 \pi}} \J_{s \sL}^z$,
where $ \J_{s \sL}^z$ is the $z$-component of the new spin current 
$\vec \J_{s } (ix) = \vec J_{s } (ix) + 2 \pi \delta (x) \vec
S$ (see %%III \Eq{al.shiftx} of 
paper~III). Thus, in the combination
$H - Y$ that occurs in the density matrix $\rho$,
$e V$ simply plays the 
role of a {\em bulk}\/ magnetic field in the 
$z$-direction, which can be gauged
away exactly by a gauge
transformation \cite[eq.~(3.37)]{AL91b}.
Hence in this model, non-equilibrium
properties {\em can}\/ be calculated exactly
using CFT.}
can strictly speaking not be solved exactly by CFT for $V \neq 0 $,
unless one neglects all effects of $Y_o$ on 
the fixed point physics. 

The form (\ref{cl.Yoevenoddapp}) for $Y_o$ leads us to conjecture
that the leading boundary operator  appearing in 
the action  when $V \neq 0$ will have the form\footnote{
It is easy to check that the operator  $J_{eo}$ 
is indeed allowed at the boundary:
it must be the product $\Phi_e \Phi_o$ of
boundary operators in the even and odd sectors,
with quantum numbers (charge,spin,flavor)= 
$(Q_e, j_e, f_e) $ = $(- Q_o, j_o, f_o)$ =
$(\pm 1, \half, \half)$. $\Phi_o$, which lives
on a free boundary (since the odd sector is free),
is simply the free fermion field $\psi_{o \alpha i}$ in  the odd sector.
$\Phi_e$ must live
on a Kondo boundary, which indeed does
allow a boundary operators with the quantum
quantum numbers $(\pm 1, \half, \half)$, as may be checked
by AL's double fusion procedure (see table~1c of \cite{Lud94a}).}
\widetext
\be
\label{cl.YoboundaryII}
         \delta S_{\sss V}
         = \lambda_{10} 
         { V \over \Tk }
        \sum_{\bar \alpha \bar i }  \int_0^{\beta}
         d \tau 
        \left[ \lpsi_{e \bar \alpha \bar i }^{\dagger} (\tau)
         \lpsi_{o \bar \alpha \bar i } (\tau) ) \; + \; 
         \lpsi_{o \bar \alpha \bar i }^{\dagger} (\tau)
         \lpsi_{e \bar \alpha \bar i } (\tau )  \right] \; 
\\
          \equiv 
         \; \lambda_{10} 
         { V \over \Tk } \int_0^{\beta}          d \tau 
         J_{eo} (\tau) \; .     
\ee
\narrowtext
Since the ``even-odd'' current $ J_{eo}$ defined
on the right-hand-side has scaling
dimension $1$ (i.e.\ $\bar \alpha_1 = 1, \alpha_0 = 0$), 
$\delta S_{\sss V}$ 
has scaling dimension
zero (see %%XXX \Eq{ft.Obound} of 
paper~III),
 and is therefore a marginal perturbation. 
This means that even for $V/\Tk \ll 1$,
if $T/V$ is made sufficiently small, the system will
eventually flow away from the \nfl\ fixed point,
at a cross-over temperature  $T^\ast_{\sss V}$, say.
 However, since
this perturbation is marginal, it  only grows logarithmically slowly
as $T$ is decreased, so that $T^\ast_{\sss V}$
will be {\em very}\/ small. Therefore, the \nflr,
in which one has both $V,T \ll \Tk$ and $T>T_{\sss V}$, can be rather large.
The lack of deviations from scaling in the data 
for the low-$T$ regime (see Section~VI of paper~I)
indicate that $T^\ast_{\sss V}$
is smaller than the lowest temperatures obtained in the experiment.

How does $\delta S_{\sss V}$ affect the scattering amplitudes?
First note that the $V$-dependence of $\delta S_{\sss V}$
enters in a very simple way, namely as a ``parameter''
that governs the strength of the perturbation.
 Therefore, the methods
of section~\ref{sec:UfromGF}, which  extract
$\tilde U_{\tilde\eta \eta} (\ve)$ from an {\em equilibrium}\/  CFT
Green's function, are still applicable.

Standard CFT arguments show that the effect of $\delta S_{\sss V}$ on
$\overline G_{\bar \eta \bar \eta'}$ is to simply cause a 
rotation\footnote{See, for example,
\protect\cite{CKLM94}\protect. At  $T = 0$,
one can prove that 
 $\delta S_{\sss V}$ generates such a rotation by
closing the $\int_{-\infty}^{\infty} d \tau$ integral 
along an infinite semi-circle in the lower half-plane
(this is allowed, because   $J_{eo} (z) \sim z^{-2} \to 0$ along 
such a contour \cite[Eq.(2.19)]{KZ84}); having closed the contour,
$\delta S_{\sss V}$ has precicely the form
 required for  a generator of $e/o$ rotations.}
of the $\bar \sigma = e/o$ 
indices of the outgoing $\bar \eta$-fields relative
to the incident $\bar \eta'$-fields by
\bea
\label{cl.Vrotate}
        \overline R_{\bar \eta \bar \eta'} (V)  &=& 
        \delta_{\bar \alpha \bar \alpha'}
        \delta_{\bar i \bar i'}
        \left( \begin{array}{cc}
                \cos \theta_{\sss V} &  - i \sin \theta_{\sss V} \\
                - i \sin \theta_{\sss V} & \cos \theta_{\sss V}
                \end{array}  \right)_{\bar \sigma \bar \sigma'} \; ,
\\
  \nonumber
   \theta_{\sss V} & \equiv &
        \arctan\left( {c V \over \Tk} \right) \; .
\eea             
Here $\theta_{\sss V}$ is simply a convenient
way to parametrize the rotation \cite{CKLM94},
and $c$ is a constant.
Thus, the effect of  $\delta S_{\sss V}$ can be incorporated
by replacing the scattering amplitude 
        $\tilde {\overline U}_{\bar \eta \bar \eta'} (\ve)$
of \Eq{Ueo} by $\overline R_{\bar \eta \bar \eta''} (V) 
       \tilde   {\overline U}_{\bar \eta'' \bar \eta'} (\ve)$.
Evidently,  the final scattering amplitude
$\tilde U_{\tilde\eta \eta} (\ve)$ of \Eq{cl.UN}
will now be $V$-dependent. 

It turns out that for the simple form (\ref{cl.VKroha})
used for the backscattering matrix $V_{\sigma \sigma'}$,
this extra $V$-dependence ``accidentally'' cancels out\footnote{This
can be seen from by replacing $\overline U$ by
$\overline R \overline U$ in \protect\Eq{cl.P} for $P_\eta (\ve)$,
and checking that $\overline R^\dagger N \sigma^z N^\dagger \overline
R = \overline R^\dagger \sigma^x \overline R = \sigma^x$,
which is independent of $V$
because $\overline R$ generates rotations around the
$x$-axis in the $e/o$ indices. However, if $V$ and $N$ are
more complicated than in \Eqs{cl.VKroha} and (\ref{cl.Nexp}),
the $V$-dependence will clearly not cancel out.} in
\Eq{cl.P} for $P_{\eta} (\ve)$, so that \Eq{cl.P} remains
valid as written.  However,
for more general forms of $V_{\sigma \sigma'}$,
it survives. To lowest order in $V / \Tk$,
there   will then be a contribution to the conductance
of the form $(V/\Tk) T^{1/2} \Gamma_1 (V/T) \equiv
 T^{3/2} \Gamma_2 (V/T)$. However, this is evidently 
only a {\em subleading}\/  correction to the scaling function
of \Eq{cl.condGscale}. It is of the same order 
as corrections arising from subleading irrelevant
operators of the equilibrium theory, that
we have argued in section~\ref{p:fitting} would  not be worth while
calculating since there are too many independent ones.

To summarize the results of this appendix: when $V \neq 0$,
corrections to the scattering amplitudes
$ \tilde U_{\eta \eta'}(\ve')$
 of order $V/\Tk$ can arise; however,
they only give rise to {\em subleading}\/ corrections  to the
scaling function $\Gamma (v)$.

%\newpage 

%\newpage 

%\newpage 

%\begin{figure}
%\centerline{\psfig{figure=4_1.eps,height=2cm}}
%\caption[A one-dimensional scattering problem:
%Free fields are incident from the right,
%scattered at the origin and outgoing to the left.]{[[5.1]]
%A  one-dimensional scattering problem:
%Free fields are incident from the right,
%scattered at the origin and outgoing to the left.
%\label{fig:scatt}}
%\end{figure}

%\newpage 

%\newpage 

%\newpage 

%\newpage 

%\newpage 
\begin{figure}
\vphantom{.}
\vspace{1cm}
\centerline{\psfig{figure=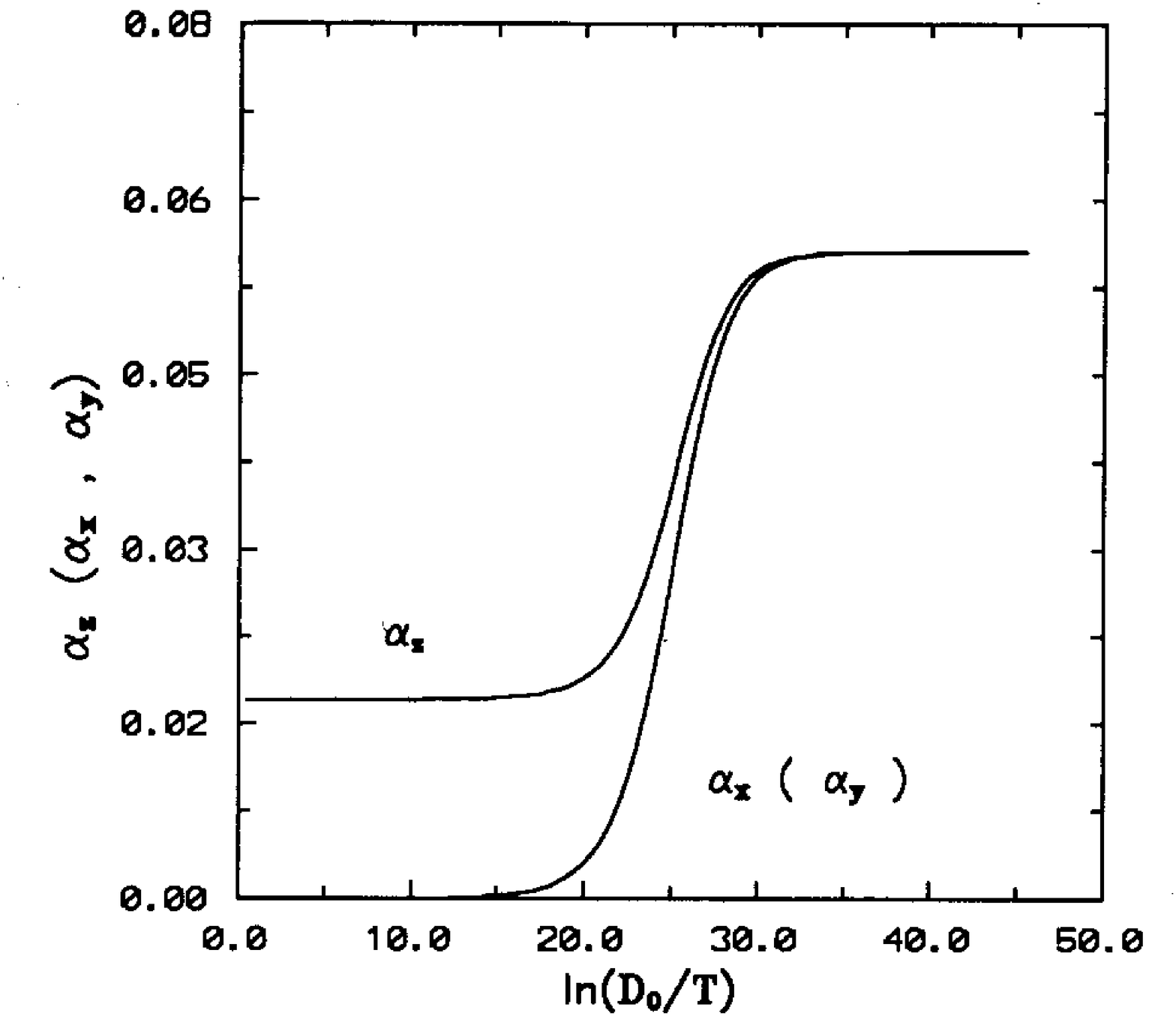,width=0.9\linewidth}}
\vphantom{.}
\vspace{1cm}
\caption[Scaling trajectories of the
coupling constants for the non-magnetic
Kondo problem.]{\label{fig:isotropicscaling}
%[[4.5]] 
Scaling trajectories of the matrix norms
${\mbox{\protect\boldmath{$\alpha$}}}^A \equiv \mbox{Tr} 
[ ({\mbox{\underline{v}}}^A)^2  ] 
$ ($A=x,y,z$),
calculated numerically for the case $ N_f=3 $. 
All three norms tend to
the same value, in accord with eq.~(\protect\ref{bk.isotropic}\protect).
Consult \protect\cite{Zar95}\protect,
from which this figure was taken, for details regarding
the initial parameters used.
}
\end{figure}

%\begin{figure}
%\centerline{\psfig{figure=flow.eps,height=8cm}}
%\caption[Flow-diagram in $\Delta-T$ plane]{
%[[new]]
%\label{fig:Deltaflow}
%Schematic flow-diagram for the flow of the 2-channel
%Kondo model in the $\Delta-T$ plane. The arrows indicate
%the direction of flow as $T$ is decreased. The origin
%corresponds to the non-Fermi-liquid fixed point,
%with $(v_{\sss K}, \Delta) = ( v_{\sss K}^\ast, 0)$.
% $\Delta = \pm \infty$ correspond
%to Fermi-liquid fixed points.
%}
%\end{figure}

%\newpage 
\begin{figure}
\vphantom{.}
\vspace{1cm}
\centerline{\psfig{figure=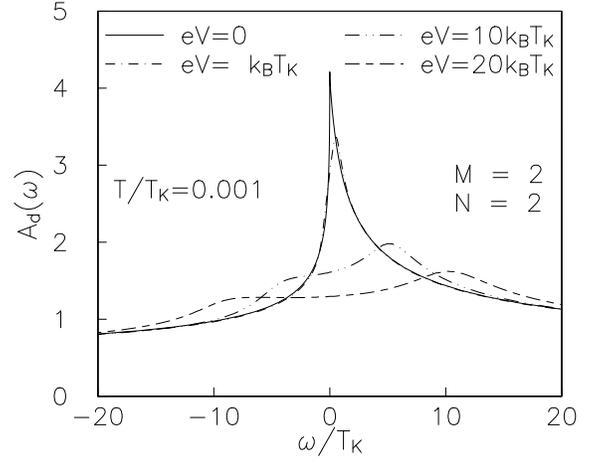,width=0.7\linewidth}}
\vphantom{.}
\vspace{1cm}
\caption[NCA results for the Kondo resonance in the
impurity spectral function $A_d (\omega)$, 
calculated at  $V = 0$ and  $T/ \Tk = 0.001$.]{
%[[9.3]] 
\label{fig:NCAVT} The Kondo resonance in the
impurity spectral function $A_d (\omega)$, 
calculated  $T/ \Tk = 0.001$
using the NCA \protect\cite{HKH95}. For our
purposes the most important feature of this figure is that
the Kondo peak does not start to split for $eV < \Tk$.}
\end{figure}

%\begin{figure}
%\vspace{8cm}
%\caption[NCA results for $\Sigma (\omega, T)$,
%the imaginary part of the retarded self-energy for
%conduction electrons at $V= 0$.]{\label{fig:selfenergy}
%%[[9.2]] 
%Here $\Sigma (\omega, T)$ is the
%imaginary part of the retarded self-energy for
%conduction electrons at $V= 0$, 
%calculated using CFT and the NCA \protect\cite{HKH95}. For the NCA curves,
%temperatures are given in units of $.001 \Tk$.
%The CFT curve corresponds to $T / \Tk \to 0$. The asymmetry
%in the NCA curves is a result of using the asymmetric
%Anderson model. The CFT and lowest-$T$ NCA
%curves have been rescaled such that the asymptotic slope
%of the $\omega < 0$ function is 1, i.e.\ the
%CFT curve corresponds to the function $( \tilde \Gamma 
%(\tilde \gamma_1 x) /  \tilde \Gamma (0) -1) $
%of \protect\Eq{ft.tildeGx}, with $\tilde \gamma_1$ an
%appropriately chosen constant.
%}
%\end{figure}

\widetext
\end{document}